\documentclass[twocolumn,superscriptaddress,floatfix,preprintnumbers,aps,nofootinbib,hyperref]{revtex4-2}
\usepackage[utf8]{inputenc}
\usepackage{epsf}
\usepackage{latexsym,amssymb,euscript}
\usepackage[dvips]{graphicx}
\usepackage{amsmath}
\usepackage{nicefrac}
\usepackage{slashed}
\usepackage{booktabs}
\usepackage[linktocpage]{hyperref}
\usepackage{braket}
\usepackage{chngcntr}
\usepackage{bbold}
\usepackage{graphics}
\usepackage{graphicx}
\usepackage{mciteplus}
\usepackage{caption}
\usepackage{subcaption}
\usepackage{adjustbox}
\usepackage{ragged2e}
\usepackage{placeins}
\usepackage[titletoc]{appendix}
\usepackage[normalem]{ulem}
\graphicspath{{./figures/}}
\hypersetup{
 linktocpage = false,
 urlcolor = urlblue,
 colorlinks = true,
 linkcolor = urlblue,
 anchorcolor = citegreen,
 citecolor = citegreen,
 pdfstartview = {XYZ null null 1.25} 
           }
\usepackage[left=2cm, right=2cm]{geometry}
\usepackage{pstricks}
\usepackage{color}
\usepackage{xcolor}
\definecolor{urlblue}{rgb}{0.2,0.4,0.7}
\definecolor{citegreen}{rgb}{0,0.4,0.2}
\definecolor{linkred}{rgb}{0.9,0.2,0.1}
\usepackage{float}
\usepackage{academicons}
\definecolor{orcidlogocol}{HTML}{A6CE39}
\usepackage{changepage}
\usepackage{fancyhdr}
\usepackage{epsfig}
\usepackage{letltxmacro}
\LetLtxMacro{\oldcite}{\cite}
\renewcommand{\cite}[1]{\mbox{\oldcite{#1}}}

\usepackage{orcidlink}

\newcommand{\drv}{{\rm d}}

\newcommand{\as}{\alpha_s}

\newcommand{\MSb}{\overline{\rm MS}}

\newcommand{\LL}{{\rm LL/LO}}

\newcommand{\NLLp}{{\rm NLL/NLO^+}}

\newcommand{\HENLOp}{{\rm HE}\mbox{-}{\rm NLO^+}}

\newcommand{\DY}{\Delta Y}

\newcommand{\F}{{\cal F}}

\newcommand{\Jpsi}{J/\psi}
\newcommand{\Yps}{\Upsilon}
\newcommand{\BCs}{B_c(^1S_0)}
\newcommand{\Bss}{B_c(^3S_1)}

\newcommand{\QXQq}{X_{Q\bar{Q}q\bar{q}}}

\newcommand{\TQQ}{T_{4Q}}
\newcommand{\TQc}{T_{4c}}
\newcommand{\TQcZpp}{T_{4c}(0^{++})}
\newcommand{\TQcOpm}{T_{4c}(1^{+-})}
\newcommand{\TQcTpp}{T_{4c}(2^{++})}
\newcommand{\TQb}{T_{4b}}
\newcommand{\TQbZpp}{T_{4b}(0^{++})}
\newcommand{\TQbOpm}{T_{4b}(1^{+-})}
\newcommand{\TQbTpp}{T_{4b}(2^{++})}

\newcommand{{\HFNRevo}}{\tt HF-NRevo}

\newcommand{{\Jethad}}{\tt JETHAD}
\newcommand{{\symJethad}}{\tt symJETHAD}
\newcommand{{\psymJethad}}{\tt (sym)JETHAD}
\newcommand{{\Hell}}{\tt HELL}
\newcommand{{\RadISH}}{\tt RadISH}
\newcommand{{\Pegasus}}{\tt QCD-PEGASUS}
\newcommand{{\HOPPET}}{\tt HOPPET}
\newcommand{{\QCDNUM}}{\tt QCDNUM}
\newcommand{{\APFEL}}{\tt APFEL}
\newcommand{{\APFELpp}}{\tt APFEL++}
\newcommand{{\APFELppp}}{\tt APFEL(++)}
\newcommand{{\EKO}}{\tt EKO}
\newcommand{{\FeynCalc}}{\tt FeynCalc}

\allowdisplaybreaks

\bibpunct{[}{]}{,}{n}{}{,}

\begin{document}

\title{\mbox{All-charm tetraquarks at hadron colliders: A high-precision fragmentation perspective}}

\author{Francesco~Giovanni~Celiberto\,\orcidlink{0000-0003-3299-2203}} 
\email{francesco.celiberto@uah.es}
\affiliation{Universidad de Alcal\'a (UAH), Departamento de F\'isica y Matem\'aticas, E-28805 Alcal\'a de Henares, Madrid, Spain}

%==========================

\begin{abstract}
We present the {\tt TQ4Q2.0} fragmentation functions for the production of all-heavy (fully heavy) $S$-wave tetraquarks ($T_{4Q}$) with scalar ($0^{++}$), axial-vector ($1^{+-}$), and tensor ($2^{++}$) quantum numbers in high-energy hadronic collisions.
This work extends the previous {\tt TQ4Q1.1} framework by incorporating nonconstituent heavy-quark contributions and introducing a replica-based uncertainty-quantification strategy derived from multiscale variations (MHOUs).
The construction follows a nonrelativistic QCD factorization approach, combining gluon- and heavy-quark-initiated fragmentation channels at leading power.
Initial-scale inputs are modeled through updated potential-inspired wave functions, while the subsequent Dokshitzer-Gribov-Lipatov-Altarelli-Parisi (DGLAP) evolution is performed via the threshold-aware {\HFNRevo} scheme.
A comprehensive systematic analysis of uncertainties is carried out, with contributions from color-composite long-distance matrix elements (LDMEs) and perturbative multiscale inputs.
The resulting {\tt TQ4Q2.0} grids, publicly released in {\tt LHAPDF6} format, provide the first complete phenomenological set for all-heavy exotics, enabling precise studies of all-charm tetraquark production and jet-associated observables within the {\tt JETHAD} environment.
This article completes the high-energy resummation-driven generation of the {\tt TQ4Q} program and establishes a definitive baseline for future collider-oriented analyses of all-heavy multiquark dynamics.
\end{abstract}

\maketitle

\setcounter{tocdepth}{3}
\begingroup
\renewcommand{\baselinestretch}{0.75}\footnotesize
\parskip 3.0pt
\tableofcontents
\renewcommand{\baselinestretch}{1.0}\normalsize
\endgroup

\parskip 6pt

%==========================
\section{Hors d'{\oe}uvre}
\label{sec:introduction}
%==========================

The structure of hadronic matter extends far beyond the minimal quark-antiquark and three-quark configurations that define conventional mesons and baryons. 
Over the past two decades, a growing body of experimental evidence has revealed the existence of multiquark states whose internal organization cannot be captured within the traditional quark model. 
These exotic configurations, including tetraquarks and pentaquarks, expose a regime where the mechanisms of confinement and hadron formation operate in a more intricate and less understood way. 
In systems containing heavy quarks, this complexity becomes more tractable: their large masses naturally suppress relativistic effects and enable a controlled separation of dynamical scales, making them ideal probes of strong-interaction dynamics. 
In this sense, exotic heavy hadrons can be viewed as a \textit{portal} into the core of the strong force, where perturbative and nonperturbative phenomena coexist and shape the formation of complex bound states.

The exploration of this regime is driven by high-energy collider experiments, most notably at the Large Hadron Collider (LHC), and will be further advanced by future facilities such as the Electron-Ion Collider (EIC)~\cite{AbdulKhalek:2021gbh,Khalek:2022bzd,Hentschinski:2022xnd,Amoroso:2022eow,Abir:2023fpo,Allaire:2023fgp} and the Future Circular Collider (FCC)~\cite{FCC:2025lpp,FCC:2025uan,FCC:2025jtd}. 
These machines provide access to extreme kinematic conditions where heavy quarks are abundantly produced and subsequently hadronize into a wide spectrum of bound states, including exotic ones. 
Within this environment, multiquark systems are not rare anomalies but natural outcomes of high-energy QCD dynamics, allowing for systematic investigations of their production patterns and internal structure.

In parallel, theoretical developments have significantly improved our ability to describe heavy-hadron production in a quantitatively reliable way. 
Advances in QCD factorization, together with the resummation of large logarithmic contributions at all orders, have opened the possibility of performing precision studies of collider observables involving heavy flavors~\cite{Feng:2020riv,Feng:2020qee,Feng:2023agq,Feng:2023ghc,Bai:2024ezn,Bai:2024flh,Nejad:2021mmp,Celiberto:2023rzw,Celiberto:2024mab,Celiberto:2024beg,Celiberto:2025dfe,Celiberto:2025ipt,Celiberto:2025ziy,Celiberto:2025vra,Wang:2025hex,Liu:2025mxv,Feng:2026orq}. 
This progress enables a consistent treatment of multiscale dynamics, where perturbative radiation and nonperturbative hadronization effects are intertwined, and provides the necessary framework to connect theoretical predictions with experimental measurements at an unprecedented level of accuracy.

Exotic hadrons can be broadly classified into two main categories: gluon-rich configurations, such as hybrids and glueballs~\cite{Kou:2005gt,Braaten:2013boa,Berwein:2015vca,Minkowski:1998mf,Mathieu:2008me,Chen:2021cjr,Csorgo:2019ewn,D0:2020tig}, and multiquark states, including tetraquarks, pentaquarks, and more complex bound systems~\cite{Gell-Mann:1964ewy,Jaffe:1976ig,Jaffe:1976ih,Jaffe:1976yi,Ader:1981db,Rosner:1985yh,Pepin:1998ih,Vijande:2011im,Esposito:2016noz,Lebed:2016hpi,Guo:2017jvc,Lucha:2017mof,Ali:2019roi}. 
While the former explicitly involve gluonic degrees of freedom, the latter are typically described in terms of multiquark Fock states with nonminimal valence content, whose internal organization remains an active area of investigation.

The modern era of exotic spectroscopy was triggered by the discovery of the $X(3872)$ by the Belle Collaboration in 2003~\cite{Belle:2003nnu}, later confirmed by several experiments~\cite{CDF:2003cab,LHCb:2013kgk,CMS:2021znk,Swanson:2006st}. 
This state, commonly interpreted as a hidden-charm tetraquark candidate~\cite{Chen:2016qju,Liu:2019zoy}, exhibits decay patterns that challenge a simple quarkonium interpretation, pointing instead to a more intricate internal structure. 
Subsequent observations, such as the $X(2900)$ reported by LHCb~\cite{LHCb:2020bls}, have further enriched the spectrum of exotic candidates, extending it to systems with open heavy flavor.

A major step forward was achieved with the observation of the doubly charmed tetraquark $T_{cc}^+$~\cite{LHCb:2021vvq,LHCb:2021auc}, whose proximity to the $DD$ threshold suggests a molecularlike configuration described within effective field theory approaches such as XEFT~\cite{Fleming:2021wmk,Dai:2023mxm,Hodges:2024awq,Mehen:2015efa,Braaten:2020iye}. 
In parallel, the observation of structures in the di-$\Jpsi$ invariant-mass spectrum, notably the $X(6900)$~\cite{LHCb:2020bwg}, has provided the first compelling evidence for all-heavy tetraquark candidates, possibly corresponding to scalar or tensor $|c\bar{c}c\bar{c}\rangle$ configurations~\cite{Chen:2022asf}, and potentially connected to higher excited states or alternative molecular interpretations in the same mass region~\cite{Agaev:2023rpj}.

Recent experimental developments have further sharpened this picture. 
The CMS Collaboration has performed the first spin-parity analysis of all-charm (fully charmed) tetraquark candidates observed in di-quarkonium channels, favoring a $J^{PC}=2^{++}$ assignment for the leading structure~\cite{CMS:2023owd,CMS:2025fpt,CMS:2026tiu,Zhu:2024swp}. 
This result disfavors loosely bound molecular scenarios and supports a compact diquark-antidiquark interpretation, providing valuable constraints on the internal dynamics of all-heavy multiquark systems. 
A comprehensive overview of current experimental progress and future prospects is given in Ref.~\cite{Nogga:2025qcm}.

Additional insight into the production environment of heavy quarkonia is provided by measurements of quarkonium-pair production at hadron colliders, which probe multiple-parton interaction (MPI) dynamics. 
ATLAS has measured prompt $\Jpsi$-pair production at $\sqrt{s}=8$TeV, extracting effective cross sections for double-parton scattering (DPS)~\cite{ATLAS:2016ydt}, while CMS has provided total and differential cross sections at $\sqrt{s}=7$TeV~\cite{CMS:2014cmt}. 
More recently, CMS reported the first observation of double $\Jpsi$ production in proton-lead collisions at $\sqrt{s}=8.16$TeV, separating single- and double-parton contributions across different kinematic regions~\cite{CMS:2024wgu}. 
These measurements offer an essential benchmark for understanding correlated heavy-quark production mechanisms in QCD.

From a theoretical perspective, systems containing two or more heavy quarks, such as doubly heavy $\QXQq$ and fully heavy $\TQQ$ states, provide particularly clean laboratories for investigating the strong interaction. 
In $\QXQq$ states, heavy quarks interact with light degrees of freedom in a controlled nonrelativistic regime, often forming diquarklike substructures. 
In contrast, $\TQQ$ configurations involve only heavy quarks and antiquarks in a $|Q\bar{Q}Q\bar{Q}\rangle$ Fock state, without valence light quarks, and can thus be regarded as a doubled, exotic analog of quarkonium (see, \emph{e.g.}, Ref.~\cite{Agaev:2023wua} for all-heavy tetraquarks and Ref.~\cite{Oncala:2025mqj} for related hybrid configurations with explicit gluonic degrees of freedom).

The nonrelativistic nature of these systems enables the application of theoretical tools originally developed for quarkonium physics. 
While charmonium is often described as the QCD analog of the hydrogen atom~\cite{Pineda:2011dg}, multiquark configurations such as $\QXQq$ and $\TQQ$ can be interpreted as more complex QCD nuclei or molecularlike systems, depending on the underlying dynamical picture~\cite{Maiani:2019cwl}.

Despite the substantial progress achieved in the study of their spectroscopy and decay properties, the production mechanisms of exotic hadrons remain far less understood. 
Only a limited number of approaches have been proposed, typically relying on model-dependent frameworks such as color evaporation~\cite{Maciula:2020wri} or hadron-quark duality~\cite{Berezhnoy:2011xy,Karliner:2016zzc,Becchi:2020mjz}. 
Other studies have investigated the role of multiparticle interactions in tetraquark production~\cite{Carvalho:2015nqf,Abreu:2023wwg} or explored possible signatures of high-energy dynamics in exotic formation~\cite{Cisek:2022uqx}, while complementary analyses have considered exclusive decays~\cite{Feng:2020qee} and photoproduction processes~\cite{Feng:2023ghc}. 
More recently, machine-learning-based approaches to multiquark bound states have also been proposed~\cite{Wu:2025wvv}.

In the bottom sector, experimental information is still scarce. 
The Belle Collaboration reported charged bottomoniumlike structures in $\Yps(5S)$ decays~\cite{Belle:2011aa}, suggesting the presence of exotic contributions, but no fully bottom or bottom-light tetraquark states have yet been firmly established. 
On the other hand, a resonance observed by the ANDY Collaboration at RHIC near $18.15$~GeV~\cite{ANDY:2019bfn} has been interpreted as compatible with predictions for $\TQb$ states~\cite{Vogt:2021lei}, while lattice QCD studies have explored bottom-charmed and doubly bottomed tetraquarks in Refs.~\cite{Francis:2018jyb,Padmanath:2023rdu,Bicudo:2015vta,Leskovec:2019ioa,Alexandrou:2024iwi}.

A particularly striking indication of the relevance of production dynamics is provided by the unexpectedly large cross sections observed for the $X(3872)$ at high transverse momentum by ATLAS, CMS, and LHCb~\cite{CMS:2013fpt,ATLAS:2016kwu,LHCb:2021ten}. 
These results point toward a production mechanism dominated by parton fragmentation, highlighting the need for a consistent theoretical framework capable of describing exotic hadron formation within high-energy QCD.

A systematic program for modeling all-heavy tetraquark fragmentation has been developed in a series of recent works.
The first-generation {\tt TQ4Q1.0} sets~\cite{Celiberto:2024mab} introduced collinear fragmentation functions (FFs) for scalar ($J^{PC}=0^{++}$) and tensor ($J^{PC}=2^{++}$) all-charm states, combining gluon- and charm-initiated channels through a hybrid construction based on NRQCD inputs and kinematic modeling.
This framework was subsequently upgraded to the {\tt TQ4Q1.1} family~\cite{Celiberto:2025dfe,Celiberto:2025ziy}, where a fully NRQCD-consistent treatment of short-distance coefficients was implemented, extended to include axial-vector ($J^{PC}=1^{+-}$) configurations, and generalized to both all-charm and all-bottom systems within a variable-flavor number scheme (VFNS)~\cite{Mele:1990cw,Cacciari:1993mq,Buza:1996wv} and the \emph{Heavy-flavor nonrelativistic-evolution} {\HFNRevo} methodology~\cite{Celiberto:2025euy,Celiberto:2024mex,Celiberto:2024bxu,Celiberto:2024rxa,Celiberto:2025xvy,Celiberto:2026rzi,Celiberto:2026zss}.
In this context, color-composite LDMEs~\cite{Feng:2020riv,Bai:2024ezn} derived from potential models were supplemented with systematic uncertainty estimates and consistently propagated to the FF level, resulting in the first fully uncertainty-aware description of tetraquark fragmentation, including both perturbative and nonperturbative contributions.

Building on these developments, the present work introduces the {\tt TQ4Q2.0} sets, which mark a decisive step toward a complete and high-precision description of all-heavy tetraquark production.
For the first time, all partonic fragmentation channels are included at the initial scale, encompassing not only gluon and constituent heavy-quark contributions, but also nonconstituent light- and heavy-quark channels.
As emphasized in Ref.~\cite{Bai:2024flh}, such contributions can become phenomenologically relevant in differential observables at collider energies and are therefore essential for a consistent precision-oriented framework.

A further key advancement is the implementation of a replica-based treatment of perturbative uncertainties (F-MHOUs), constructed through multiscale variations.
This strategy provides a dynamically correlated estimate of missing higher-order effects and defines a robust baseline for future extractions of nonperturbative parameters from data.
In addition, it opens the way to \emph{multimodal} approaches~\cite{Celiberto:2025ipt,Celiberto:2026rdk} and the application of modern statistical and machine-learning techniques in the analysis of fragmentation processes, where replica ensembles can be directly interfaced with data-driven methodologies.
This approach is inspired by, and runs in parallel with, well-established methodologies developed in the study of collinear and multidimensional hadron structure~\cite{Kassabov:2022orn,Harland-Lang:2018bxd,Ball:2021icz,McGowan:2022nag,NNPDF:2024dpb,Pasquini:2023aaf}.
Altogether, it represents a first step toward a data-driven determination of tetraquark fragmentation dynamics.

The resulting {\tt TQ4Q2.0} grids are publicly released in {\tt LHAPDF6} format~\cite{Buckley:2014ana}, providing the first fully differential, uncertainty-aware, and multichannel phenomenological set for all-heavy tetraquark production.
This release enables direct integration into collider analyses and facilitates systematic studies of jet-associated observables and event yields at the (HL-)LHC and future FCC facilities.

Altogether, these developments signal the transition from exploratory modeling to a precision-driven description of tetraquark fragmentation.
In this sense, the {\tt TQ4Q2.0} framework aims to establish a new reference standard for the study of all-heavy multiquark dynamics in high-energy hadronic collisions.

To validate our fragmentation framework in a realistic collider environment, we investigate the inclusive production of all-charm tetraquarks in association with a jet in proton-proton collisions at HL-LHC and FCC energies.
The reference process is analyzed within a $\NLLp$ hybrid formalism (HyF), combining next-to-leading order (NLO) collinear factorization with the resummation of high-energy logarithms beyond NLL accuracy.

This setup enables direct access to observables that are particularly sensitive to high-energy dynamics, such as rapidity intervals, jet correlations, and, crucially, expected event yields.
The latter play a central role in guiding experimental searches, as they provide actionable benchmarks for experimental analyses and help define realistic discovery prospects for all-charm tetraquark signals in collider environments.

Our predictions target current and future facilities, including the (HL-)LHC and projected FCC configurations, and are obtained within the {\Jethad} numerical framework, supplemented by its symbolic extension {\symJethad}~\cite{Celiberto:2020wpk,Celiberto:2022rfj,Celiberto:2023fzz,Celiberto:2024mrq,Celiberto:2024swu,Celiberto:2025csa,Celiberto:2026ooh}.
In this context, the publicly released {\tt TQ4Q2.0} grids provide the first complete phenomenological set for all-heavy exotics, enabling precision studies of all-charm tetraquark production and jet-associated observables.

This work completes the high-energy, resummation-driven development of our program and establishes a robust baseline for future collider-oriented analyses of all-heavy multiquark dynamics.
The {\tt TQ4Q2.0} framework provides a unified and phenomenologically grounded description of all-charm tetraquark production, bridging first-principles QCD dynamics with experimentally accessible observables.

The paper is organized as follows.
Section~\ref{sec:FFs} introduces the theoretical framework for heavy-flavor fragmentation, outlining its extension from quarkonium systems to all-heavy tetraquarks within NRQCD, and presents the construction of the {\tt TQ4Q2.0} functions for both charmed and bottomed states.
Section~\ref{sec:phenomenology} is devoted to the phenomenological analysis of tetraquark-jet production at HL-LHC and FCC energies within the HyF approach, with particular emphasis on event yields as a key guide for experimental searches.
Finally, Sec.~\ref{sec:conclusions} summarizes our findings and discusses future directions towards high-precision fragmentation studies.

%==========================
\section{Fragmentation of all-heavy tetraquarks}
\label{sec:FFs}
%==========================

This section begins with a concise overview of the main features of heavy-flavor fragmentation, covering heavy-light hadrons, quarkonium systems, and exotic bound states (Sec.~\ref{ssec:FFs_intro}). 
We then discuss the implementation of NRQCD in the construction of initial-scale inputs for both gluon- and constituent-heavy-quark fragmentation channels into $\TQQ$ tetraquarks (Secs.~\ref{ssec:FFs_NRQCD} and~\ref{ssec:FFs_initial_scale}). 
The section concludes with the timelike DGLAP evolution of the full {\tt TQ4Q2.0} FF set within the {\HFNRevo} framework (Sec.~\ref{ssec:FFs_TQ4Q20}), illustrating a modular and internally consistent approach to all-heavy tetraquark fragmentation.

%==========================
\subsection{Heavy-flavor fragmentation: Key concepts}
\label{ssec:FFs_intro}
%==========================

Unlike light hadrons, whose FFs are fully nonperturbative, heavy-flavor hadrons involve a fragmentation mechanism that is partly perturbative. 
The large heavy-quark mass introduces a short-distance scale within the perturbative QCD domain, so initial-scale heavy-hadron FFs necessarily contain perturbative ingredients.

For singly heavy-flavored hadrons such as $D$, $B$, and $\Lambda_{c,b}$, fragmentation is commonly described as a two-step process~\cite{Cacciari:1996wr,Cacciari:1993mq,Jaffe:1993ie,Kniehl:2005mk,Helenius:2018uul,Helenius:2023wkn,Generet:2023vte}. 
In the first step, an energetic parton $i$ produces a heavy quark $Q$, a subprocess that can be computed in perturbative QCD because the coupling at the heavy-quark mass scale is weak. 
The associated short-distance coefficient (SDC) for the $[i \to Q]$ transition develops on a timescale shorter than hadronization. 
The first NLO calculations of these SDCs were presented in Refs.~\cite{Mele:1990yq,Mele:1990cw}, and were later extended up to next-to-NLO in Refs.~\cite{Rijken:1996vr,Mitov:2006wy,Blumlein:2006rr,Melnikov:2004bm,Mitov:2004du,Biello:2024zti}.

The second step is genuinely nonperturbative and describes the conversion of the heavy quark into the observed hadron. 
It is usually modeled through phenomenological parameterizations~\cite{Kartvelishvili:1977pi,Bowler:1981sb,Peterson:1982ak,Andersson:1983jt,Collins:1984ms,Colangelo:1992kh} or effective-theory approaches~\cite{Georgi:1990um,Eichten:1989zv,Grinstein:1992ss,Neubert:1993mb,Jaffe:1993ie}. 
A full VFNS FF set is then obtained by evolving these initial conditions through DGLAP equations at the chosen perturbative accuracy.

A similar picture holds for quarkonia, although the presence of a $[Q\bar{Q}]$ pair already in the leading Fock state makes the dynamics more structured. 
This feature is naturally described in NRQCD~\cite{Caswell:1985ui,Thacker:1990bm,Bodwin:1994jh,Cho:1995vh,Cho:1995ce,Leibovich:1996pa,Bodwin:2005hm}; comprehensive discussions of its formal basis and phenomenology can be found in Refs.~\cite{Grinstein:1998xb,Kramer:2001hh,QuarkoniumWorkingGroup:2004kpm,Lansberg:2005aw,Lansberg:2019adr}. 
Within NRQCD, heavy quarks are treated as nonrelativistic particles, so production is factorized into perturbative SDCs for the creation of the $[Q\bar{Q}]$ pair and nonperturbative LDMEs for its hadronization. 
Physical quarkonia are represented as superpositions of Fock components, organized through a double expansion in $\alpha_s$ and in the relative velocity $v_{\cal Q}$.

NRQCD can describe both low- and high-$p_T$ production. 
At low $p_T$, the dominant mechanism is direct short-distance production of the $[Q\bar{Q}]$ pair, followed by hadronization. 
At large $p_T$, single-parton fragmentation becomes dominant. 
The latter naturally belongs to a VFNS with timelike DGLAP evolution, whereas the former is more closely related to a fixed-flavor number scheme (FFNS)~\cite{Alekhin:2009ni}, with two-parton fragmentation and higher-power contributions~\cite{Fleming:2012wy,Kang:2014tta,Echevarria:2019ynx,Boer:2023zit,Celiberto:2024mex,Celiberto:2024bxu,Celiberto:2024rxa}.

The first leading-order (LO) analyses of gluon and heavy-quark fragmentation into $S$-wave vector quarkonia in color-singlet channels were given in Refs.~\cite{Braaten:1993rw,Braaten:1993mp}. 
Subsequent NLO calculations~\cite{Zheng:2019gnb,Zheng:2021sdo} enabled the construction of the first VFNS DGLAP-evolved sets, the {\tt ZCW19$^+$} family~\cite{Celiberto:2022dyf,Celiberto:2023fzz}, later extended to $\BCs$ and $\Bss$ mesons through the {\tt ZCFW22} functions~\cite{Celiberto:2022keu,Celiberto:2024omj}. 
More recently, the first dedicated release of collinear FFs for pseudoscalar heavy quarkonia, the {\tt NRFF1.0} set, was presented in Ref.~\cite{Celiberto:2025euy}. 
Predictions based on {\tt ZCFW22} were found to agree with LHCb data~\cite{LHCb:2014iah,LHCb:2016qpe,Celiberto:2024omj}, showing that $\BCs$ production remains below 0.1\% of singly bottomed $B$-meson yields~\cite{Celiberto:2024omj}, thus supporting the VFNS description at large $p_T$.

Recent investigations indicate that NRQCD factorization can be extended to exotic states, including the double $\Jpsi$ structures observed at the LHC~\cite{LHCb:2020bwg,ATLAS:2023bft,CMS:2023owd}, often interpreted as compact tetracharms~\cite{Zhang:2020hoh,Zhu:2020xni}. 
In this case, $\TQc$ production originates from the short-distance creation of two charm and two anticharm quarks at a scale $\sim 1/m_c$, followed by nonperturbative binding. 
The first NRQCD-based initial condition for gluon fragmentation into color-singlet $S$-wave $\TQc$ states was derived in Ref.~\cite{Feng:2020riv}. 
That calculation motivated the first VFNS FF family for heavy-light tetraquarks, {\tt TQHL1.0}, introduced in Ref.~\cite{Celiberto:2023rzw} and reviewed in~\cite{Celiberto:2024mrq}. 
A further step was taken in Ref.~\cite{Celiberto:2024beg}, where the {\tt TQ4Q1.1} and {\tt TQHL1.1} sets were released, including NRQCD-based inputs also for the $[Q \to \TQQ]$ channel~\cite{Bai:2024ezn}, a refined treatment of doubly-heavy tetraquark fragmentation, and bottomoniumlike states. 
The {\tt TQ4Q1.1} functions were subsequently employed as a benchmark case study for the phenomenological description of all-heavy tetraquark production in Ref.~\cite{Celiberto:2025ziy}. 
A dedicated analysis of axial-vector ($1^{+-}$) tetraquarks followed in Ref.~\cite{Celiberto:2025dfe}, where composite LDME uncertainties were propagated into the FFs for the first time (see Ref.~\cite{Celiberto:2026zed} for a review). 
The {\tt TQ4Q1.1} family was later employed to study indirect searches for charmed tetraquarks in Higgs and electroweak decays~\cite{Ma:2025ryo}, while the gluon FF was analyzed separately in Ref.~\cite{Nakhaei:2025zty}, and subsequently used in forward-rapidity studies to probe the sensitivity of all-charm tetraquark production to intrinsic charm components in the proton~\cite{Celiberto:2025vra}.

Along the same line of research, FFs for all-heavy pentaquarks and rare $\Omega$ baryons were developed in Refs.~\cite{Celiberto:2025ipt,Celiberto:2026rdk,Celiberto:2026ooh} and~\cite{Celiberto:2025ogy,Celiberto:2025csa}, leading to the {\tt PQ5Q1.0} and {\tt OMG3Q1.0} releases.

%==========================
\subsection{NRQCD from quarkonia to tetraquarks}
\label{ssec:FFs_NRQCD}
%==========================

In this subsection, we summarize the NRQCD-based construction of collinear FFs, emphasizing its role as a systematic tool to describe the transition from partonic degrees of freedom to physical hadrons. 
For clarity, we first review the quarkonium case, which provides the baseline for extending the formalism to exotic multiquark systems.

The derivation of tetraquark FFs follows the approach developed in Refs.~\cite{Feng:2020riv,Bai:2024ezn}, adapted here to the specific goals of our analysis. 
We refer to those works for a detailed account of the NRQCD calculations and the treatment of both short- and long-distance contributions.

Within NRQCD, the FF of a parton $i$ into a quarkonium state $\cal Q$ with momentum fraction $z$ at the initial scale $\mu_{F,0}$ reads
\begin{equation}
\label{FFs_NRQCD_onium}
D_i^{\cal Q}(z, \mu_{F,0}) = \sum_{[n]} {\cal D}_i^{Q\bar{Q}}(z, [n]) \langle {\cal O}^{\cal Q}([n]) \rangle \;.
\end{equation}
The coefficients ${\cal D}_i^{Q\bar{Q}}(z, [n])$ are perturbative SDCs, computable as expansions in $\alpha_s$ and containing DGLAP-type logarithms, while $\langle {\cal O}^{\cal Q}([n]) \rangle$ denote the nonperturbative LDMEs, scaling with the relative velocity $v_{\cal Q}$. 
The spectroscopic label $[n] \equiv \,^{2S+1}L_J^{(c)}$ specifies spin, orbital, and color quantum numbers.

Equation~\eqref{FFs_NRQCD_onium} embodies two key NRQCD principles: quarkonia are described as superpositions of Fock states $[n]$, and contributions are organized in a double expansion in $\alpha_s$ and $v_{\cal Q}$.

A comparison with open heavy-flavor fragmentation highlights both similarities and essential differences. 
For singly heavy hadrons $h_Q$, the FF at the initial scale involves a convolution between a perturbative kernel and a nonperturbative transition function describing $Q \to h_Q$~\cite{Cacciari:1996wr,Cacciari:1993mq,Jaffe:1993ie}. 
This reflects momentum sharing with soft degrees of freedom and induces a nontrivial $z$ dependence.

In contrast, NRQCD expresses the $[i \to {\cal Q}]$ FF as a sum over Fock states, where each term is the product of a perturbative SDC and a constant LDME. 
Since the heavy quarks are produced at short distances, no additional momentum redistribution is required, and LDMEs are $z$-independent. 
This explains why open-flavor FFs require phenomenological $z$-dependent models, whereas NRQCD FFs are linear combinations of perturbative functions weighted by scalar coefficients.

Although LDMEs are not calculable from first principles, they can be constrained phenomenologically or estimated via lattice QCD or potential models. 
A commonly adopted simplification is the \emph{vacuum saturation approximation} (VSA)~\cite{Shifman:1978bx,Gilman:1979bc,Bodwin:1994jh}, which suppresses intermediate states beyond the vacuum by powers of $v_{\cal Q}$. 
Under this assumption, the LDME factorizes as
\begin{equation}
\label{VSA}
\hspace{-0.45cm}
\langle 0 | \chi^\dagger \Pi_n \psi \, \mathcal{P}_{\cal Q} \, \psi^\dagger \Pi_n^\prime \chi | 0 \rangle
\simeq
\langle 0 | \chi^\dagger \Pi_n \psi | {\cal Q} \rangle \langle {\cal Q} | \psi^\dagger \Pi_n^\prime \chi | 0 \rangle \;
\end{equation}
where $\Pi_n$ and $\Pi_n^\prime$ are spin-color projectors, $\mathcal{P}_{\cal Q}$ projects onto the physical state, and $\chi$, $\psi$ are nonrelativistic Pauli fields.

Extending Eq.~\eqref{FFs_NRQCD_onium} to all-heavy tetraquarks, the initial-scale FF of a parton $i$ into $\TQQ$ can be written as
\begin{equation}
\label{FFs_TQc_general}
D_i^{\TQQ}(z, \mu_{F,0}) = \sum_{[n]} {\cal D}_i^{4Q}(z,[n]) \langle {\cal O}^{\TQQ}([n]) \rangle \;,
\end{equation}
where ${\cal D}_i^{4Q}(z,[n])$ describes the perturbative production of a compact $|Q\bar{Q}Q\bar{Q}\rangle$ state with quantum numbers $[n]$, and $\langle {\cal O}^{\TQQ}([n]) \rangle$ encodes its nonperturbative transition into the physical tetraquark.

The main difference with respect to quarkonia lies in the structure of the intermediate state. 
Instead of a two-body system, tetraquarks originate from four-body configurations with richer spin-color couplings and symmetry constraints. 
This leads to a larger set of contributing Fock states and a more involved pattern of LDMEs.

While weaker binding and multiple configurations may affect the convergence of the velocity expansion, low-lying $S$-wave compact tetraquarks are still expected to obey the factorized structure of Eq.~\eqref{FFs_TQc_general}, providing a consistent starting point for initial-scale inputs and their subsequent DGLAP evolution.

For a all-heavy tetraquark $\TQQ(J^{PC})$, with quantum numbers $J^{PC} = 0^{++}$, $1^{+-}$, or $2^{++}$, and retaining only leading terms in the velocity expansion, Eq.~\eqref{FFs_TQc_general} can be rewritten as
\begin{equation}
\begin{split}
 \label{TQQ_FF_initial-scale}
 D^{\TQQ(J^{PC})}_i(z,\mu_{F,0}) \, &= \,
 \frac{1}{m_Q^9}
 \sum_{[n]} 
 \tilde{\cal D}^{(J^{PC})}_i(z,[n]) \\
 &\times \, \langle {\cal O}^{\TQQ(J^{PC})}([n]) \rangle
 \;,
\end{split}
\end{equation}
where $m_Q = m_c = 1.5$~GeV ($m_Q = m_b = 4.9$~GeV) denotes the heavy-quark mass. 
The composite index $[n]$ spans the configurations $[3,3]$, $[6,6]$, $[3,6]$, and $[6,3]$. 
We define dimensionless SDCs through
\begin{equation}
\begin{split}
 \label{TQQ_FF_SDCs_tilde}
 \tilde{\cal D}^{(J^{PC})}_i(z,[n]) \rangle
 \, \equiv \,
 m_Q^9 \,
 {\cal D}^{(J^{PC})}_i(z,[n])
\;,
\end{split}
\end{equation}
and the symmetry relations
\begin{equation}
\begin{split}
 \label{TQQ_FF_initial-scale_symmetry}
 \tilde{\cal D}^{(J^{PC})}_i(z,[3,6]) \, &= \, \tilde{\cal D}^{(J^{PC})}_i(z,[6,3]) \;,
 \\
 \langle {\cal O}^{\TQc(J^{PC})}([3,6]) \rangle \, &= \, \langle {\cal O}^{\TQQ(J^{PC})}([6,3]) \rangle^*
 \;.
\end{split}
\end{equation}

The fragmentation framework employed here is expected to dominate at large transverse momentum, typically $p_T \gtrsim 3 M_{\TQQ}$, with $M_{\TQQ}$ the tetraquark mass. 
This condition marks the onset of the collinear regime, where logarithms $\ln(p_T^2/M_{\TQQ}^2)$ become sizeable, power corrections $\mathcal{O}(M_{\TQQ}^2/p_T^2)$ are suppressed, and nonfragmentation contributions are subleading~\cite{Cacciari:1995yt,Artoisenet:2007xi,Ma:2014svb}.

Early studies of quarkonium production~\cite{Cacciari:1994dr,Cacciari:1995yt,Cacciari:1996dg} indicate that gluon fragmentation dominates for $p_T \gtrsim 10 \div 15$ GeV in $\Jpsi$ production, while analyses of charmed $B$ mesons~\cite{Kolodziej:1995nv,Artoisenet:2007xi} suggest thresholds up to $\sim 80$~GeV. 
By analogy, for tetraquarks with $M_{\TQc} \sim 4 m_c \simeq 6.5$ GeV and $M_{\TQb} \sim 4 m_b \simeq 18$ GeV, we conservatively identify the fragmentation region as $p_T \gtrsim 20$ GeV and $p_T \gtrsim 50$ GeV, respectively, which defines the kinematic domain of reliability for our FFs.

At moderate and low $p_T$, fragmentation competes with nonfragmentation single-parton scattering (SPS) mechanisms and, potentially, with MPIs, in particular DPS, where two independent hard scatterings occur within the same proton-proton collision.
DPS has been extensively investigated in processes such as double-jet production~\cite{Ducloue:2015jba} and multiquarkonium final states, including $[\Jpsi{+}\Jpsi]$~\cite{Lansberg:2014swa,Lansberg:2020rft}, $[\Jpsi{+}\Yps]$~\cite{Lansberg:2020rft}, $[\Jpsi{+}Z]$~\cite{Lansberg:2016rcx}, $[\Jpsi{+}W]$~\cite{Lansberg:2017chq}, and triple-$\Jpsi$ production~\cite{dEnterria:2016ids,Shao:2019qob}, where it can significantly affect low-$p_T$ observables.

In the context of tetraquarks, early studies based on double color-evaporation models predict rapidly increasing cross sections with energy in both proton-proton and proton-ion collisions, with possible enhancements scaling with the nuclear mass number $A$ or larger~\cite{Carvalho:2015nqf,Abreu:2023wwg}. 
Comparisons between single-parton and DPS production of $|c\bar{c}c\bar{c}\rangle$ states relevant for $T_{4c}$ show that DPS contributions can exceed single-parton ones by over an order of magnitude, depending on the kinematics~\cite{Maciula:2020wri}.

These results indicate that DPS can provide a substantial, and sometimes dominant, contribution at low and intermediate $p_T$. 
However, DPS decreases more rapidly at large $p_T$ and does not exhibit the logarithmic enhancements characteristic of collinear fragmentation. 
In the high-$p_T$ region considered here, $p_T \gtrsim 3 M_{\TQQ}$, fragmentation is therefore expected to dominate, while DPS effects can be neglected at leading power, although they should be included in future analyses targeting lower transverse momenta.

%==========================
\subsection{Initial-scale FF inputs}
\label{ssec:FFs_initial_scale}
%==========================

\begin{figure*}[!t]
\centering
\includegraphics[width=0.325\textwidth]{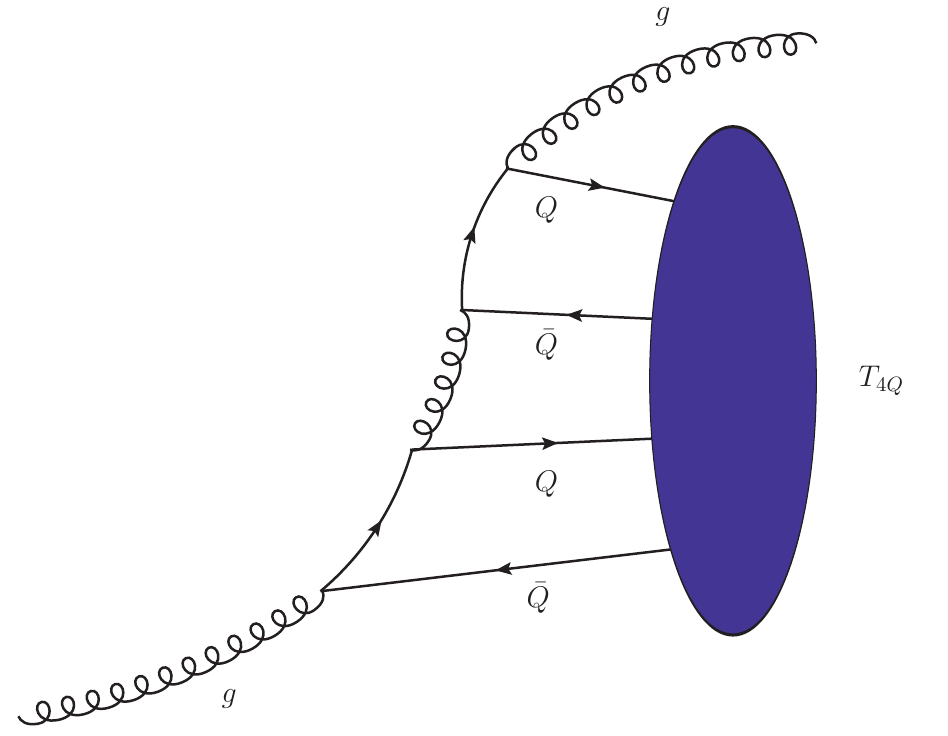}
\hspace{0.00cm}
\includegraphics[width=0.325\textwidth]{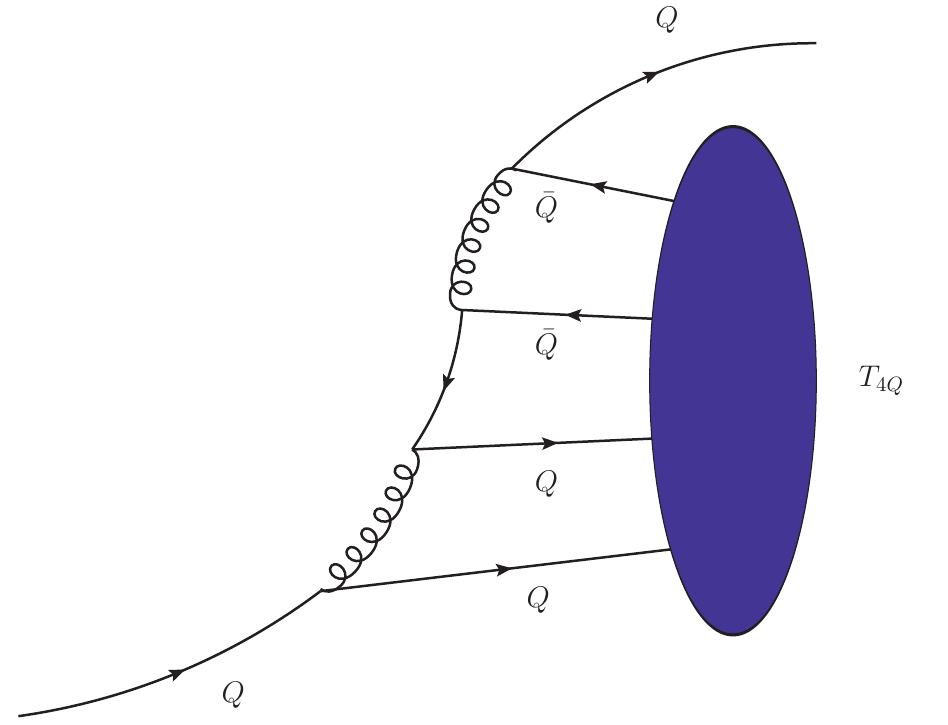}
\hspace{0.00cm}
\includegraphics[width=0.325\textwidth]{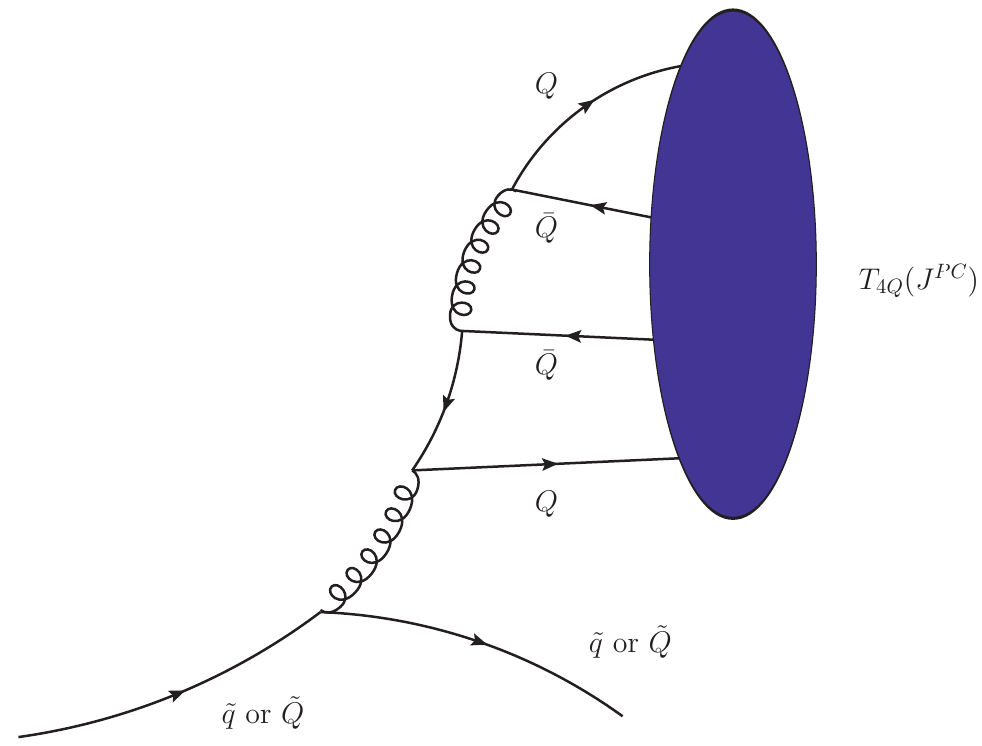}

\caption{
\justifying
\noindent
Leading-order diagrams for the collinear fragmentation of a gluon (left), a heavy quark (center), and a nonconstituent quark (right) into a all-heavy $S$-wave tetraquark in a color-singlet state. 
The partonic subprocesses on the left of each diagram encode the perturbative short-distance coefficients (SDCs), while the dark blue blobs on the right represent the nonperturbative long-distance matrix elements (LDMEs) governing the hadronization into the final state.
}
\label{fig:FF_diagrams}
\end{figure*}

For clarity, we adopt the same organizational scheme already introduced in our previous release of the {\tt TQ4Q1.1} FF set~\cite{Celiberto:2025ziy}. 
This section is structured as follows: we begin by introducing the dimensionless SDCs, then outline the construction of color-composite LDMEs, discuss the role of color-octet mechanisms, and finally present the LDME values together with their associated uncertainties.

\vspace{1em}
\noindent
\textbf{The dimensionless SDCs.} 
The SDCs entering Eq.~\eqref{TQQ_FF_initial-scale} are computed following the standard perturbative matching between QCD and NRQCD matrix elements, as implemented in Refs.~\cite{Feng:2020riv,Bai:2024ezn,Bai:2024flh}. 
Within this framework, SDCs are extracted by matching perturbative QCD amplitudes to NRQCD matrix elements evaluated on fictitious, free multiquark states carrying the same quantum numbers as the physical tetraquark~\cite{Bodwin:1994jh,Petrelli:1997ge}. 
At present, analytic results for all parton-initiated channels into $S$-wave $\TQQ$ states are available only at LO in both $\alpha_s$ and the relative velocity $v_{\cal Q}$.

As in the quarkonium case, SDCs are insensitive to long-distance hadronization dynamics and can therefore be computed by replacing the physical state $\TQQ$ with a free four-quark configuration $|[QQ][\bar{Q}\bar{Q}]\rangle$ with identical quantum numbers. 
The matching is performed at LO in $\alpha_s$ and $v_{\cal Q}$, retaining only the lowest-order NRQCD operators and neglecting derivative terms. 
The tetraquark is described within a diquark--antidiquark picture, where projection onto definite spin, parity, and color is achieved through suitable NRQCD four-quark operators. 
This construction effectively reduces the four-body system to two-body building blocks, facilitating the projection onto the relevant quantum numbers and the definition of gauge-invariant operators used to extract the corresponding SDCs~\cite{Feng:2020riv,Bai:2024ezn,Bai:2024flh}.
Both $[3 \otimes \bar{3}]$ and $[6 \otimes \bar{6}]$ color configurations are included.

The allowed quantum numbers are constrained by the interplay between Fermi--Dirac statistics and the $S$-wave orbital structure. 
For scalar ($0^{++}$) and tensor ($2^{++}$) states, both color configurations contribute, although the $[6 \otimes \bar{6}]$ component enters only in the spin-0 channel. 
In contrast, for the axial-vector ($1^{+-}$) state, Fermi--Dirac symmetry forbids the $[6 \otimes \bar{6}]$ configuration, leaving only the $[\bar{3} \otimes 3]$ channel. 
Moreover, gluon fragmentation is absent at LO due to the Landau--Yang theorem, while nonconstituent (light or heavy) quark channels $[\tilde{q}, \tilde{Q} \to \TQQ(1^{+-})]$ are excluded by charge-conjugation ($C$-parity) invariance. 
Consequently, only the constituent heavy-quark channel contributes in this case.

Analytic expressions for the dimensionless SDCs are collected in Appendix~\hyperlink{app:A}{A}. 
All partonic channels, originally derived in Refs.~\cite{Feng:2020riv,Bai:2024ezn,Bai:2024flh} and later independently reproduced with the {\psymJethad} symbolic engine~\cite{Celiberto:2020wpk,Celiberto:2022rfj,Celiberto:2023fzz,Celiberto:2024mrq,Celiberto:2024swu,Celiberto:2025csa,Celiberto:2026ooh}, are provided at LO accuracy for color-singlet configurations (see Fig.~\ref{fig:FF_diagrams}).

\vspace{1em}
\noindent
\textbf{The color-composite LDMEs.} 
As outlined above, our framework delivers analytic SDCs at the initial scale, which are subsequently matched to the corresponding NRQCD LDMEs. 
These matrix elements encode the nonperturbative transition from the $| [QQ][\bar{Q}\bar{Q}] \rangle$ configuration to the physical tetraquark state and must be determined either from data or through phenomenological modeling.

In close analogy with quarkonium, a common simplification is provided by the VSA, whereby LDMEs are approximated as products of vacuum-to-Fock-state matrix elements. 
For tetraquarks, however, the presence of correlated diquark and antidiquark subsystems leads to a more involved structure. 
Following Refs.~\cite{Feng:2020riv,Bai:2024ezn}, we employ color-composite LDMEs, expanding the physical state in a diquark--antidiquark basis and projecting onto color-singlet operators.

This construction allows one to organize the LDMEs according to the color configurations of the constituent clusters, namely $[3,3]$, $[6,6]$, and mixed $[3,6]$/$[6,3]$ components, as introduced in Eq.~\eqref{TQQ_FF_initial-scale}. 
These quantities represent the probability amplitudes for a perturbatively produced four-quark system to hadronize into the physical bound state, and must be specified for each spin and color channel. 
For completeness, we list below the color-composite operators associated with the $\TQQ$ LDMEs, referring to Sec.~IV of Ref.~\cite{Feng:2020riv} for further technical details:
\begin{align}
\label{TQQ_FF_LDMEs_operators}
{\cal O}^{(0)}_{\bar{3} \otimes 3} &= - \frac{1}{\sqrt{3}} 
\left[
\psi^T_a (i \sigma^2) \sigma^i \psi_b \right]
\left[
\chi^\dagger_c \sigma^i (i \sigma^2) \chi^\ast_d 
\right]
{\cal C}^{ab;cd}_{\bar{3} \otimes 3} \;,
\nonumber \\[0.20cm]
{\cal O}^{\alpha\beta;(2)}_{\bar{3} \otimes 3} &= 
\left[
\psi^T_a (i \sigma^2) \sigma^m \psi_b \right]
\left[
\chi^\dagger_c \sigma^n (i \sigma^2) \chi^\ast_d 
\right]
\Gamma^{\alpha\beta;mn}
{\cal C}^{ab;cd}_{\bar{3} \otimes 3} \;,
\nonumber \\[0.20cm]
{\cal O}^{(0)}_{6 \otimes \bar{6}} &= 
\left[
\psi^T_a (i \sigma^2) \psi_b \right]
\left[
\chi^\dagger_c (i \sigma^2) \chi^\ast_d 
\right]
{\cal C}^{ab;cd}_{6 \otimes \bar{6}} \;,
\end{align}
where $\sigma^2$ denotes the second Pauli matrix, and $\psi$ and $\chi$ are the standard NRQCD fields defined in Eq.~\eqref{VSA}. 
The rank-4 Lorentz tensor is given by
\begin{equation}
\label{Gamma_klmn_rank4}
 \Gamma^{kl;mn} \equiv \frac{1}{2} \left( \delta^{km} \delta^{ln} + \delta^{kn} \delta^{lm} - \frac{2}{3} \delta^{kl} \delta^{mn} \right) \,,
\end{equation}
while the color tensors read
\begin{equation}
\begin{split}
\label{C_abcd}
 C_{\bar{3} \otimes 3}^{ab;cd} &= \frac{1}{(\sqrt{2})^2} \, \epsilon^{abm} \epsilon^{cdn} \, \frac{\delta^{mn}}{\sqrt{N_c}} \\
 &=\, \frac{1}{2\sqrt{3}} \left( \delta^{ac} \delta^{bd} - \delta^{ad} \delta^{bc} \right) \;, \\[0.20cm]
 C_{6 \otimes \bar{6}}^{ab;cd} &= \frac{1}{2\sqrt{6}} \left( \delta^{ac} \delta^{bd} + \delta^{ad} \delta^{bc} \right) \;,
\end{split}
\end{equation}
where $\epsilon^{ijk}$ and $\delta^{mn}$ denote the Levi-Civita and Kronecker tensors used to construct color-singlet combinations.

Finally, we remark that the diquark--antidiquark basis adopted here is formally equivalent, through Fierz transformations, to a molecularlike decomposition in terms of $[1 \otimes 1]$ or $[8 \otimes 8]$ color structures. 
This choice does not imply a literal bound-state interpretation, but rather reflects the working assumptions of our proxy model, where tetraquark formation is driven by dominant spin and color correlations between tightly bound diquark and antidiquark constituents. 
Such a description is fully consistent with a fragmentation-based production mechanism, which naturally favors the direct formation of compact multiquark states.

\vspace{1em}
\noindent
\textbf{Color-octet suppression.}
The inclusion of color-octet channels in NRQCD-based initial-scale FFs remains conceptually and phenomenologically challenging. 
Following Refs.~\cite{Feng:2020riv,Bai:2024ezn}, we restrict the present analysis to color-singlet contributions, due to the lack of reliable constraints on octet LDMEs. 
A consistent treatment of these terms would require dedicated nonperturbative input, and is therefore beyond the current level of control.

Guidance can be drawn from the quarkonium production puzzle (see Refs.~\cite{Brambilla:2010cs,Andronic:2015wma}), where color-octet mechanisms are essential for vector states such as $\Jpsi$~\cite{Braaten:1993rw,Cho:1995vh,Cho:1995ce,Beneke:1996tk,Butenschoen:2010rq,Chao:2012iv,Gong:2012ug}, but suppressed for pseudoscalar quarkonia and bottomonia~\cite{Braaten:1993rw,Han:2014jya,LHCb:2014oii,LHCb:2019zaj,LHCb:2022byt,Artoisenet:2008fc,Gong:2010bk}. 

A similar hierarchy was observed in our previous {\tt TQ4Q1.1} analysis~\cite{Celiberto:2025ziy}, and persists here. 
The relevance of octet contributions depends strongly on $J^{PC}$. 
The scalar ($0^{++}$) channel involves multiple color configurations ($[3,3]$, $[6,6]$), enhancing model dependence, while the tensor ($2^{++}$) case is structurally simpler but still sensitive to wave-function modeling. 
In contrast, the axial-vector ($1^{+-}$) channel admits a single color-spin configuration, leading to particularly stable singlet-dominated predictions~\cite{Bai:2024ezn,Celiberto:2025dfe}.

A parametric estimate of octet effects can be obtained from NRQCD scaling arguments. 
In analogy to pseudoscalar quarkonia, singlet contributions scale as $v_{\cal Q}^3$, while octet terms scale as $v_{\cal Q}^7$ and carry an additional $\alpha_s$ suppression~\cite{Bodwin:1994jh,Braaten:1994vv}. 
This yields
\begin{equation*}
\frac{\text{octet}}{\text{singlet}} \sim \alpha_s \, v_{\cal Q}^4 \;.
\end{equation*}
For $T_{4c}$, taking $v_{{\cal Q}c}^2 \sim 0.3$ and $\alpha_s \sim 0.2$~\cite{Bodwin:1994jh}, one finds
\begin{equation}
\alpha_s \, v_{{\cal Q}c}^4 \sim 0.02 \;,
\end{equation}
while for $T{4b}$, with $v_{{\cal Q}b}^2 \sim 0.1$~\cite{Brambilla:2010cs},
\begin{equation}
\alpha_s \, v_{{\cal Q}_b}^4 \sim 10^{-3} \;,
\end{equation}
indicating a strong suppression.

These conclusions remain valid upon inclusion of nonconstituent quark fragmentation channels. 
Although these enter at the same perturbative order, their impact is dynamically suppressed at the observable level, due to a less favorable convolution with parton distribution functions (PDFs) and a reduced overlap with the tetraquark bound-state configuration. 
Consequently, they do not modify the qualitative hierarchy among channels nor the robustness of the singlet approximation, especially in the axial-vector case.

\vspace{1em}
\noindent
\textbf{LDMEs and uncertainty estimates.}
The $\langle {\cal O}^{\TQQ(J^{PC})}([n]) \rangle$ LDMEs provide the nonperturbative input to the initial-scale FFs. 
In the absence of experimental or lattice determinations for all-heavy tetraquarks, we follow the strategy adopted in our previous {\tt TQ4Q1.1} analysis~\cite{Celiberto:2025ziy}, which we briefly summarize here for completeness.

The construction relies on potential-model estimates of the wave function at the origin, following Ref.~\cite{Feng:2020riv}, where all-charm systems are described via a Schr{\"o}dinger equation with Cornell-like potentials~\cite{Eichten:1974af,Eichten:1978tg}, and mapped to LDMEs through the VSA. 
Among the models proposed in Refs.~\cite{Zhao:2020nwy,Lu:2020cns,liu:2020eha}, we retain the relativistic Model~II~\cite{Lu:2020cns}, as Models~I and~III either overestimate cross sections or lead to numerical instabilities.

Further refinements from Ref.~\cite{Bai:2024ezn} introduce additional LDME sets (Models~IV and~V). 
Since Model~IV~\cite{Yu:2022lak} is close to Model~II, while Model~V~\cite{Wang:2019rdo} is strongly suppressed, we follow Refs.~\cite{Celiberto:2025dfe,Celiberto:2025ziy} and adopt the average of Models~II and~IV for the axial-vector channel, using their spread as uncertainty.

The dependence on the LDME model is strongly channel dependent. 
The axial-vector ($1^{+-}$) state is particularly stable, as it involves a single $[3,3]$ configuration fixed by Fermi-Dirac symmetry, leading to robust predictions~\cite{Bai:2024ezn,Celiberto:2025dfe}. 
In contrast, the scalar ($0^{++}$) channel involves multiple interfering components ($[3,3]$, $[6,6]$, and mixed terms), while the tensor ($2^{++}$) state, although simpler, remains sensitive to the wave-function normalization. 
To account for this enhanced model dependence, we conservatively assign to these channels an uncertainty equal to twice that extracted for the axial-vector case.

For all-bottom tetraquarks, we adopt a Coulomb-inspired scaling Ansatz~\cite{Feng:2023agq}, relating $\TQb$ and $\TQc$ LDMEs through
\begin{equation}
\label{sLDMEs_T4b}
\hspace{-0.40cm}
 \frac{\langle {\cal O}^{\TQb(J^{PC})}([n]) \rangle}{\langle {\cal O}^{\TQc(J^{PC})}([n]) \rangle} 
 = 
 \frac{{\langle \cal O}^{\TQb}{\rm [C]} \rangle}{{\langle \cal O}^{\TQc}{\rm [C]}\rangle} 
 \simeq
 \left( \frac{m_b \as^{[b]}}{m_c \as^{[c]}} \right)^9 
 \simeq \,  
 400
 \;.
\end{equation}
Here $\as^{[Q]}$ is evaluated at the scale $m_Q v_{\cal Q}$, and the $[{\rm C}]$ label denotes Coulomb-dominated dynamics. 
This scaling follows from $|{\cal R}_{\cal Q}(0)|^2 \propto (\alpha_s m_Q)^3$ for each diquark pair, yielding an overall exponent of 9, consistent with quarkonium expectations~\cite{Bodwin:1994jh,Eichten:2019gig}.

To estimate model uncertainties, we consider alternative potentials such as Cornell~\cite{Eichten:1978tg} and Buchm{\"u}ller-Tye~\cite{Buchmuller:1980su}. 
These typically reduce $|{\cal R}_{\cal Q}(0)|$ by $\sim 2\%$, leading to
\begin{equation}
\label{sLDMEs_T4b_Delta}
 \left( \frac{\Delta_{\rm model} {\cal R}_{\cal Q}(0)}{{\cal R}_{\cal Q}(0)} \right)^4 \simeq (0.25)^4 = 0.4\% \;.
\end{equation}
This effect is negligible compared to the $\sim 20$--$25\%$ uncertainty inherited from the $\TQc$ sector, and is therefore not propagated further.

The numerical values of the color-composite LDMEs used in our analysis are reported in Tables~\ref{tab:T4c_LDMEs} and~\ref{tab:T4b_LDMEs}.

\begin{table}[t]
\centering
\begin{tabular}{c|c|c|c}
\toprule
$[n]$ & $\TQcZpp$ [GeV$^9$] & $\TQcOpm$ [GeV$^9$] & $\TQcTpp$ [GeV$^9$] \\
\midrule
$[3,3]$ & $0.0347 \pm 0.0076$ & $0.0878 \pm 0.0098$ & $0.0720 \pm 0.0158$ \\
$[6,6]$ & $0.0128 \pm 0.0028$ & $0$                 & $0$                 \\
$[3,6]$ & $0.0211 \pm 0.0046$ & $0$                 & $0$                 \\
\bottomrule
\end{tabular}
\caption{
\justifying
\noindent
Values of color-composite long-distance matrix elements, $\langle {\cal O}^{\TQc(J^{PC})}([n]) \rangle$, for the $\TQcZpp$, $\TQcOpm$, and $\TQcTpp$ states. 
Uncertainties follow the procedure described in the main text.
}
\label{tab:T4c_LDMEs}
\end{table}

\begin{table}[t]
\centering
\begin{tabular}{c|c|c|c}
\toprule
$[n]$ & $\TQbZpp$ [GeV$^9$] & $\TQbOpm$ [GeV$^9$] & $\TQbTpp$ [GeV$^9$] \\
\midrule
$[3,3]$ & $13.88 \pm 3.05$ & $35.1 \pm 3.9$ & $28.80 \pm 6.34$ \\
$[6,6]$ & $5.12 \pm 1.13$  & $0$            & $0$             \\
$[3,6]$ & $8.44 \pm 1.86$  & $0$            & $0$             \\
\bottomrule
\end{tabular}
\caption{
\justifying
\noindent
Color-composite long-distance matrix elements, $\langle {\cal O}^{\TQb(J^{PC})}([n]) \rangle$, for the $\TQbZpp$, $\TQbOpm$, and $\TQbTpp$ states at the initial scale. 
The values are obtained through color-Coulomb scaling from the corresponding $\TQc$ values, as detailed in the main text.
}
\label{tab:T4b_LDMEs}
\end{table}

%==========================
\subsection{{\tt TQ4Q2.0} FFs from {\HFNRevo}}
\label{ssec:FFs_TQ4Q20}
%==========================

 %%%%%%%%%%%%%%%%%%%%%%%%%%%%%%%%%%
\begin{table*}
 \begin{center}
 \begin{tabular}[c]{|c|c||c|c|c|c||c|}
 \toprule
 $\TQQ$ & $J^{PC}$ & $\mu_{F,0}(g \to \TQQ)$ & $\mu_{F,0}(d,u,s \to \TQQ)$ & $\mu_{F,0}(Q \to \TQQ)$ & $\mu_{F,0}(\tilde{Q} \to \TQQ)$ & $Q_0 \equiv \max \left( \{ \mu_{F,0} \} \right)$ \\
 \midrule
 $\TQc$ & $0^{++}$ & $4m_c$ & $m_{d,u,s}+4m_c$ & $5m_c$ & $\boldsymbol{m_b+4m_c}$ & $\boldsymbol{m_b+4m_c}$ \\
 %\midrule
 $\TQc$ & $1^{+-}$ &  &  & $\boldsymbol{5m_c}$ &  & $\boldsymbol{5m_c}$ \\
 %\midrule
 $\TQc$ & $2^{++}$ & $4m_c$ & $m_{d,u,s}+4m_c$ & $5m_c$ & $\boldsymbol{m_b+4m_c}$ & $\boldsymbol{m_b+4m_c}$ \\
 %\midrule
 $\TQb$ & $0^{++}$ & $4m_b$ & $m_{d,u,s}+4m_b$ & $\boldsymbol{5m_b}$ & $m_c+4m_b$ & $\boldsymbol{5m_b}$ \\
 %\midrule
 $\TQb$ & $1^{+-}$ &  &  & $\boldsymbol{5m_b}$ &  & $\boldsymbol {5m_b}$ \\
 %\midrule
 $\TQb$ & $2^{++}$ & $4m_b$ & $m_{d,u,s}+4m_b$ & $\boldsymbol{5m_b}$ & $m_c+4m_b$ & $\boldsymbol{5m_b}$ \\
 \bottomrule
  \end{tabular}
 \caption{
\justifying
\noindent
 Central columns: initial factorization scale, $\mu_{F,0}$, for the fragmentation of a given parton species (gluon, light quark, constituent and nonconstituent heavy quark) to a color-singlet,  $S$-wave $\TQQ$ tetraquark.
 Rightmost column: evolution-ready energy scale, $Q_0$, set to the maximum of $\mu_{F,0}$ values.
 For clarity, values of $Q_0$ are highlighted in bold font.
 Due to the symmetry of the final-state tetraquark, FFs from quark and antiquark of the same flavor are identical.
 }
 \label{tab:muF0_Q0}
 \end{center}
\end{table*}

The construction of the {\tt TQ4Q2.0} collinear FFs is completed by evolving the initial-scale inputs through DGLAP equations.
In contrast to light-hadron fragmentation, all partonic channels feature nontrivial evolution thresholds, dictated by the kinematics of the underlying perturbative splittings shown in Fig.~\ref{fig:FF_diagrams} and encoded in the corresponding SDCs.

Kinematic constraints fix the minimum invariant mass of the underlying splittings and thus determine the corresponding evolution thresholds.
For gluon fragmentation, one has $\mu_{F,0}(g \to \TQQ) = 4 m_Q$, while for the heavy-quark channel $\mu_{F,0}(Q \to \TQQ) = 5 m_Q$.
For nonconstituent partons, namely a light quark $\tilde{q}$ or a heavy quark $\tilde{Q}$, the threshold reads $\mu_{F,0}(\tilde{q}, \tilde{Q} \to \TQQ) = m_{\tilde{q},\tilde{Q}} + 4 m_Q$.

The corresponding antiquark channels share identical FFs and thresholds, as implied by charge-conjugation symmetry and the identical kinematics of quark and antiquark splittings.
All threshold scales $\mu_{F,0}$, together with the evolution-ready scale $Q_0$ defined as the maximum among them for a given $\TQQ$ state, are summarized in Table~\ref{tab:muF0_Q0}.

To properly account for the presence of multiple partonic thresholds in the DGLAP evolution, we adopt a dedicated strategy based on the {\HFNRevo} framework~\cite{Celiberto:2025euy,Celiberto:2024mex,Celiberto:2024bxu,Celiberto:2024rxa,Celiberto:2025xvy,Celiberto:2026rzi,Celiberto:2026zss}.

This approach is tailored to the evolution of heavy-hadron FFs constructed from nonrelativistic inputs, and is organized around three main ingredients: physical interpretation, threshold-aware evolution dynamics, and uncertainty quantification.
The interpretation step connects the low-transverse-momentum production mechanism to an effective two-parton fragmentation picture, as discussed in Sec.~\ref{ssec:FFs_intro}, enabling a consistent matching between FFNS and VFNS schemes.
The evolution component dynamically activates the different partonic channels at their natural kinematic scales, ensuring a consistent and physically motivated treatment across thresholds.
Finally, the uncertainty quantification provides a systematic assessment of MHOUs through controlled multiscale variations.

Conceived as a bridge between precision QCD and a hadron-structure-oriented description of quarkonium fragmentation~\cite{Celiberto:2025euy}, the {\HFNRevo} framework has rapidly evolved into a powerful tool for the study of rare and exotic hadronic systems.
It has been successfully applied to scalar ($0^{++}$) and tensor ($2^{++}$) tetraquarks through the {\tt TQ4Q1.0}~\cite{Celiberto:2024mab} and {\tt TQ4Q1.1}~\cite{Celiberto:2025dfe,Celiberto:2025ziy} sets---which lay the groundwork for the {\tt TQ4Q2.0} functions introduced here---as well as to triply heavy baryons such as ${\rm \Omega}_{3Q}$ via the {\tt OMG3Q1.0} determinations~\cite{Celiberto:2025ogy}.
In this broader context, {\HFNRevo} emerges as a versatile evolution framework capable of consistently handling FFs with initial-scale inputs from all partonic species.
This multichannel structure requires a dedicated treatment of evolution thresholds, specific to each contribution.

Within the {\HFNRevo} framework, the DGLAP evolution in the presence of multiple partonic thresholds is organized into two consecutive stages.
The first stage, denoted as {\tt EDevo}, implements a semianalytical, expanded evolution from the lowest activation scale up to the highest threshold.
In this regime, partonic channels are dynamically switched on at their respective kinematic thresholds, allowing for an independent treatment of each contribution across distinct evolution domains.
The second stage, referred to as {\tt AOevo}, consists of a fully numerical, all-order evolution starting from the highest scale $Q_0$ and extending to larger energies.
This two-step construction provides a controlled matching across thresholds while preserving perturbative stability over the entire evolution range.

\begin{figure*}[t]
\centering

 \includegraphics[scale=0.18,clip]{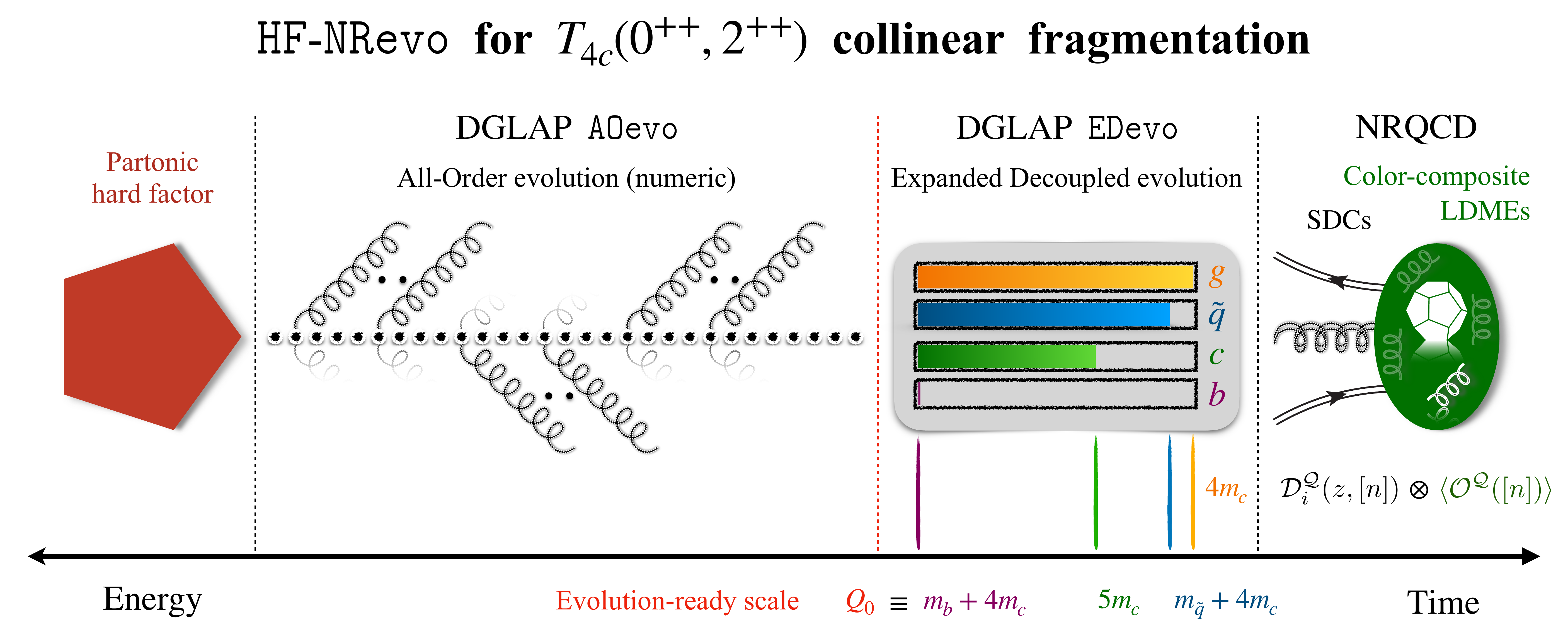}

 \vspace{0.55cm}

 \includegraphics[scale=0.18,clip]{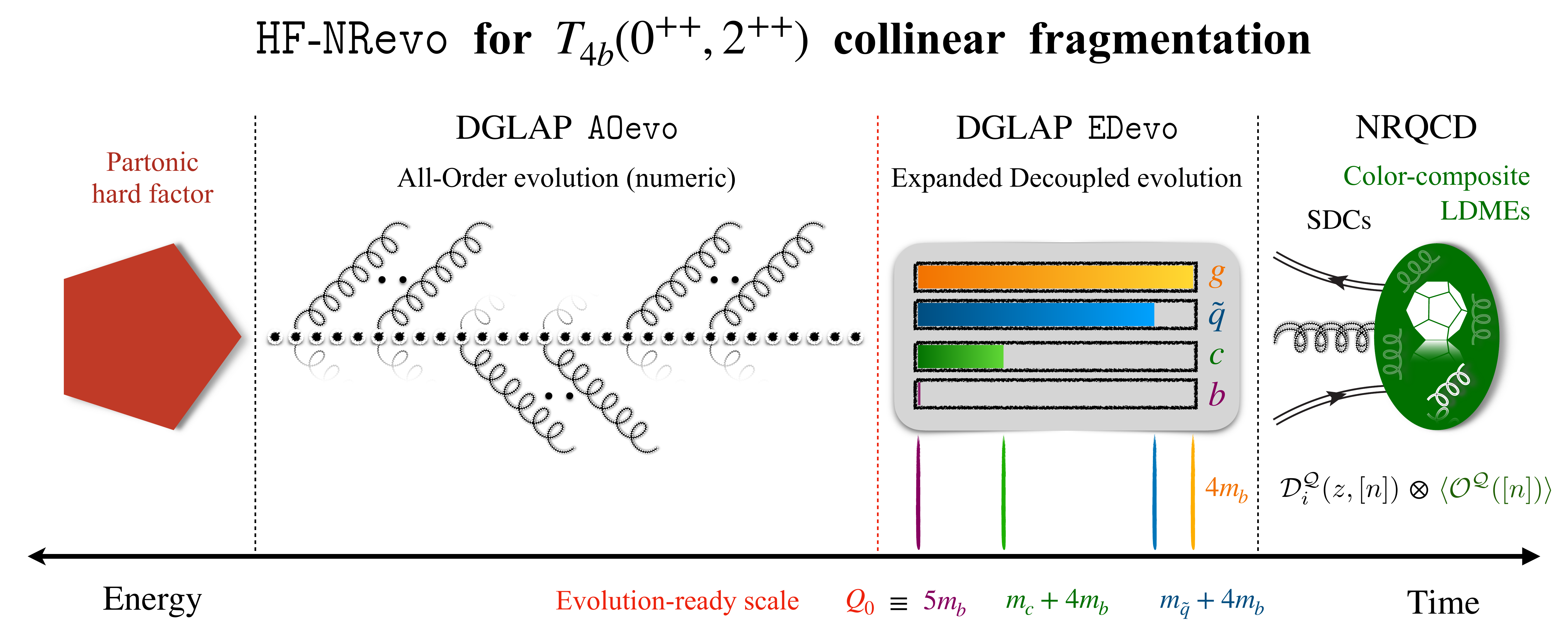}

\caption{
\justifying
\noindent
Two-stage evolution pipeline for $\TQc(0^{++},2^{++})$ (upper panels) and $\TQb(0^{++},2^{++})$ (lower panels) fragmentation functions within the {\HFNRevo} framework. 
The scheme highlights a sequential DGLAP evolution: an initial semianalytical {\tt EDevo} stage, starting at the lowest threshold and extending up to the evolution-ready scale $Q_0 = m_b + 4m_c$ (upper) or $Q_0 = 5m_b$ (lower), where partonic channels are progressively activated, followed by a fully numerical {\tt AOevo} stage completing the evolution.
}
\label{fig:T4Q-0pp-2pp_cs_HFNRevo_sketch}
\end{figure*}

To illustrate the {\HFNRevo} approach to $\TQQ$ fragmentation, the upper panel of Fig.~\ref{fig:T4Q-0pp-2pp_cs_HFNRevo_sketch} displays a conceptual flowchart for the two-stage evolution of all-charm scalar ($0^{++}$) and tensor ($2^{++}$) states.
The horizontal axis represents the flow of energy from infrared to ultraviolet scales (right to left), opposite to the chronological evolution of the process.

When read from right to left, the diagram describes the construction of the FFs.
The workflow starts from NRQCD inputs, obtained by convoluting SDCs with LDMEs (rightmost block).
The semianalytical {\tt EDevo} stage (center-right) then initiates the evolution at the lowest threshold, $\mu_{F,0}(g \to \TQc) = 4m_c$, and proceeds by dynamically activating successive channels.
The light nonconstituent quark threshold, $\mu_{F,0}(\tilde{q} \to \TQc) = m_{\tilde{q}} + 4m_c$, is followed by the charm channel at $\mu_{F,0}(c \to \TQc) = 5m_c$, and finally by the bottom threshold, $\mu_{F,0}(b \to \TQc) = m_b + 4m_c$, which defines the evolution-ready scale $Q_0$.

Beyond $Q_0$, the numerical {\tt AOevo} stage (center-left) performs all-order NLO DGLAP evolution with the full set of active flavors, currently implemented via {\APFELpp}~\cite{Bertone:2013vaa,Carrazza:2014gfa,Bertone:2017gds} and to be extended to {\tt EKO}~\cite{Candido:2022tld,Hekhorn:2023gul}.
The evolved FFs are then convoluted with the hard-scattering kernel to obtain physical observables.

The lower panel of Fig.~\ref{fig:T4Q-0pp-2pp_cs_HFNRevo_sketch} shows the corresponding workflow for all-bottom $\TQb(0^{++},2^{++})$ states.
The structure is analogous, but with a different ordering of thresholds.
The evolution starts at $\mu_{F,0}(g \to \TQb) = 4m_b$, followed by the light-quark channel, $\mu_{F,0}(\tilde{q} \to \TQb) = m_{\tilde{q}} + 4m_b$, then the charm threshold, $\mu_{F,0}(c \to \TQb) = m_c + 4m_b$, and finally the bottom channel, which fixes $Q_0 = 5m_b$.
The subsequent {\tt AOevo} stage evolves the system to higher scales in the same manner as in the charmed case.

As shown in Table~\ref{tab:muF0_Q0}, axial-vector states ($1^{+-}$), both in the all-charm and all-bottom sectors, exhibit a suppression of several initial fragmentation channels at LO.
In particular, the gluon-initiated contribution is forbidden by the Landau-Yang theorem, while nonconstituent quark channels are excluded by charge-conjugation ($C$-parity) conservation.

As a consequence, only the constituent heavy-quark channel contributes at the initial scale, fixed at $\mu_{F,0} = 5m_Q$.
In this configuration, the semianalytical {\tt EDevo} stage is effectively absent, and the evolution is initiated directly through the {\tt AOevo} module at $Q_0 = 5m_Q$.
All other partonic channels are then dynamically generated by DGLAP evolution, without requiring explicit initial-scale inputs.

The {\tt TQ4Q2.0} set represents, to the best of our knowledge, the first systematic study of all-heavy tetraquark fragmentation in which all partonic channels are consistently included at the level of NRQCD initial conditions.
This construction parallels the strategy adopted for pseudoscalar quarkonia in the {\tt NRFF1.0} framework~\cite{Celiberto:2025euy}, where all partonic inputs are available at NLO accuracy.
In the present case, the SDCs entering the {\tt TQ4Q2.0} functions are currently known at LO, while the DGLAP evolution is implemented at NLO.

It is worth emphasizing, however, that the perturbative complexity of the LO SDC calculation for all-heavy tetraquarks is already comparable to that of higher-order computations in quarkonium production.
In particular, the evaluation of LO tetraquark SDCs involves multiheavy-quark topologies of complexity analogous to the real-emission contributions entering the NLO calculation of pseudoscalar quarkonium SDCs.
A direct comparison between Fig.~\ref{fig:FF_diagrams} and Fig.~1 of Ref.~\cite{Celiberto:2025euy} illustrates this point.

\begin{figure*}[t]
\centering

   \hspace{-0.00cm}
   \includegraphics[scale=0.670,clip]{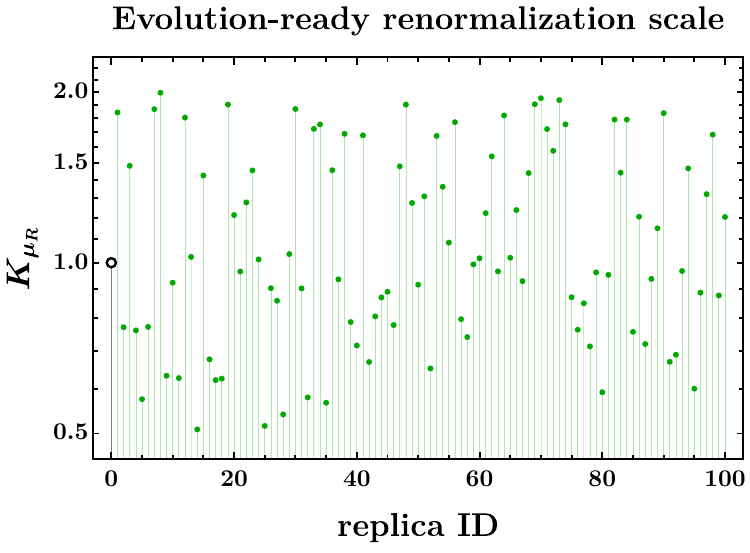}
   \hspace{0.20cm}
   \includegraphics[scale=0.670,clip]{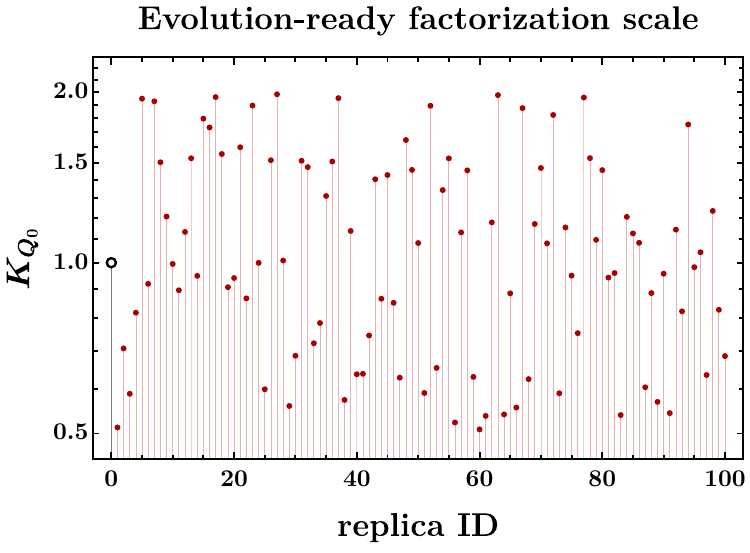}

\caption{
\justifying
\noindent
Replicalike sampling of perturbative F-MHOUs in the two-dimensional scale space.
$\mathcal{O}(100)$ configurations are generated by varying the renormalization scale $\mu_R$ and the evolution-ready scale $Q_0$ independently through pseudorandom factors $K_{\mu_R}$ and $K_{Q_0}$, sampled logarithmically in the range $[1/2,\,2]$ around their natural values.
The resulting ensemble defines the uncertainty band via its envelope.
}
\label{fig:er_T4Q_K}
\end{figure*}

In previous implementations of the {\HFNRevo} scheme, such as the {\tt TQ4Q1.1} study~\cite{Celiberto:2025ziy} and the pseudoscalar quarkonium analysis leading to the {\tt NRFF1.0} set~\cite{Celiberto:2025euy}, perturbative-fragmentation scale uncertainties (F-MHOUs) were estimated through one-dimensional variations of the evolution-ready scale $Q_0$ around its natural value, typically chosen as $Q_0 = 5 m_Q$.
This procedure provided a first assessment of missing higher-order effects in the FF initial conditions by generating a set of replicas associated with scale variations in the range $[1/2,\,2]$.

In the present work, we extend this strategy to a fully multidimensional treatment of F-MHOUs, consistent with the general philosophy of the {\HFNRevo} scheme.
As illustrated in Fig.~\ref{fig:er_T4Q_K}, we perform a replicalike sampling of the perturbative parameter space by simultaneously and independently varying the renormalization scale $\mu_R$ and the evolution-ready scale $Q_0$.
For each configuration, pseudorandom factors $K_{\mu_R}$ and $K_{Q_0}$ are generated with logarithmic sampling in the range $[1/2,\,2]$ around their natural values, leading to an ensemble of $\mathcal{O}(100)$ FF replicas.

The resulting set defines an uncertainty band through its envelope, providing a robust and dynamically correlated estimate of missing higher-order contributions.
Unlike the previous one-dimensional approach, this method captures the interplay between different perturbative scales and enables a more faithful propagation of uncertainties to collider-level observables.

Beyond uncertainty estimation, the replica ensemble offers a natural starting point for future data-driven analyses.
In particular, it provides a flexible framework for constraining LDMEs through fits to differential distributions at high-luminosity facilities such as the HL-LHC, where sensitivity to fragmentation dynamics is expected to improve significantly.
Moreover, the replica structure is well suited for integration with modern inference techniques, including machine-learning-assisted fits, enabling efficient exploration of multidimensional parameter spaces and correlations among perturbative and nonperturbative inputs.

The {\tt TQ4Q2.0} analysis presented here constitutes the first application of the full {\HFNRevo} uncertainty-quantification strategy to bound-state fragmentation, including all-heavy exotic systems.

\begin{figure*}[!t]
\centering

   \hspace{-0.00cm}
   \includegraphics[scale=0.410,clip]{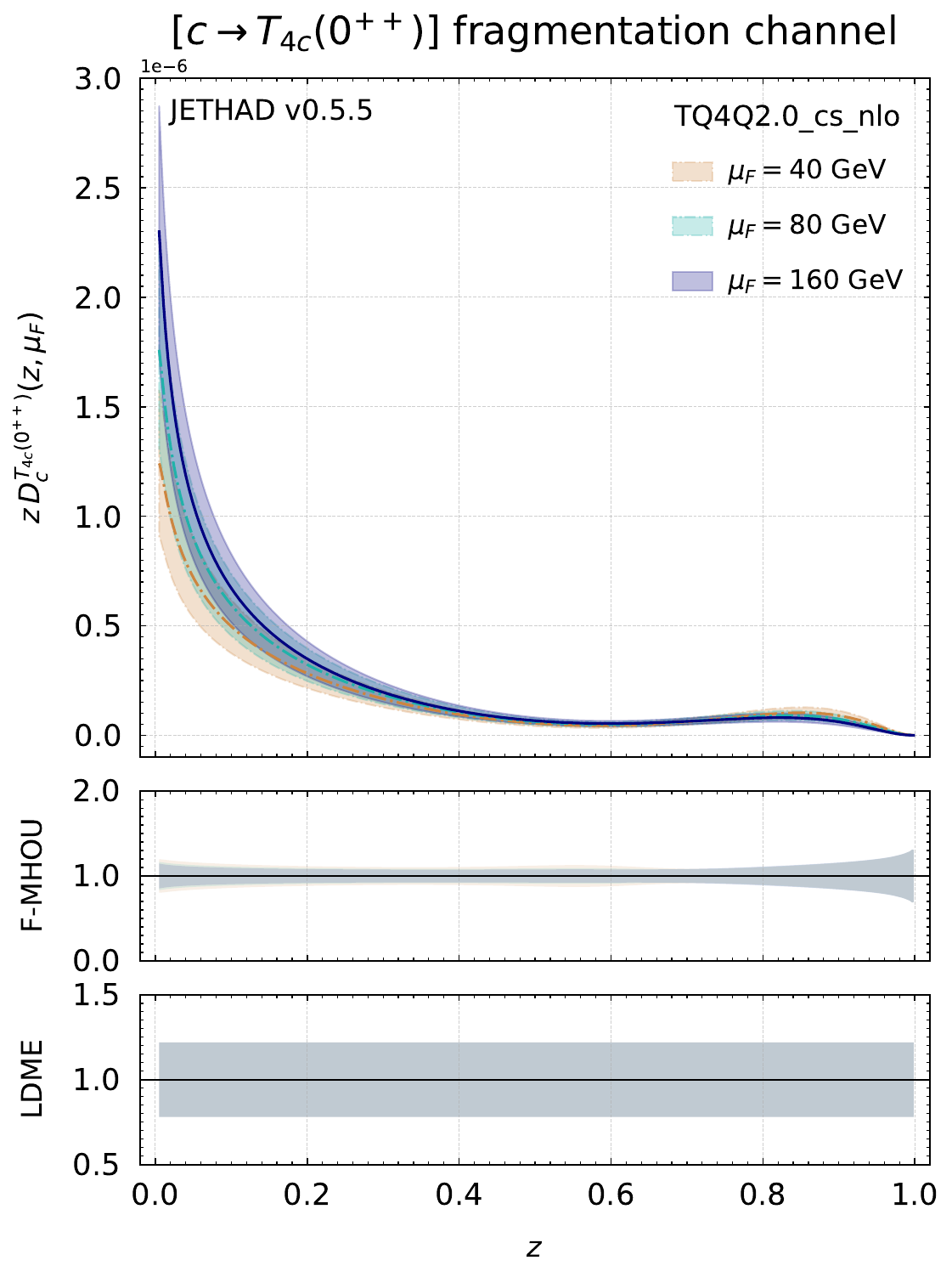}
   \hspace{0.90cm}
   \includegraphics[scale=0.410,clip]{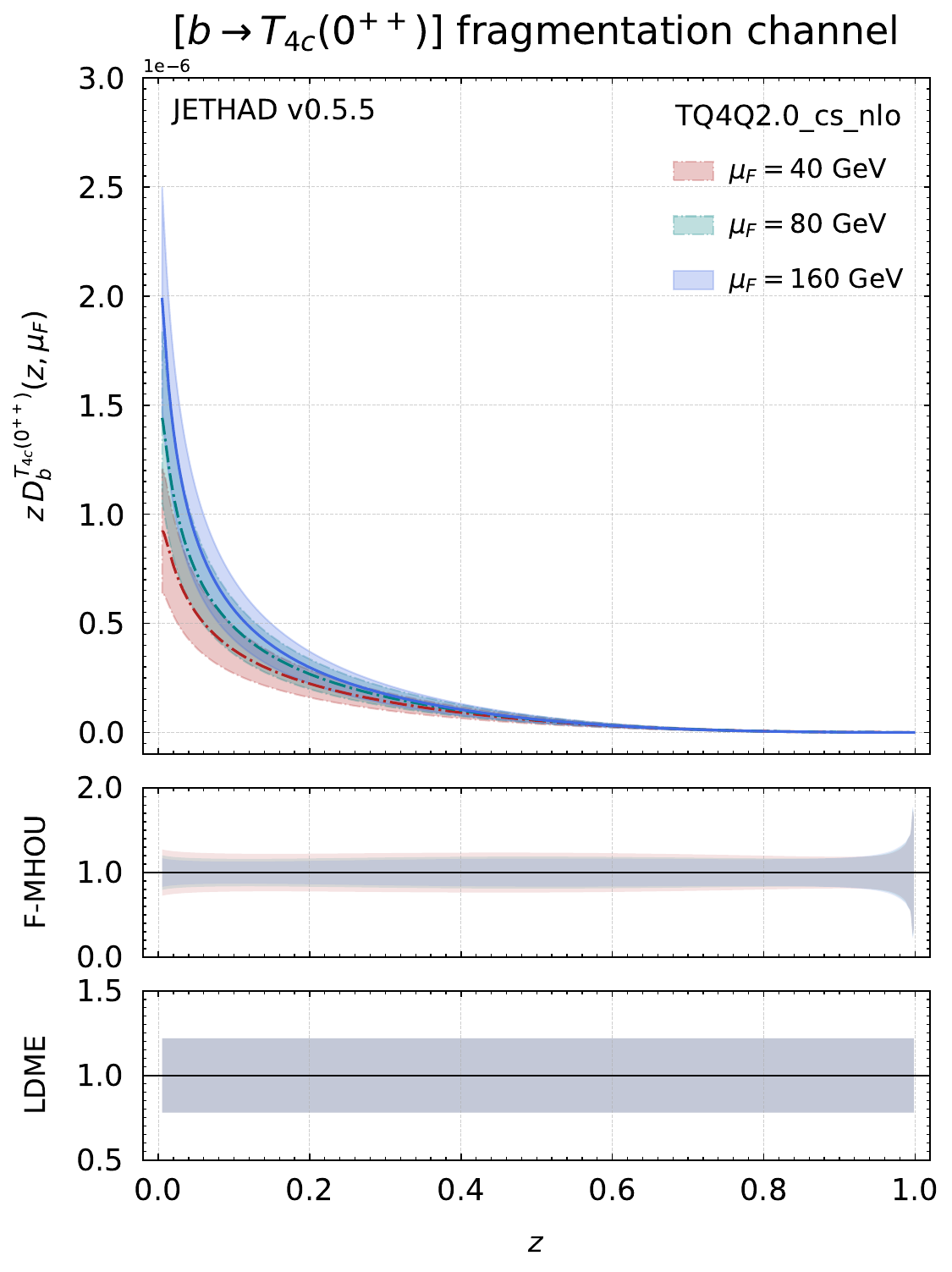}
%   \hspace{0.05cm}

   \vspace{0.25cm}

   \hspace{-0.00cm}
   \includegraphics[scale=0.410,clip]{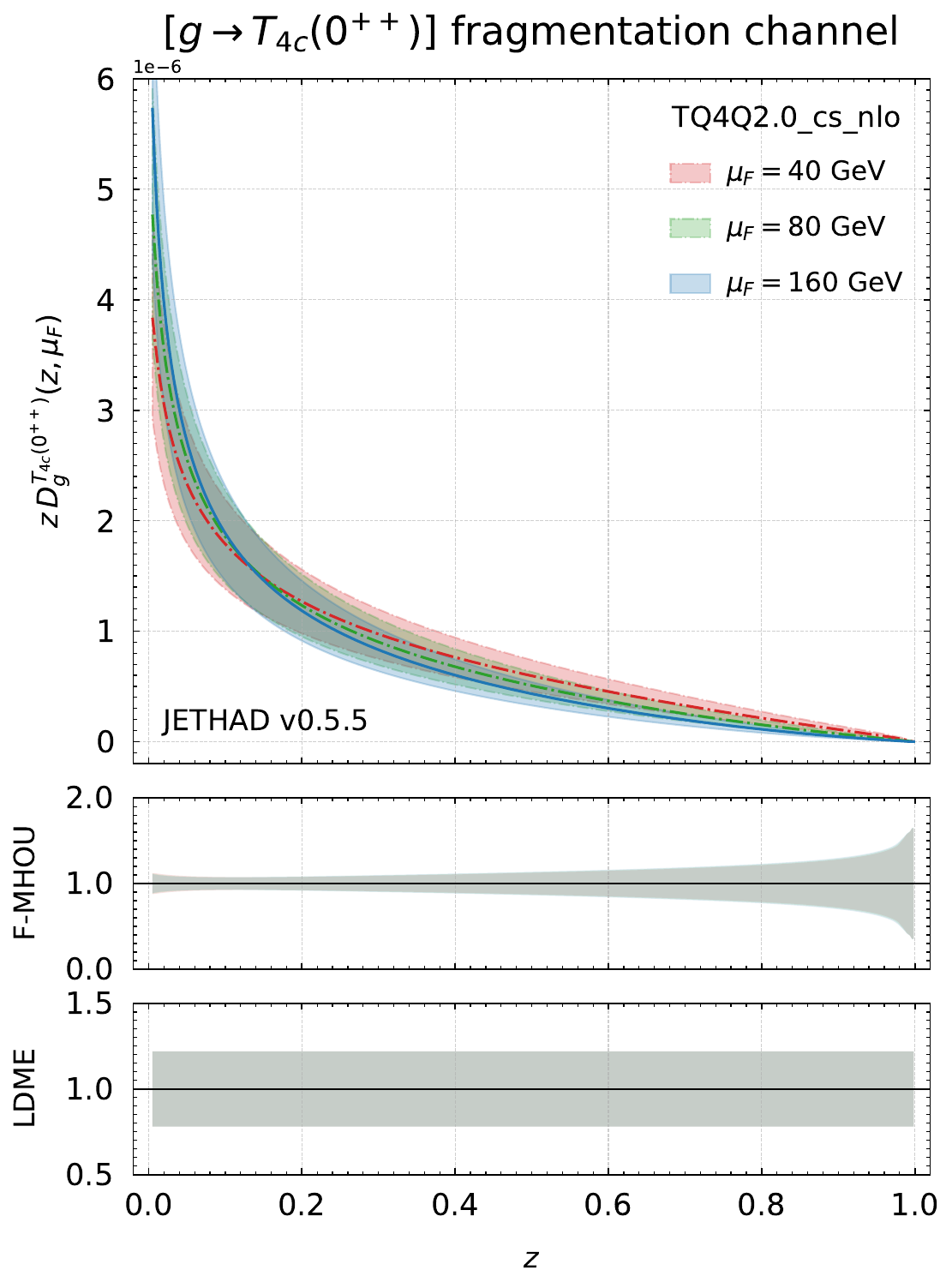}
   \hspace{0.90cm}
   \includegraphics[scale=0.410,clip]{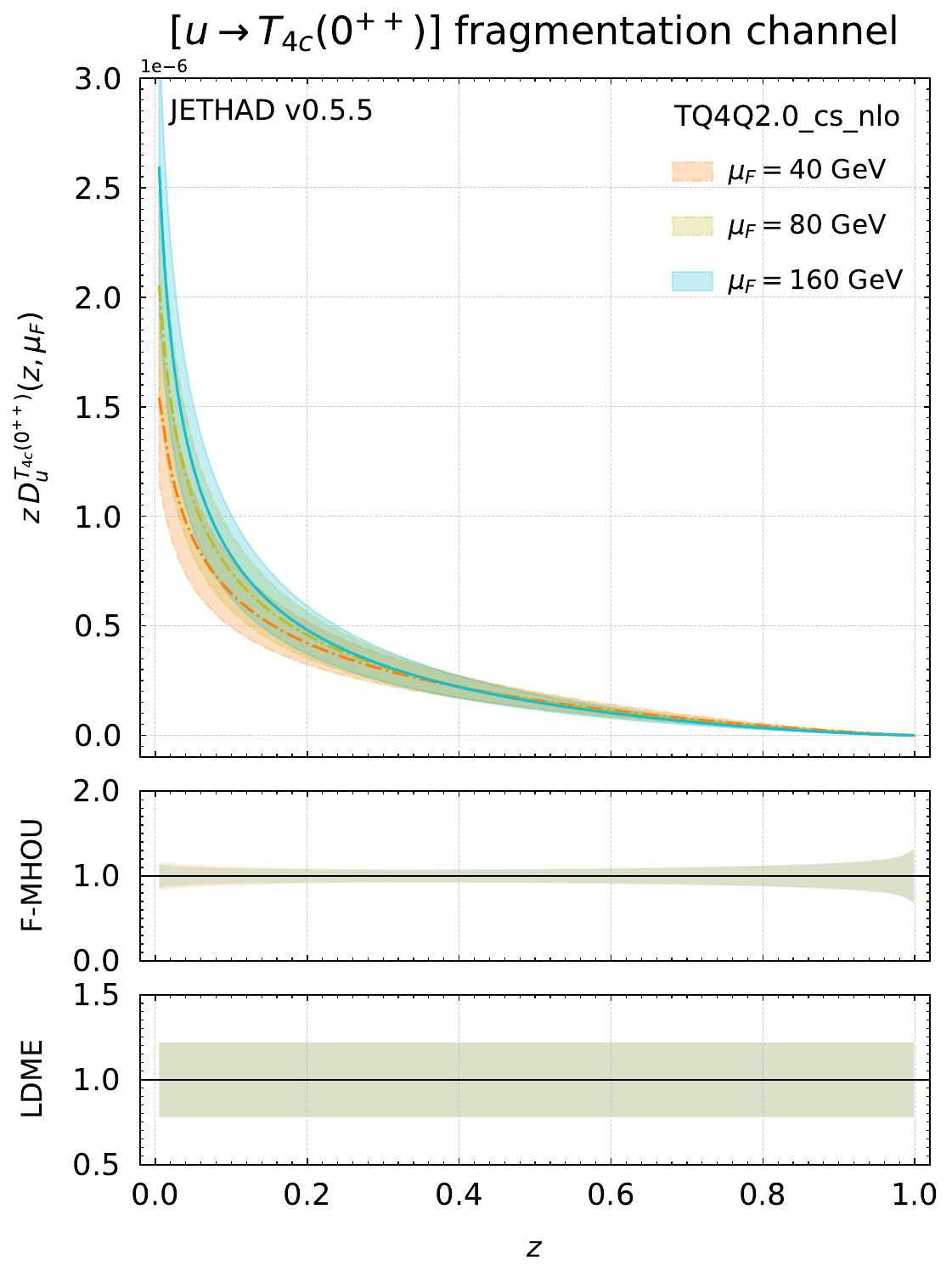}
%   \hspace{0.05cm}

\caption{
\justifying
\noindent
Momentum dependence of the {\tt TQ4Q2.0} FFs for all-charm scalar tetraquarks, $\TQcZpp$, at different energy scales.
Shaded bands in the main panels denote the total uncertainty, obtained by combining F-MHOUs and LDME variations.
The first auxiliary panel highlights the effect of F-MHOUs through the replica envelope normalized to the central prediction, while the second isolates LDME uncertainties as ratios to the central curve.}
\label{fig:FFs-z_Tc0}
\end{figure*}

\begin{figure*}[!t]
\centering

   \hspace{-0.00cm}
   \includegraphics[scale=0.410,clip]{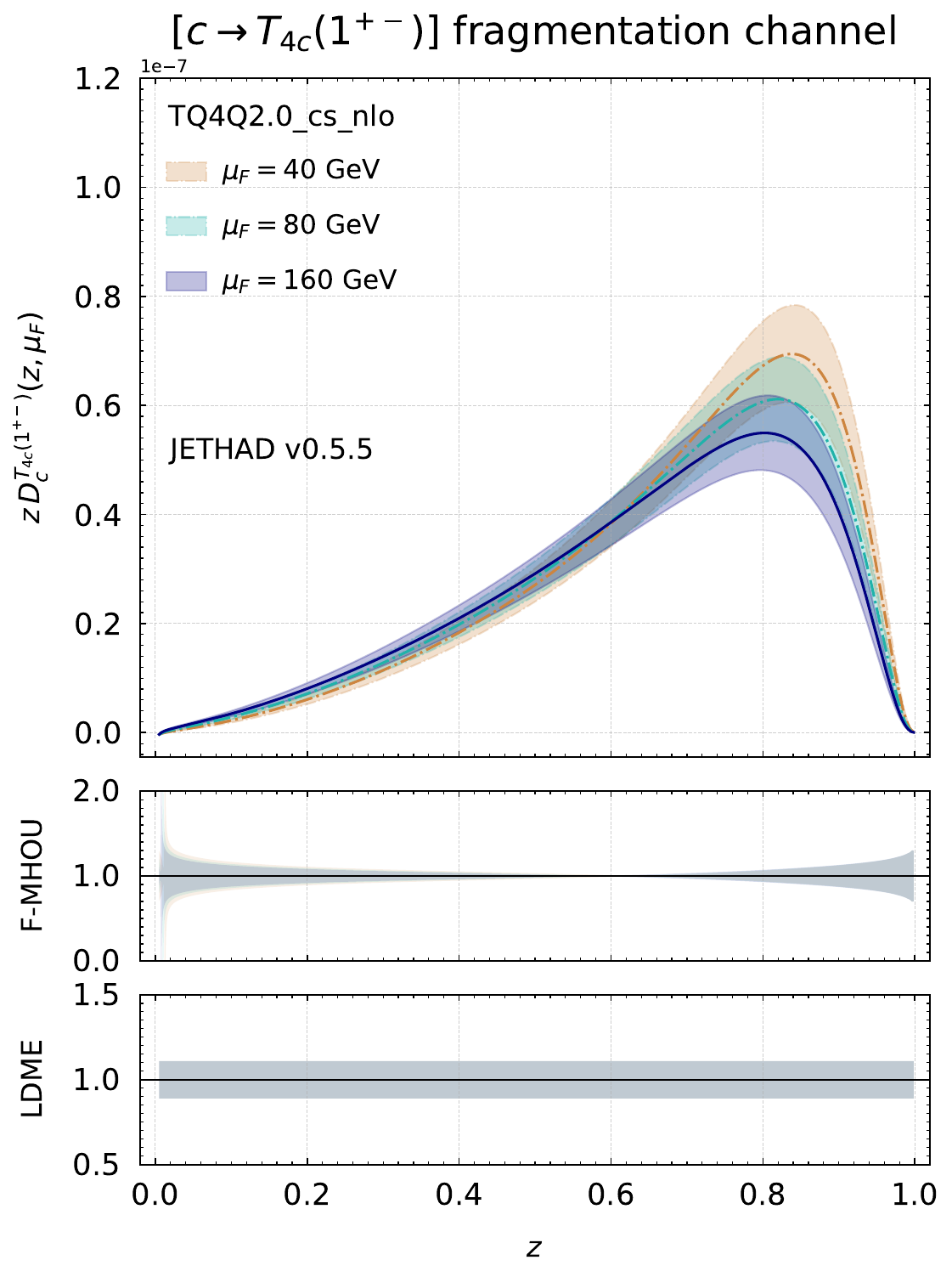}
   \hspace{0.90cm}
   \includegraphics[scale=0.410,clip]{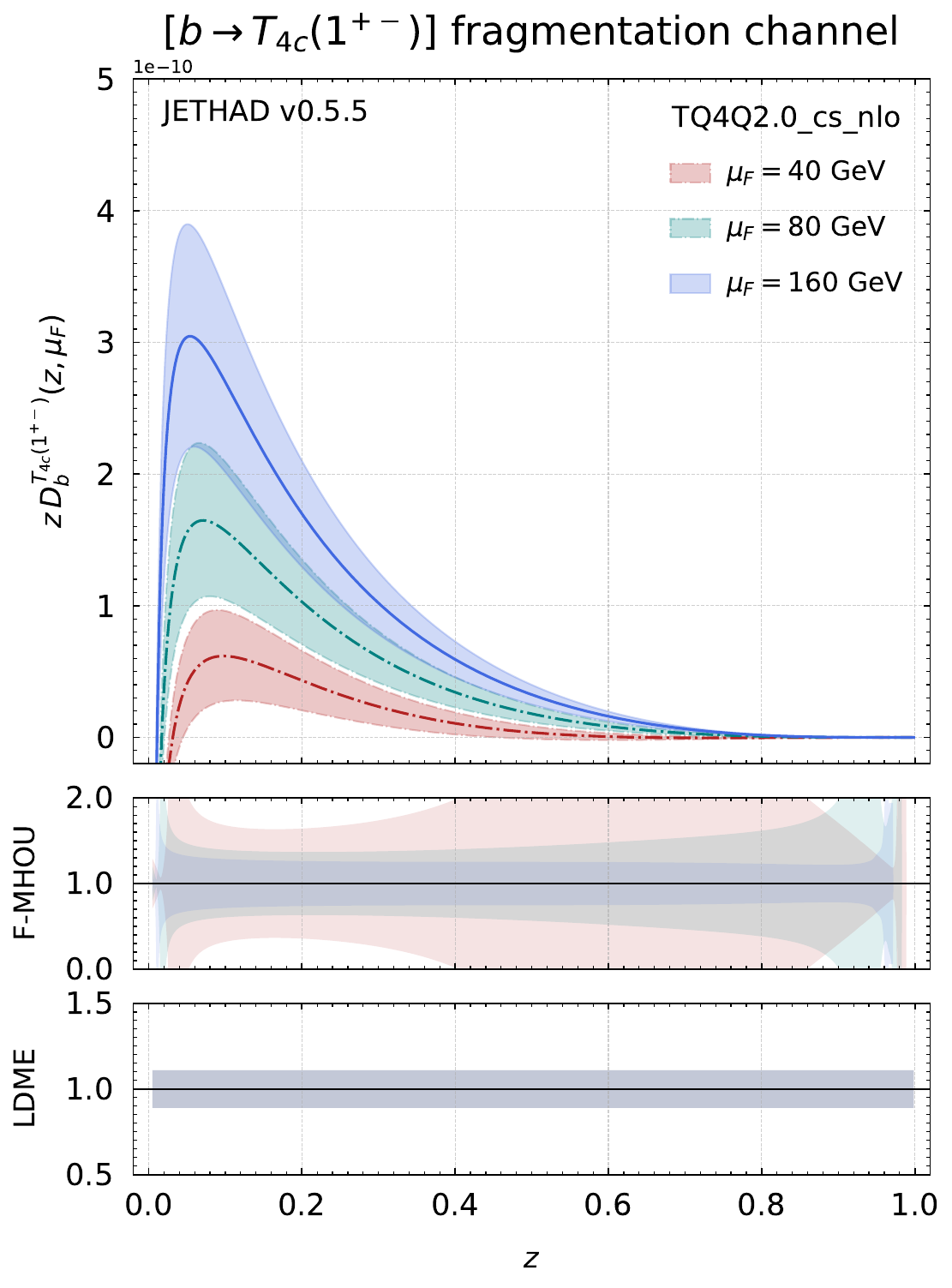}
%   \hspace{0.05cm}

   \vspace{0.25cm}

   \hspace{-0.00cm}
   \includegraphics[scale=0.410,clip]{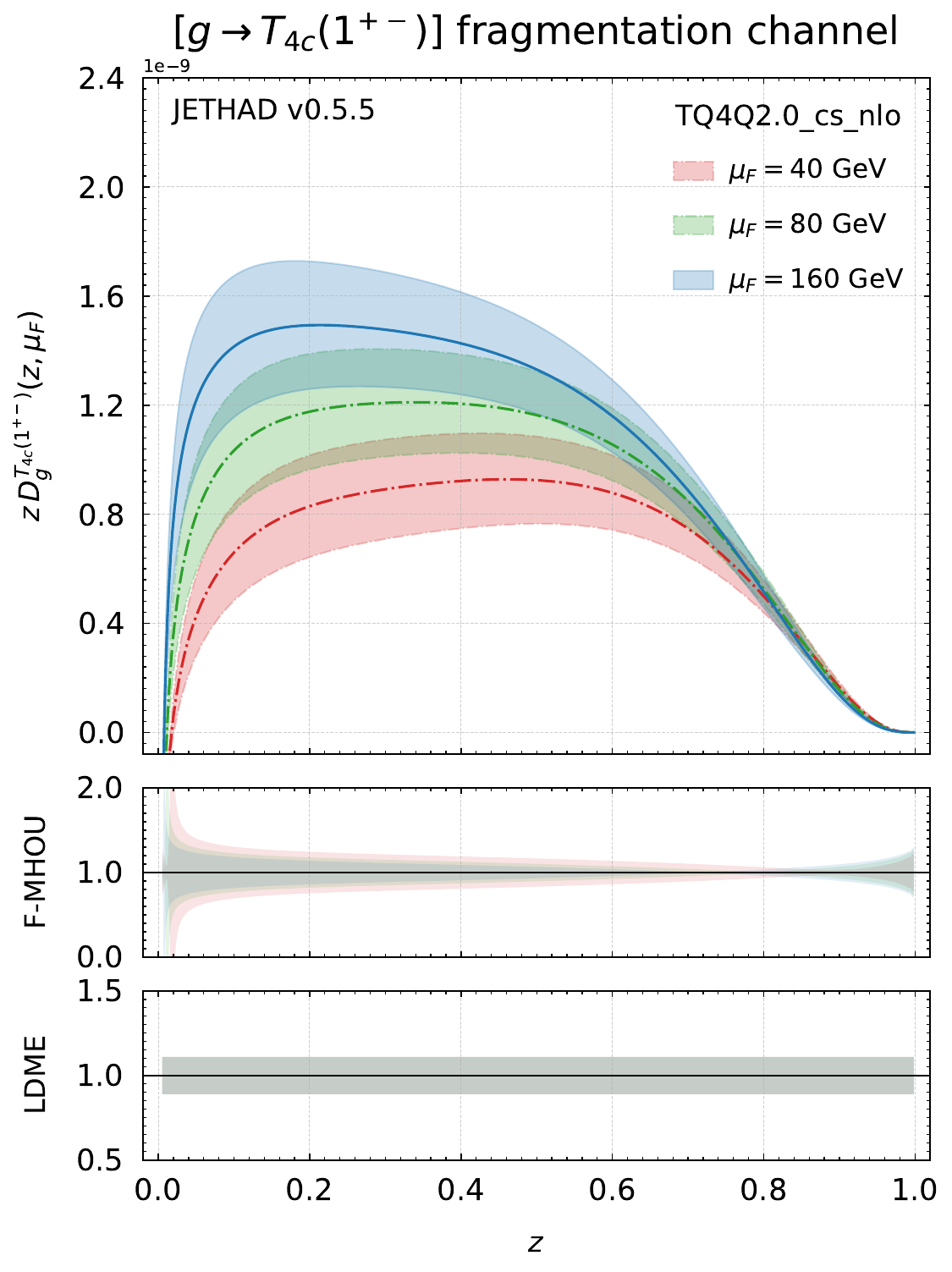}
   \hspace{0.90cm}
   \includegraphics[scale=0.410,clip]{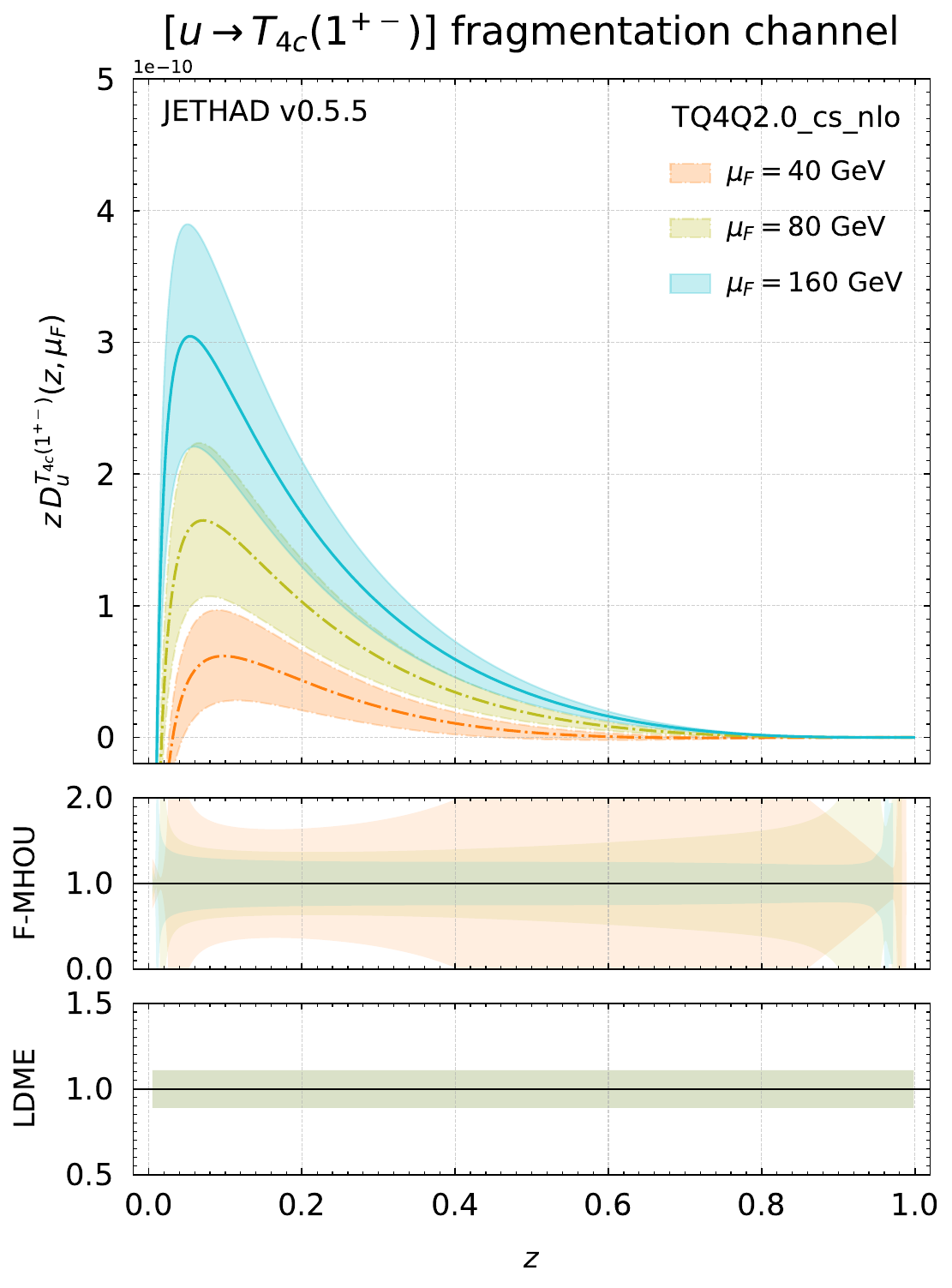}
%   \hspace{0.05cm}

\caption{
\justifying
\noindent
Momentum dependence of the {\tt TQ4Q2.0} FFs for all-charm axial-vector tetraquarks, $\TQcOpm$, at different energy scales.
Shaded bands in the main panels denote the total uncertainty, obtained by combining F-MHOUs and LDME variations.
The first auxiliary panel highlights the effect of F-MHOUs through the replica envelope normalized to the central prediction, while the second isolates LDME uncertainties as ratios to the central curve.}
\label{fig:FFs-z_Tc1}
\end{figure*}

\begin{figure*}[!t]
\centering

   \hspace{-0.00cm}
   \includegraphics[scale=0.410,clip]{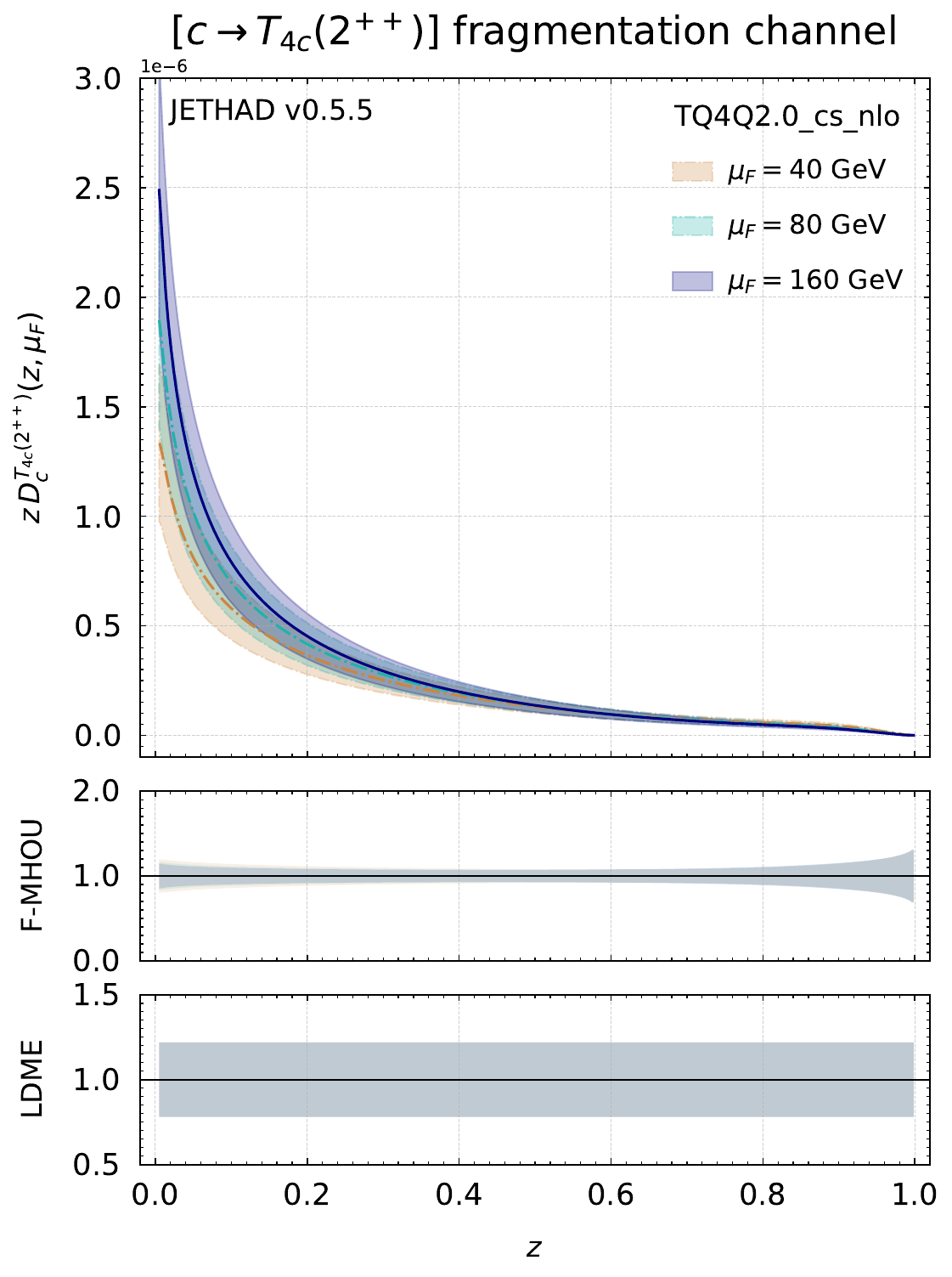}
   \hspace{0.90cm}
   \includegraphics[scale=0.410,clip]{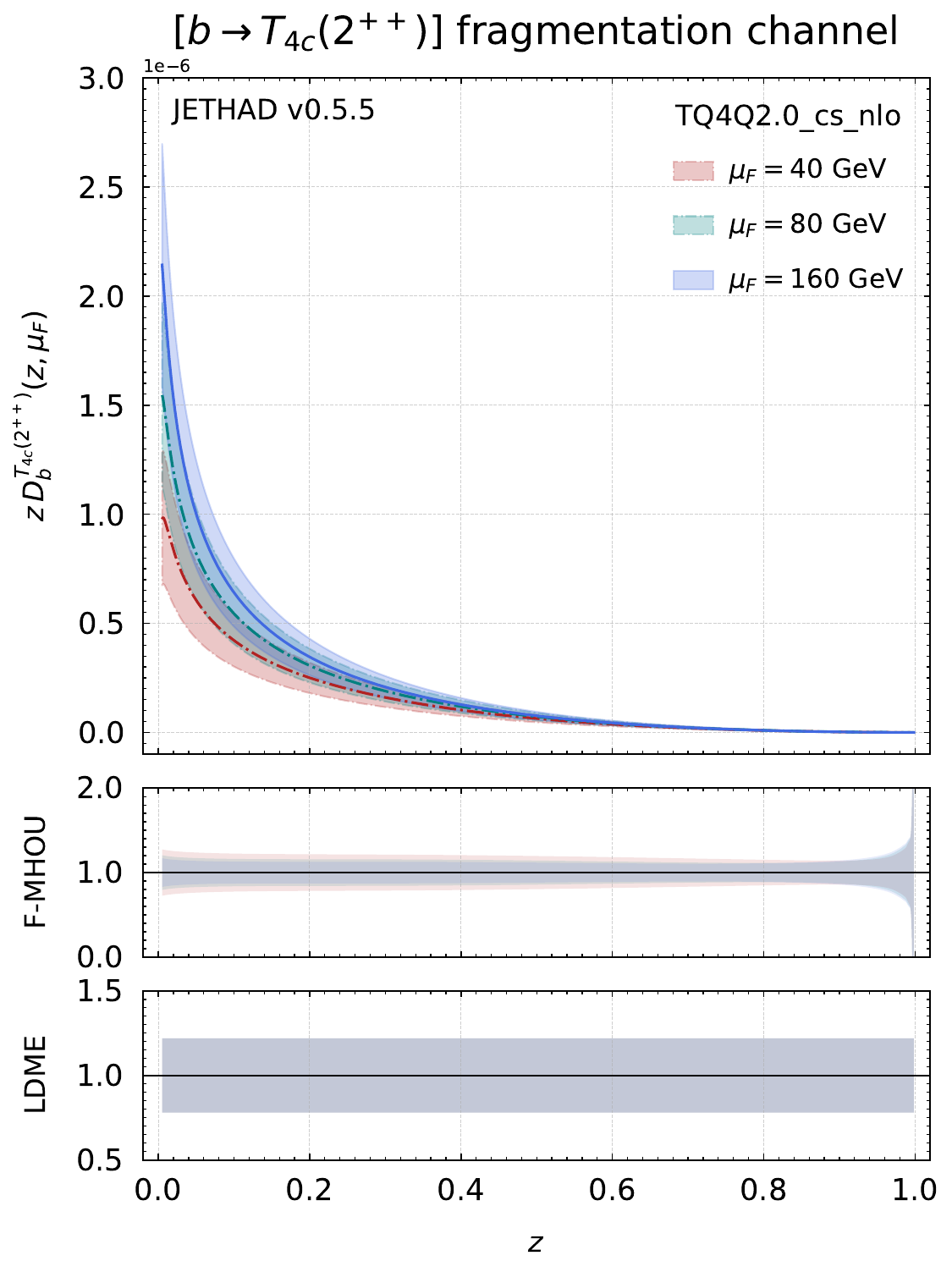}
%   \hspace{0.05cm}

   \vspace{0.25cm}

   \hspace{-0.00cm}
   \includegraphics[scale=0.410,clip]{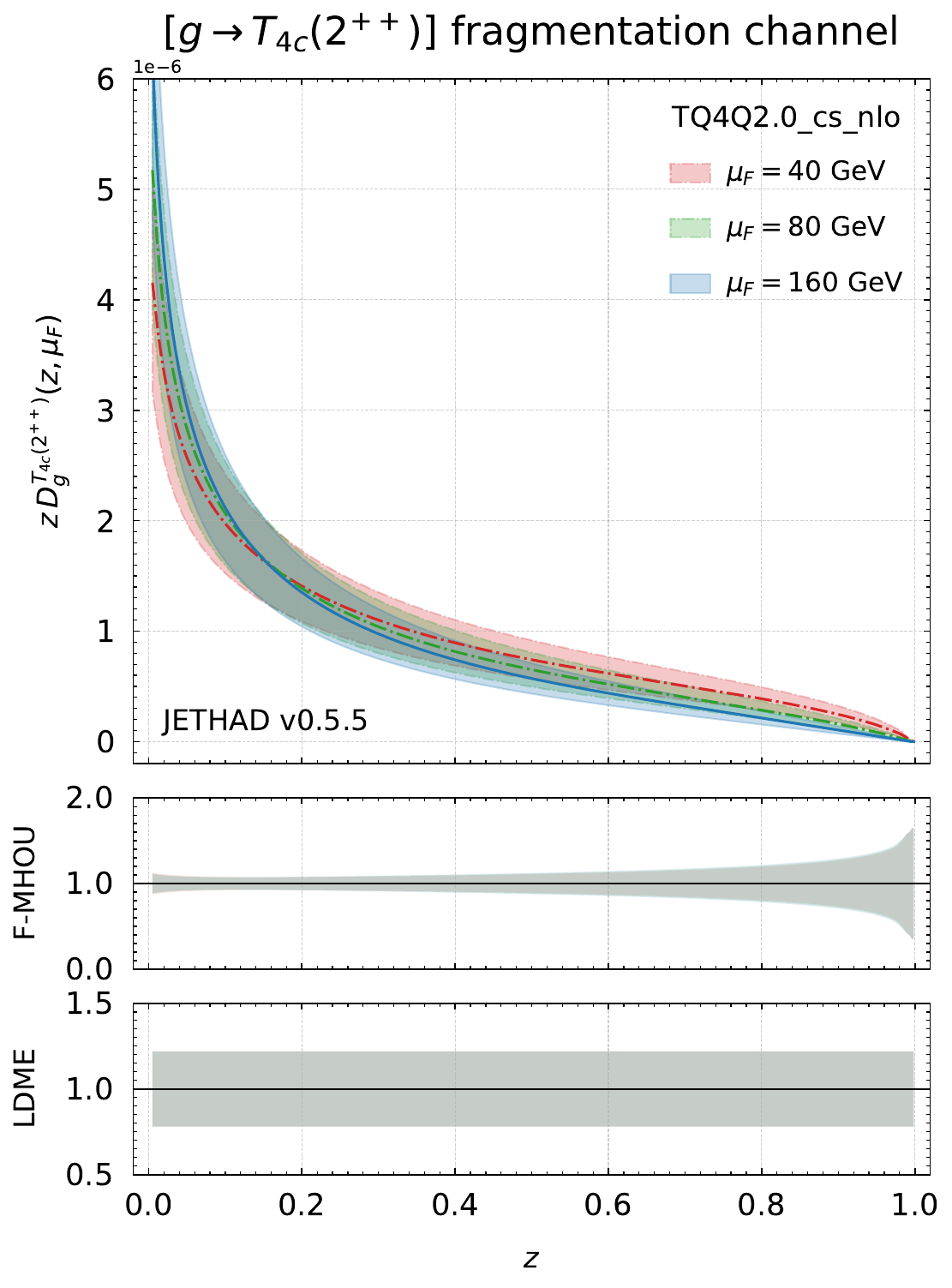}
   \hspace{0.90cm}
   \includegraphics[scale=0.410,clip]{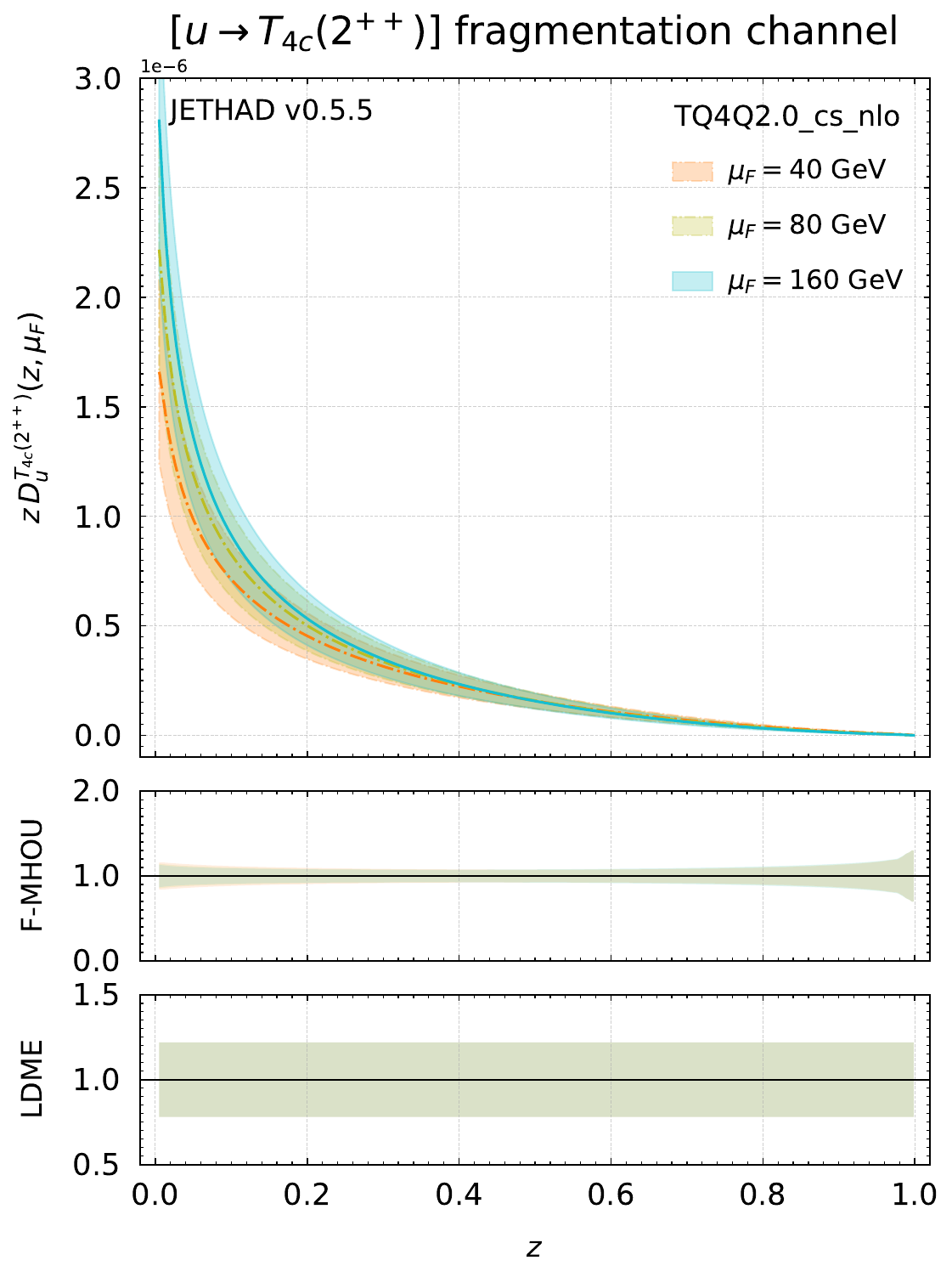}
%   \hspace{0.05cm}

\caption{
\justifying
\noindent
Momentum dependence of the {\tt TQ4Q2.0} FFs for all-charm tensor tetraquarks, $\TQcTpp$, at different energy scales.
Shaded bands in the main panels denote the total uncertainty, obtained by combining F-MHOUs and LDME variations.
The first auxiliary panel highlights the effect of F-MHOUs through the replica envelope normalized to the central prediction, while the second isolates LDME uncertainties as ratios to the central curve.}
\label{fig:FFs-z_Tc2}
\end{figure*}

\vspace{1em}
\noindent
\textbf{Momentum dependence.}
The $z$-dependence of the {\tt TQ4Q2.0} FFs, multiplied by $z$, is shown in Figs.~\ref{fig:FFs-z_Tc0},~\ref{fig:FFs-z_Tc1}, and~\ref{fig:FFs-z_Tc2} for the $0^{++}$, $1^{+-}$, and $2^{++}$ all-charm tetraquarks.
For simplicity, only the up-quark channel is displayed among light flavors, as mass differences induce negligible effects in both SDCs and evolution.
The corresponding all-bottom FFs are collected in Appendix~\hyperlink{app:B}{B}.
In all figures, the main panels display the total uncertainty from F-MHOUs and LDMEs, combined in quadrature.
The ancillary panels separate these effects: the first shows F-MHOUs via the replica envelope normalized to the central prediction, while the second reports LDME variations as ratios to the central curve.

To assess the impact of perturbative QCD evolution, we present the FFs at three representative factorization scales, $\mu_F = 40$, $80$, and $160$~GeV.

The $z$-dependence of the $[c \to \TQc]$ FFs exhibits a sharply channel-dependent behavior.
In the scalar (Figs.~\ref{fig:FFs-z_Tc0}, upper left) and tensor (Figs.~\ref{fig:FFs-z_Tc2}, upper left) channels, the distributions are dominated by a low-$z$ component: they start from a finite value at $z=0$ and decrease monotonically, with only a mild enhancement at large $z$ ($z \sim 0.8$--$0.9$).
The increasing steepness at small $z$ with growing $\mu_F$ reflects the amplification of radiative emissions driven by DGLAP evolution.
This pattern signals a soft-dominated fragmentation regime, where momentum is efficiently redistributed toward low $z$, while the residual high-$z$ tail indicates configurations in which the heavy quark retains a substantial fraction of the hadron momentum, in line with heavy-flavor expectations~\cite{Suzuki:1977km,Bjorken:1977md,Braaten:1993mp}.

A qualitatively different structure emerges in the axial-vector channel (Fig.~\ref{fig:FFs-z_Tc1}, upper left).
Here the FF vanishes at $z \to 0$ and develops a pronounced, localized peak at $0.75 \lesssim z \lesssim 0.9$, largely stable under scale evolution.
This reflects a strong bias toward hard fragmentation and a reduced phase space for soft radiation.

The gluon-initiated FFs reinforce this dichotomy.
For scalar and tensor states (Figs.~\ref{fig:FFs-z_Tc0} and~\ref{fig:FFs-z_Tc2}, lower left), the distributions are again dominated by low-$z$ contributions and decrease monotonically, without any sizable large-$z$ enhancement.
By contrast, the axial-vector gluon FF (Fig.~\ref{fig:FFs-z_Tc1}, lower left) exhibits a broad maximum in the intermediate region, $0.2 \lesssim z \lesssim 0.5$, which becomes more pronounced with increasing $\mu_F$, while vanishing at $z \to 0$ and being suppressed at large $z$.
This distinctive behavior originates from the fact that the axial-vector gluon FF is entirely generated by DGLAP evolution, with no nonperturbative input at the initial scale, thus providing a direct probe of the radiative structure of the evolution kernel.

A similar analysis can be extended to the nonconstituent quark channels, namely $[\tilde{b} \to \TQc]$ and $[u \to \TQc]$, shown in the upper-right and lower-right panels of Figs.~\ref{fig:FFs-z_Tc0} and~\ref{fig:FFs-z_Tc2}.
In the scalar and tensor cases, these contributions follow a pattern comparable to the charm channel, with a monotonic decrease in $z$ and no sizable large-$z$ enhancement, while remaining suppressed with respect to the gluon channel.
In the axial-vector configuration, both channels are entirely generated by DGLAP evolution and therefore exhibit shapes qualitatively similar to the gluon case, characterized by an intermediate-$z$ enhancement.
Their normalization is nonetheless reduced by about one order of magnitude, reflecting their purely radiative origin.

These findings are consistent with the observations of Ref.~\cite{Bai:2024flh}, where light-quark fragmentation is found to be subleading with respect to gluon-induced production, yet competitive with heavy-quark channels in certain kinematic regimes.
In our framework, this hierarchy emerges dynamically at the level of FFs, with gluon dominance, followed by nonconstituent quarks, and finally the constituent heavy-quark contribution.

Taken together, these results reveal a clear and robust pattern: all-charm tetraquark fragmentation is soft-dominated in the scalar and tensor channels, while the axial-vector configuration is intrinsically hard.
This dichotomy constitutes a distinctive signature of the production mechanism and provides a direct handle for probing fragmentation dynamics through rapidity and jet-associated observables at the HL-LHC.
The quantitative impact of including nonconstituent quark channels on collider observables will be assessed in Sec.~\ref{sec:phenomenology}.

A comparison across quantum numbers reveals a clear hierarchy in normalization, with axial-vector FFs systematically suppressed with respect to the scalar and tensor channels.
This reduction, exceeding one order of magnitude in both quark- and gluon-initiated processes, is consistent with the NRQCD structure of the $1^{+-}$ state and the reduced overlap with leading production mechanisms.

As the factorization scale increases, all FFs exhibit a progressive softening of their $z$-shape, with enhanced population at small $z$ driven by DGLAP evolution.
Despite this effect, the overall structure of the distributions remains stable, indicating that the main features of the fragmentation pattern are already established at the initial scale and preserved under evolution.

These observations further support the emerging picture of all-charm tetraquark fragmentation, where scalar and tensor states are predominantly soft-driven, while the axial-vector configuration is intrinsically hard.

\begin{figure*}[!t]
\centering

   \hspace{-0.00cm}
   \includegraphics[scale=0.410,clip]{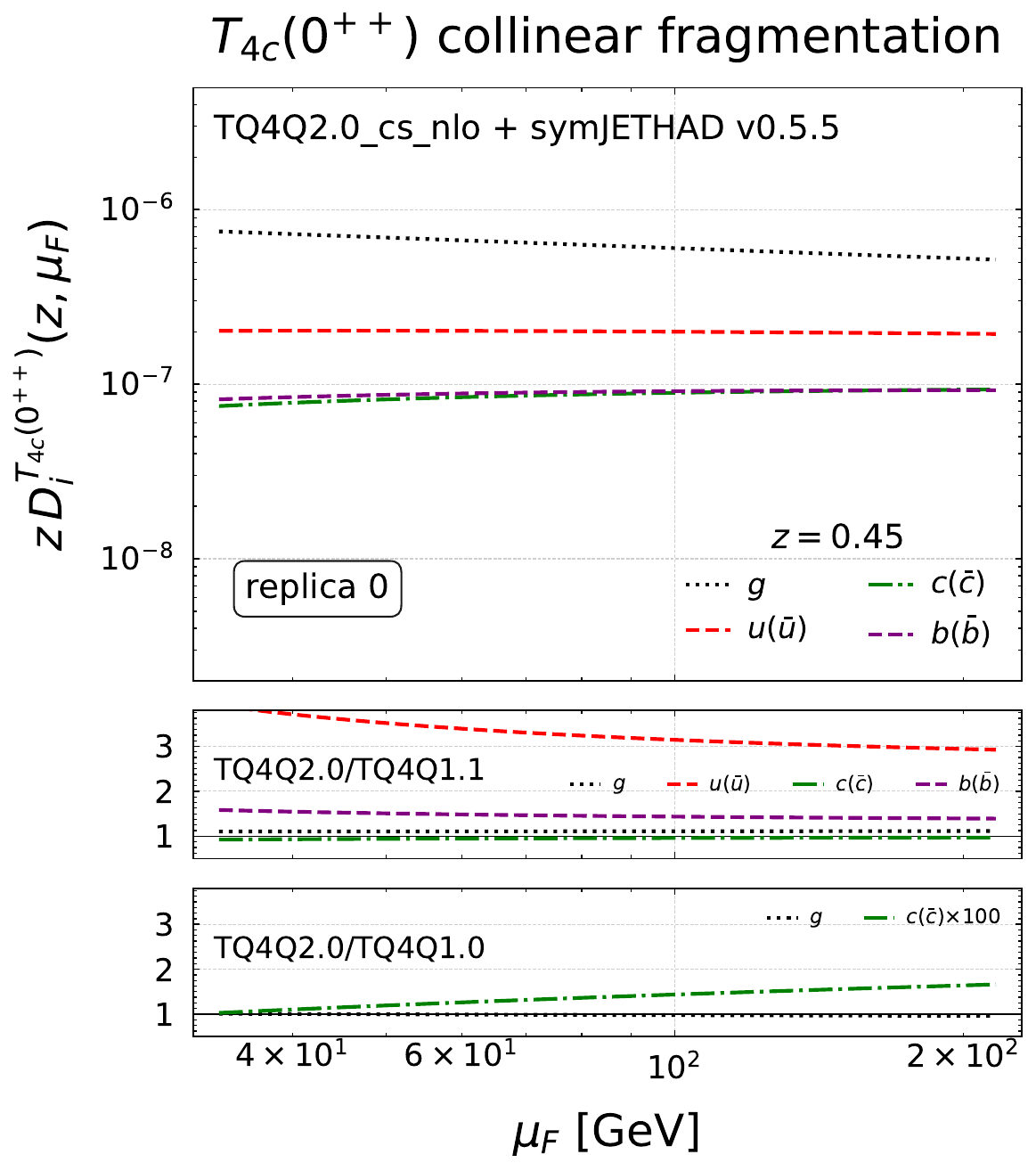}
   \hspace{0.90cm}
   \includegraphics[scale=0.410,clip]{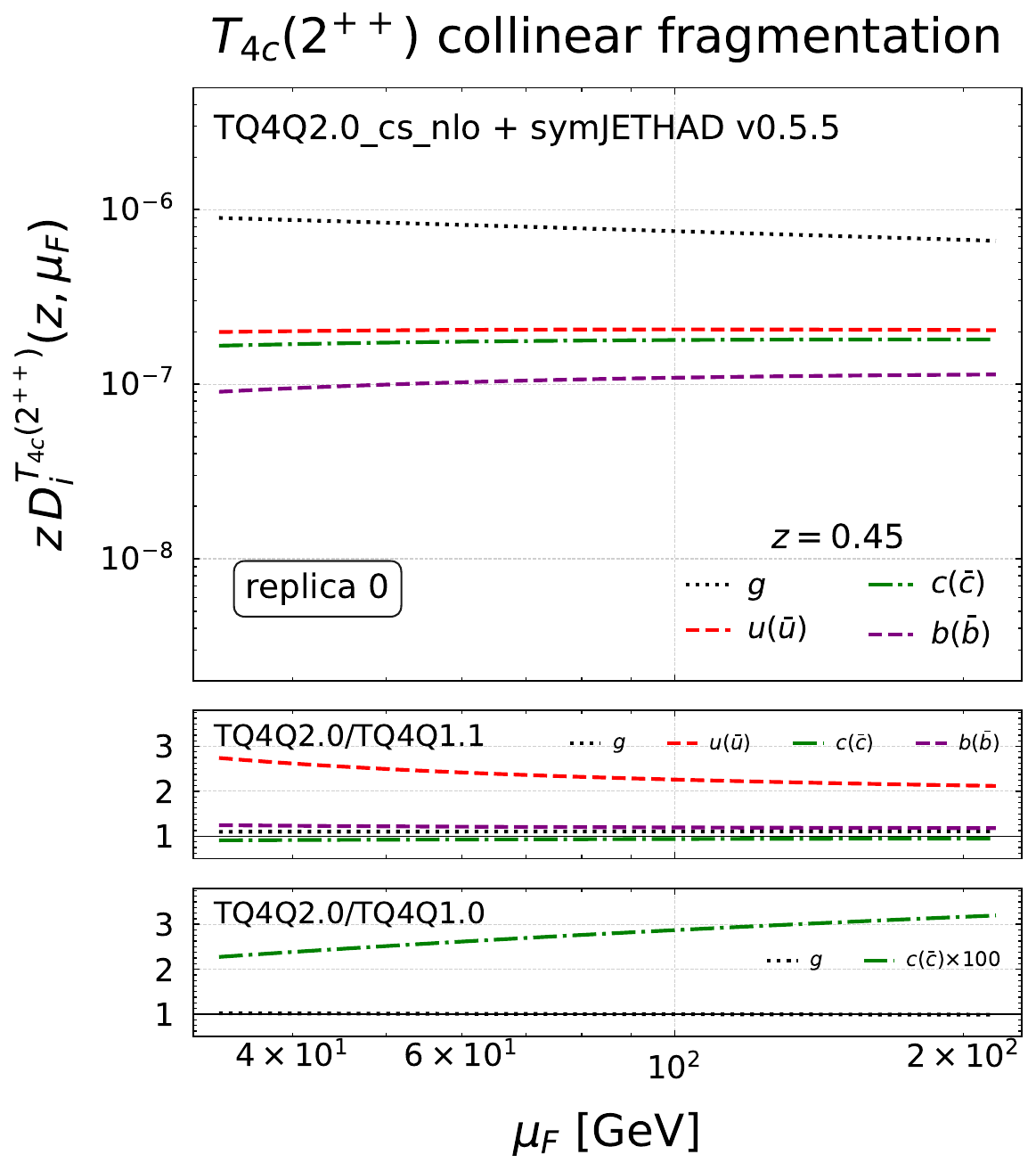}

%\captionsetup{format=plain, justification=justified, singlelinecheck=false}
\caption{
\justifying
\noindent
Energy dependence of the {\tt TQ4Q2.0} FFs for all-charm scalar ($0^{++}$, left) and tensor ($2^{++}$, right) tetraquarks, evaluated at $z = \langle z \rangle \simeq 0.45$.
Ratios of the new {\tt TQ4Q2.0} functions (this work) to the previous {\tt TQ4Q1.1}~\protect\cite{Celiberto:2025dfe,Celiberto:2025ziy} and {\tt TQ4Q1.0}~\protect\cite{Celiberto:2024mab} sets are shown in the first and second ancillary panels, respectively.
}
\label{fig:FFs-muF_TQc}
\end{figure*}

\vspace{1em}
\noindent
\textbf{Energy dependence.}
Figure~\ref{fig:FFs-muF_TQc} highlights the evolution with $\mu_F$ of the gluon-, charm-, up-, and bottom-initiated {\tt TQ4Q2.0} FFs for $\TQcZpp$ (left) and $\TQcTpp$ (right), shown as $z D(z,\mu_F)$.
The comparison is performed at $z = \langle z \rangle \simeq 0.45$, representative of high-energy hadroproduction~\cite{Celiberto:2020wpk,Celiberto:2021dzy,Celiberto:2021fdp,Celiberto:2022dyf,Celiberto:2022keu,Celiberto:2024omj}, and only central predictions are displayed.
For simplicity, we restrict the discussion to these two states, as they are the only ones for which all sets in the {\tt TQ4Q1.x} and {\tt TQ4Q2.0} families are available, allowing for a consistent comparison across releases.

The main panels display the absolute scale dependence, while the two ancillary panels show the ratios to the previous {\tt TQ4Q1.1}~\cite{Celiberto:2025ziy} and {\tt TQ4Q1.0}~\cite{Celiberto:2024mab} determinations, respectively.

A clear hierarchy among partonic channels emerges in both scalar and tensor configurations.
The gluon FF is the dominant contribution across the full $\mu_F$ range, while the up-quark channel provides the second-largest one.
The ordering of the heavy-quark channels is instead state dependent: in the scalar case, the charm and bottom FFs remain relatively close to each other, whereas in the tensor case the charm contribution becomes more clearly separated from the bottom one and moves closer to the light-quark channel.
This hierarchy is stable under evolution and reflects the interplay between initial-scale inputs and DGLAP-generated components.

The dependence on $\mu_F$ is overall mild for all channels.
The gluon FF exhibits a slow decrease with increasing scale, the light-quark contribution remains nearly flat, while the charm and bottom channels display a slight growth, consistent with timelike DGLAP evolution and the progressive buildup of quark contributions from gluon radiation.
Importantly, the relative hierarchy among channels remains essentially unchanged across the full scale range, indicating that evolution redistributes strength without altering the underlying fragmentation pattern.

The comparison with previous FF sets provides further insight.
The first ancillary panel shows that the {\tt TQ4Q2.0} functions are broadly consistent with the {\tt TQ4Q1.1} ones, with moderate deviations that are most pronounced in the nonconstituent quark channels.
In particular, the light-quark contribution is enhanced by up to a factor of a few, reflecting the improved treatment of multichannel initial conditions and scale variations.

A more informative comparison is provided by the second ancillary panel, which displays the ratio between the {\tt TQ4Q2.0} FFs and the earlier {\tt TQ4Q1.0} set~\cite{Celiberto:2024mab}, originally available only for the scalar and tensor charmed states.
In that construction, gluon and charm channels were modeled within a simplified framework, leading to a strongly suppressed charm contribution.

In contrast, the {\tt TQ4Q2.0} results exhibit a substantial enhancement of the charm FF across the full $\mu_F$ range, bringing it closer to the light-quark channel, while the gluon contribution remains comparatively stable.
This behavior reflects the transition from a Suzuki-inspired modeling~\cite{Suzuki:1977km} of heavy-quark fragmentation to a fully NRQCD-based construction with consistent multichannel inputs.
As a result, the relative weight of the charm channel is significantly increased, leading to a more balanced and dynamically consistent hierarchy among partonic contributions.

Again, these results are in line with the findings of Ref.~\cite{Bai:2024flh}, where light-quark fragmentation is found to be subleading with respect to gluon-induced production, yet competitive with heavy-quark channels.
In our case, this hierarchy is dynamically reproduced at the level of FFs, with gluon dominance followed by nonconstituent quarks and, finally, constituent heavy quarks.

An important feature of Fig.~\ref{fig:FFs-muF_TQc} is the smooth behavior of the gluon FFs as $\mu_F$ increases.
Across both scalar and tensor channels, the gluon contribution remains stable, with only mild variations over the considered range.
This regularity, already observed in other heavy-flavor systems~\cite{Celiberto:2022grc}, plays a crucial role in ensuring the perturbative stability of semi-inclusive observables.

This concept of \emph{natural stability}~\cite{Celiberto:2022grc}, first identified in heavy-meson production~\cite{Celiberto:2021dzy,Celiberto:2021fdp} and later confirmed for quarkonia~\cite{Celiberto:2022dyf,Celiberto:2025euy,Celiberto:2022keu,Celiberto:2024omj}, $B_c$-like states, and rare~\cite{Celiberto:2025ogy} or exotic hadrons~\cite{Celiberto:2023rzw,Celiberto:2024mab,Celiberto:2025dfe,Celiberto:2025ipt}, is further reinforced by the {\tt TQ4Q2.0} results.
In particular, the gluon channel, whether initialized at the evolution-ready scale $Q_0$ or generated radiatively above it, contributes in a smooth and controlled way across the full $\mu_F$ range, supporting robust predictions in the presence of higher-order corrections and MHOUs.

Overall, the energy dependence of the {\tt TQ4Q2.0} FFs provides a consistent picture of the relative strength and scale sensitivity of different fragmentation channels.
The interplay between gluon dominance and nonconstituent quark contributions emerges as a key feature, complementing the $z$-dependent analysis and setting the stage for phenomenological applications to exotic hadron production at large transverse momenta.

%==========================
\section{Hadron-collider phenomenology}
\label{sec:phenomenology}
%==========================

To support phenomenological studies, we provide predictions for rapidity and azimuthal-angle distributions relevant to inclusive tetraquark-plus-jet production at HL-LHC and FCC energies.
Our reference framework is the $\NLLp$ HyF scheme, where NLO collinear factorization, including DGLAP evolution of PDFs and FFs, is consistently supplemented by next-to-leading high-energy resummation through the NLO BFKL kernel and NLO forward-production impact factors.
This approach is naturally suited to semihard observables, which probe QCD dynamics in kinematic regions where fixed-order perturbation theory alone becomes insufficient.
The `$+$' prescription further accounts for additional subleading contributions generated by products of NLO impact factors.
A detailed description of the $\NLLp$ formalism employed in our analysis is given in Appendix~\hyperlink{app:C}{C}.

In this regime, large logarithms of energy can spoil perturbative stability, requiring dedicated resummation techniques.
The Balitsky-Fadin-Kuraev-Lipatov (BFKL) framework~\cite{Fadin:1975cb,Balitsky:1978ic} systematically resums leading and next-to-leading energy logarithms, allowing cross sections to be expressed as transverse-momentum convolutions of universal Green’s functions~\cite{Fadin:1998py,Ciafaloni:1998gs} with process-dependent emission functions embedding collinear inputs such as PDFs and FFs.
This structure naturally leads to the HyF scheme, which unifies high-energy and collinear dynamics~\cite{Colferai:2010wu,Celiberto:2015yba,Celiberto:2017ptm,Celiberto:2020wpk}.

BFKL resummation has been extensively applied to a wide class of processes, including NLO Mueller-Navelet jets \cite{Mueller:1986ey,Ducloue:2013hia,Colferai:2015zfa,Celiberto:2015yba,Celiberto:2015mpa,Celiberto:2016ygs,Celiberto:2017ius,Caporale:2018qnm,deLeon:2021ecb,Celiberto:2022gji,Baldenegro:2024ndr}, dihadron systems \cite{Celiberto:2016hae,Celiberto:2017ptm,Celiberto:2017ius,Celiberto:2020rxb,Celiberto:2022rfj}, hadron-jet correlations \cite{Bolognino:2018oth,Bolognino:2019cac,Bolognino:2019yqj,Celiberto:2020wpk,Celiberto:2020rxb,Celiberto:2022kxx}, multijet production \cite{Caporale:2015int,Caporale:2016soq,Caporale:2016xku,Celiberto:2016vhn,Caporale:2016zkc,Celiberto:2017ius}, forward Higgs \cite{Hentschinski:2020tbi,Celiberto:2022fgx,Celiberto:2020tmb,Mohammed:2022gbk,Celiberto:2023rtu,Celiberto:2023uuk,Celiberto:2023eba,Celiberto:2023nym,Celiberto:2023rqp,Celiberto:2022zdg,Celiberto:2024bbv}, Drell-Yan \cite{Celiberto:2018muu,Golec-Biernat:2018kem}, and heavy-flavor emissions \cite{Celiberto:2017nyx,Boussarie:2017oae,Bolognino:2019ouc,Bolognino:2019yls,Bolognino:2021mrc,Celiberto:2021dzy,Celiberto:2021fdp,Celiberto:2022dyf,Celiberto:2022grc,Celiberto:2022keu,Celiberto:2022kza,Celiberto:2024omj,Gatto:2025kfl}.

Complementary studies of single-forward emissions in the linear small-$x$ regime have provided insight into gluon dynamics through unintegrated gluon distributions, with analyses at HERA~\cite{Anikin:2011sa,Besse:2013muy,Bolognino:2018rhb,Bolognino:2018mlw,Bolognino:2019pba,Celiberto:2019slj,Bolognino:2021bjd,Luszczak:2022fkf} and the EIC~\cite{Bolognino:2021niq,Bolognino:2021gjm,Bolognino:2021bjd,Bolognino:2022uty,Bolognino:2022ndh}.
These developments have led to resummed PDFs~\cite{Ball:2017otu,Abdolmaleki:2018jln,Bonvini:2019wxf,Silvetti:2022hyc,Celiberto:2025nnq,Silvetti:2023suu,Rinaudo:2024hdb} and improved small-$x$ TMDs~\cite{Bacchetta:2020vty,Bacchetta:2024fci,Celiberto:2021zww}.

Heavy-flavor channels have proven particularly advantageous in this context.
Observables involving $\Lambda_c$~\cite{Celiberto:2021dzy} and $b$-hadron production~\cite{Celiberto:2021fdp} exhibit a pattern of natural stabilization~\cite{Celiberto:2022grc}, in contrast to light-hadron emissions, which are more sensitive to large higher-order corrections~\cite{Bolognino:2018oth,Celiberto:2020wpk}.
This behavior originates from the use of VFNS-based collinear fragmentation and motivates the construction of DGLAP-evolved FFs with NRQCD initial conditions~\cite{Braaten:1993mp,Zheng:2019dfk,Braaten:1993rw,Chang:1992bb,Braaten:1993jn,Ma:1994zt,Zheng:2019gnb,Zheng:2021sdo,Feng:2021qjm,Feng:2018ulg}.
Such developments have enabled precision studies ranging from quarkonia~\cite{Celiberto:2022dyf,Celiberto:2023fzz} to heavy mesons~\cite{Celiberto:2022keu,Celiberto:2024omj}, and have recently opened the way to fragmentation-based descriptions of exotic states, including doubly and fully heavy tetraquarks~\cite{Celiberto:2023rzw,Celiberto:2024beg,Celiberto:2024mab,Celiberto:2025dfe}, pentaquarks~\cite{Celiberto:2025ipt}, and triply heavy baryons~\cite{Celiberto:2025ogy}

Section~\ref{ssec:uncertainty} details the strategy adopted for a systematic assessment of theoretical uncertainties affecting the observables under study. 
Results for rapidity-differential rates and corresponding event yields are presented in Secs.~\ref{ssec:I} and~\ref{ssec:I-yld}, respectively, assuming the full integrated luminosity collected by CMS at $\sqrt{s} = 13$ TeV during Run 2 (2015--2018). 
Rapidity-differential distributions are presented for all-charm tetraquarks, whereas event yields are reported for both all-charm and all-bottom states for completeness. 
All numerical results are obtained using {\Jethad}~\cite{Celiberto:2020wpk,Celiberto:2022rfj,Celiberto:2023fzz,Celiberto:2024mrq,Celiberto:2024swu,Celiberto:2025csa,Celiberto:2026ooh}, a hybrid \textsc{Python}/\textsc{Fortran} multimodular framework for the computation, management, and postprocessing of collider observables across different theoretical approaches.

Although our predictions are fully inclusive in the final state (\emph{i.e.}, not tied to specific tetraquark decay modes), they remain compatible with reconstruction strategies based on di-$\Jpsi$ final states. 
Within this setup, observables such as rapidity separation and jet angular multiplicity can serve as effective discriminants for signal extraction when combined with double-quarkonium triggers. 
Related strategies have already been explored experimentally: CMS studied prompt double-$\Jpsi$ production at 7~TeV~\cite{CMS:2014cmt}, while D0 investigated DPS contributions to di-$\Jpsi$ production at the Tevatron~\cite{D0:2014vql}. 
Extending these analyses to all-heavy tetraquark candidates appears both feasible and well motivated.

%==========================
\subsection{Uncertainty assessment}
\label{ssec:uncertainty}
%==========================

A robust phenomenological study demands a careful and systematic evaluation of theoretical uncertainties.  
In this work, we isolate and quantify the dominant sources of uncertainty entering our framework, encompassing both perturbative and nonperturbative contributions.  
This separation enables us to assess the impact of each component individually, as well as their combined effect on collider observables.  
Specifically, we account for the following contributions:
\begin{itemize}

 \item[$(a)$]
\textbf{Perturbative H-MHOUs}. 
These correspond to missing higher-order uncertainties associated with the perturbative hard factor(s), beyond the implemented $\NLLp$ accuracy.
They are estimated through variations of the renormalization and factorization scales around their central values by factors between $1/2$ and $2$.

\item[$(b)$]
\textbf{Perturbative F-MHOUs}.
These originate from the perturbative uncertainty affecting the initial conditions of the FFs at the starting scale, as well as their residual fragmentation-scale dependence. 
As detailed in Sec.~\ref{ssec:FFs_TQ4Q20}, we adopt a replicalike uncertainty-quantification strategy based on multiscale variations, where $\mathcal{O}(100)$ configurations are generated by varying the relevant scales within the range $[1/2,\,2]$ around their natural values. 
The resulting ensemble defines an uncertainty band through its envelope, thereby providing a robust estimate of missing higher-order effects in the FF evolution affecting the perturbative fragmentation sector. 
To the best of our knowledge, this represents the first application of a replica-based MHOU framework to bound-state fragmentation, including exotic hadronic systems.

 \item[$(c)$]
 \textbf{Nonperturbative LDMEs}.  
 These encode the long-distance, hadron-specific dynamics governing the hadronization stage.  
 Their uncertainties are estimated by varying the relevant matrix elements within ranges compatible with potential-model calculations, as detailed in Sec.~\ref{ssec:FFs_initial_scale}. 
 The resulting bands quantify the sensitivity of collider observables to hadronization-model assumptions.

 \item[$(d)$]
 \textbf{Proton PDFs}. 
 An additional source of uncertainty arises from the proton PDFs, which are nonperturbative inputs extracted from global fits to experimental data, in contrast to our proxy-model FFs.  
 However, dedicated numerical studies for tetraquark-jet production indicate that variations across different PDF sets or replicas remain below the $1\%$ level.  
 For this reason, we adopt the central member of the {\tt NNPDF4.0} set~\cite{NNPDF:2021uiq,NNPDF:2021njg}, without propagating the full PDF-fit uncertainty, which is subleading with respect to MHOUs and LDME effects.

 \item[$(e)$]
 \textbf{Phase-space numerical integration}.
 The dominant numerical uncertainty originates from the multidimensional integration over the final-state phase space (see Eq.~\eqref{DY_rate}) and over the Mellin variable $\nu$ (see Eqs.~\eqref{CnNLL},~\eqref{CnLL}, and~\eqref{CnHENLO} in Appendix~\hyperlink{app:C}{C}).  
 These integrals are evaluated using the native routines of {\tt JETHAD}, with errors consistently controlled below the $1\%$ level.  
 Subleading contributions arise from the one-dimensional integrations over the partonic longitudinal momentum fractions $x$, entering the PDF--FF convolution in the LO and NLO emission functions (see Eq.~\eqref{LOHEF}).  
 Dedicated checks confirm that these effects are negligible compared to the dominant multidimensional integrations.

\end{itemize}

%==========================
\subsection{Rapidity-differential rates}
\label{ssec:I}
%==========================

We begin our phenomenological analysis by considering rapidity-interval distributions, defined as cross sections differential in the rapidity separation $\DY = y_1 - y_2$ between the two tagged final-state objects.
The observable takes the form
\begin{equation}
\label{DY_rate}
 \frac{\drv \sigma(\DY, s)}{\drv \DY} \, = \, C_{n=0} \;,
\end{equation}
where $C_{n=0}$ denotes the azimuthally averaged coefficient, obtained by integrating over the final-state phase space at fixed $\DY$ (see Appendix~\hyperlink{app:C}{C}).
By restricting to the $n=0$ conformal spin, all angular correlations are removed, isolating the dominant energy-dependent contribution and enhancing the sensitivity to high-energy resummation effects.

Rapidity-interval distributions provide a direct probe of the interplay between small-$x$ dynamics and the collinear structure of hadrons, particularly in semi-inclusive configurations involving an all-heavy tetraquark recoiling against a jet.

The kinematic setup is chosen to match the CMS detector acceptance.
We impose $|y_1| < 2.5$ for the tetraquark, consistent with barrel calorimeter coverage~\cite{Chatrchyan:2012xg}, and $|y_2| < 4.7$ for the jet, reflecting the extended reach of the end cap calorimeters~\cite{Khachatryan:2016udy}.
The transverse momentum of the $\TQQ(J^{PC})$ state is varied in the range $30$--$120$~GeV, while the associated light jet spans $50$--$120$~GeV.
The lower transverse-momentum cut on the tetraquark reflects a compromise between maintaining sufficient event statistics and suppressing the moderate-to-low transverse-momentum region, where nonfragmentation SPS mechanisms are expected to play a more substantial role and where the applicability of both the VFNS fragmentation framework and the HyF formalism becomes less controlled.
In this sense, the region around $p_T \sim 30$~GeV should be regarded as an intermediate regime, where fragmentation effects become phenomenologically relevant but may still coexist with sizable SPS contributions.
These choices are aligned with current and projected LHC analyses involving hadronic and jet final states~\cite{Khachatryan:2016udy,Khachatryan:2020mpd}.
The use of asymmetric transverse-momentum cuts enhances the sensitivity to high-energy resummation effects with respect to fixed-order predictions~\cite{Celiberto:2015yba,Celiberto:2015mpa,Celiberto:2020wpk}.

\begin{figure*}[!t]
\centering

   \hspace{0.00cm}
   \includegraphics[scale=0.415,clip]{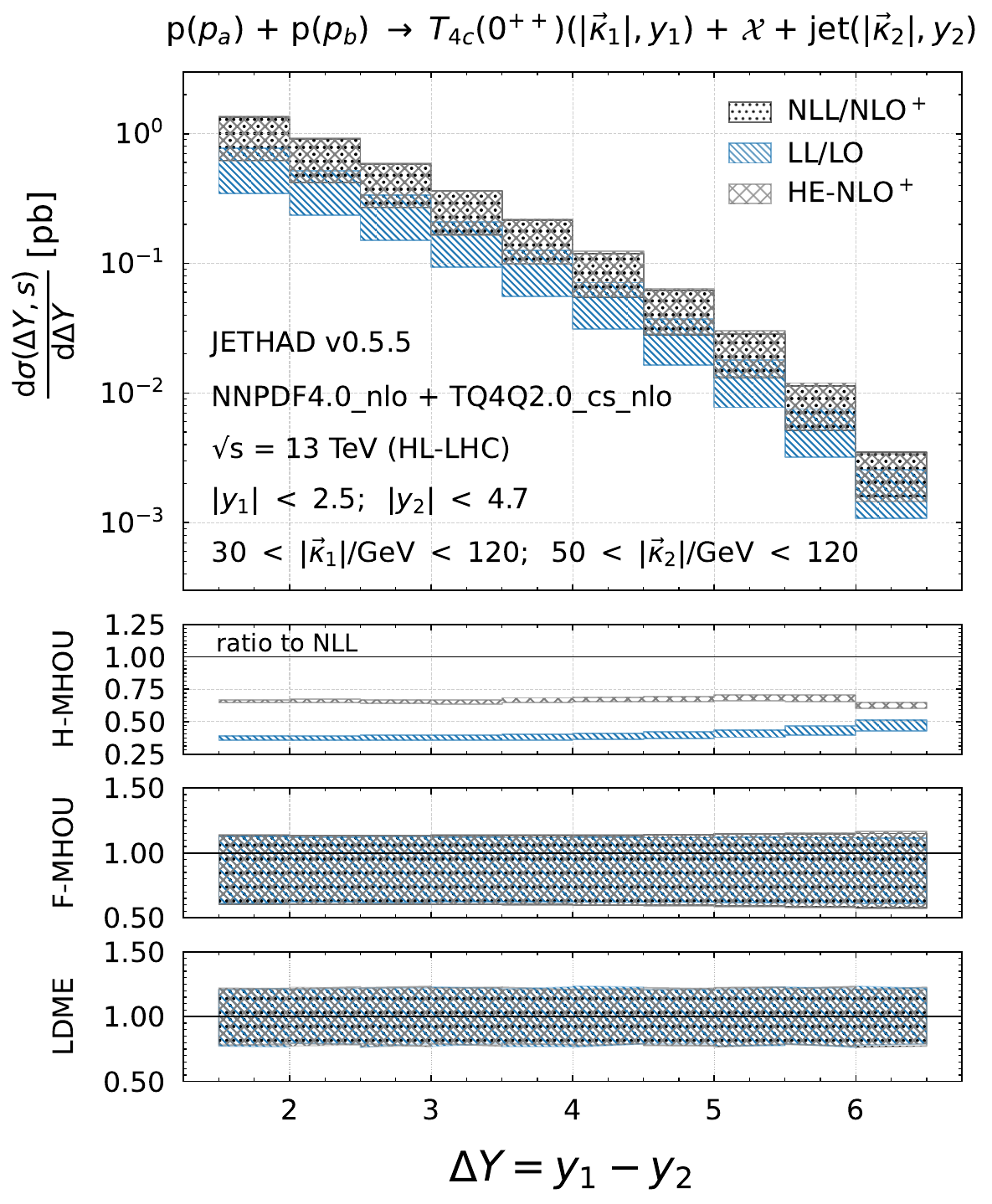}
   \hspace{-0.00cm}
   \includegraphics[scale=0.415,clip]{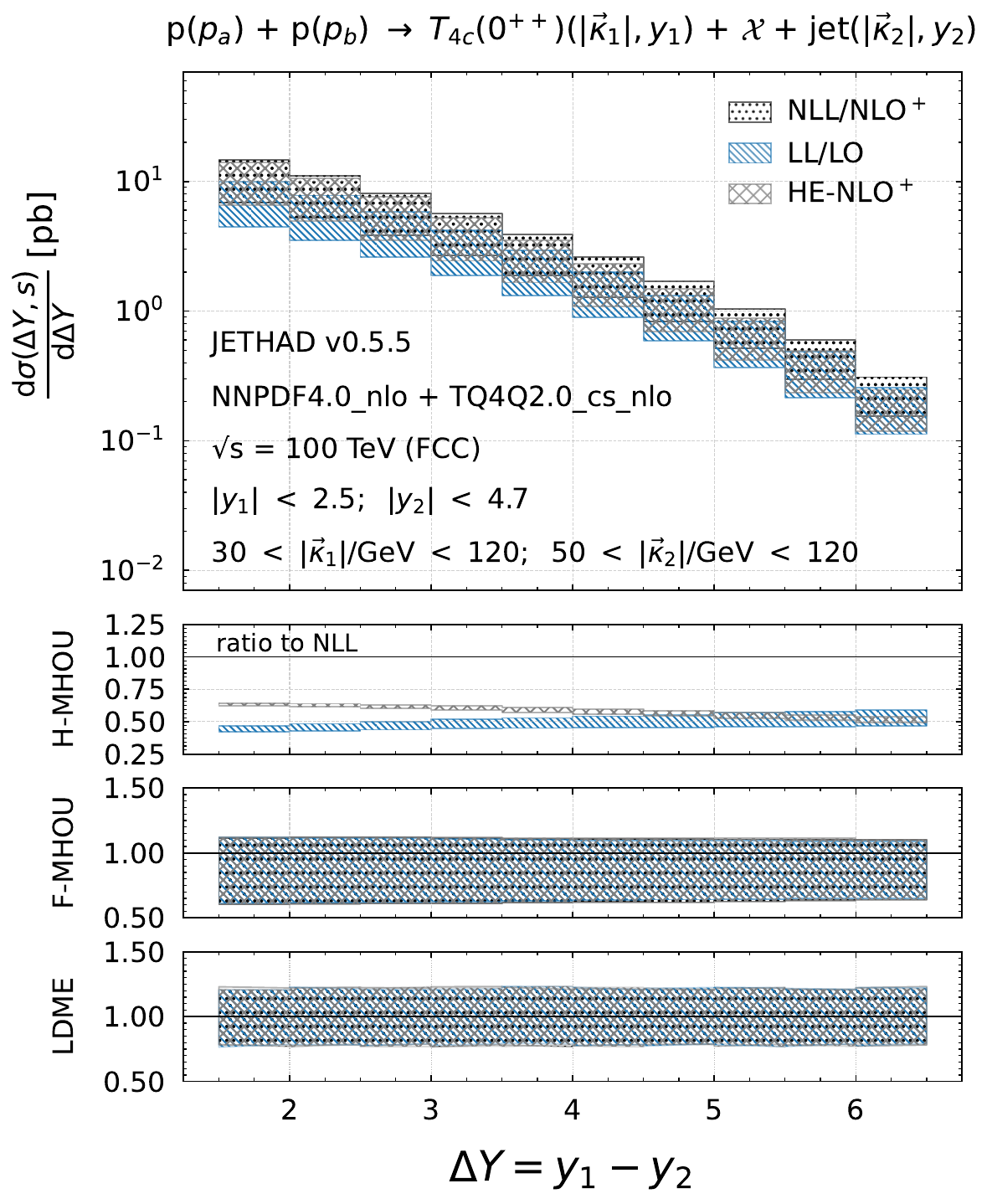}

\caption{
\justifying
\noindent
Distributions in rapidity for scalar $\TQcZpp$ tetraquarks produced in association with a jet at $\sqrt{s} = 13$ TeV (HL-LHC, left) and $100$ TeV (nominal FCC, right). 
Shaded bands in the main panels represent the total uncertainty, obtained by combining H-MHOUs, F-MHOUs, LDME variations, and phase-space integration effects. 
Ancillary panels display: $(i)$ the ratios of $\LL$ and $\HENLOp$ predictions to the $\NLLp$ baseline, including H-MHOUs only; $(ii)$ F-MHOUs shown as the replica envelope normalized to the central prediction; $(iii)$ LDME uncertainties given as ratios to the central value.
}
\label{fig:I_TQ0}
\end{figure*}

\begin{figure*}[t]
\centering

   \hspace{0.00cm}
   \includegraphics[scale=0.415,clip]{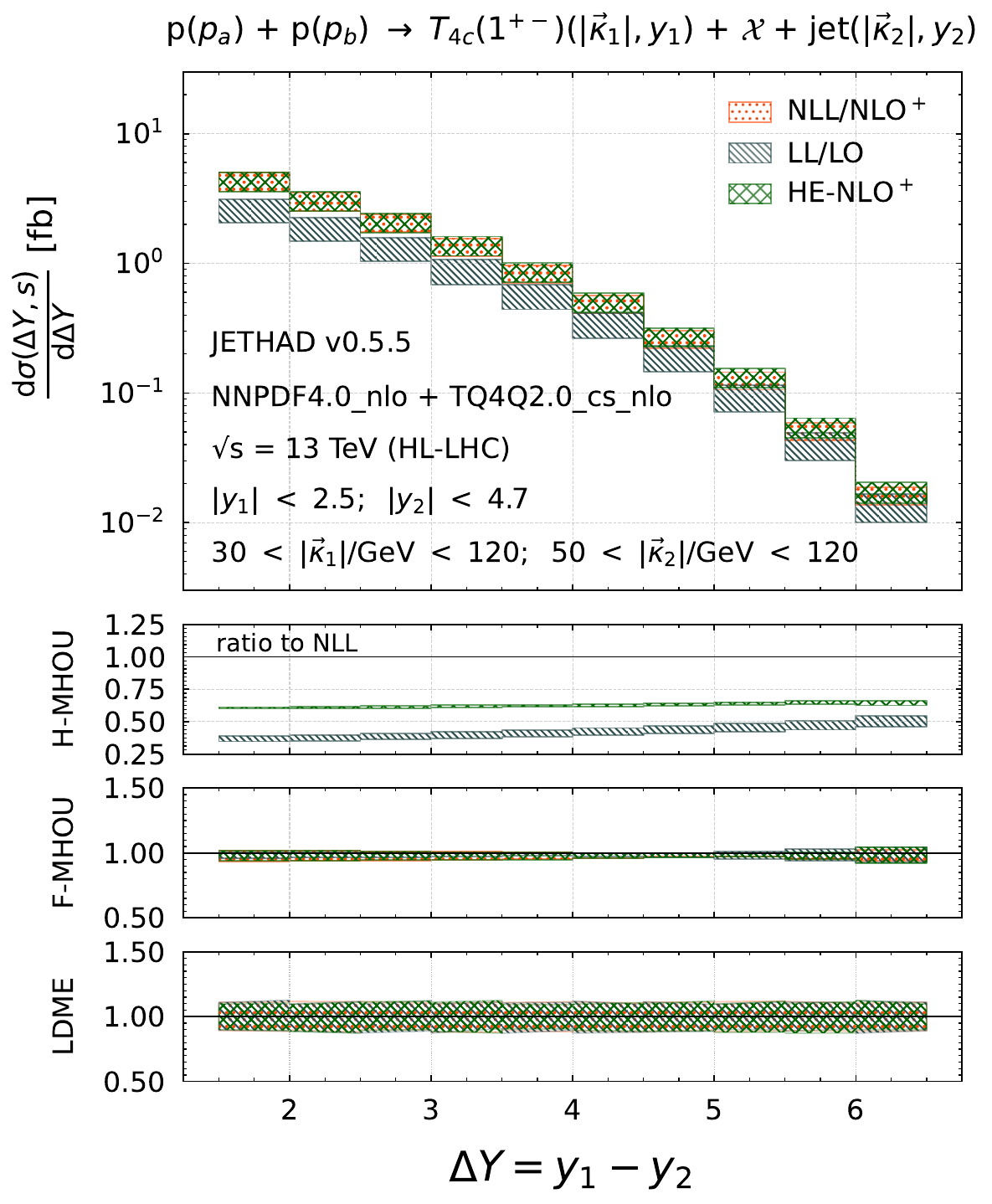}
   \hspace{-0.00cm}
   \includegraphics[scale=0.415,clip]{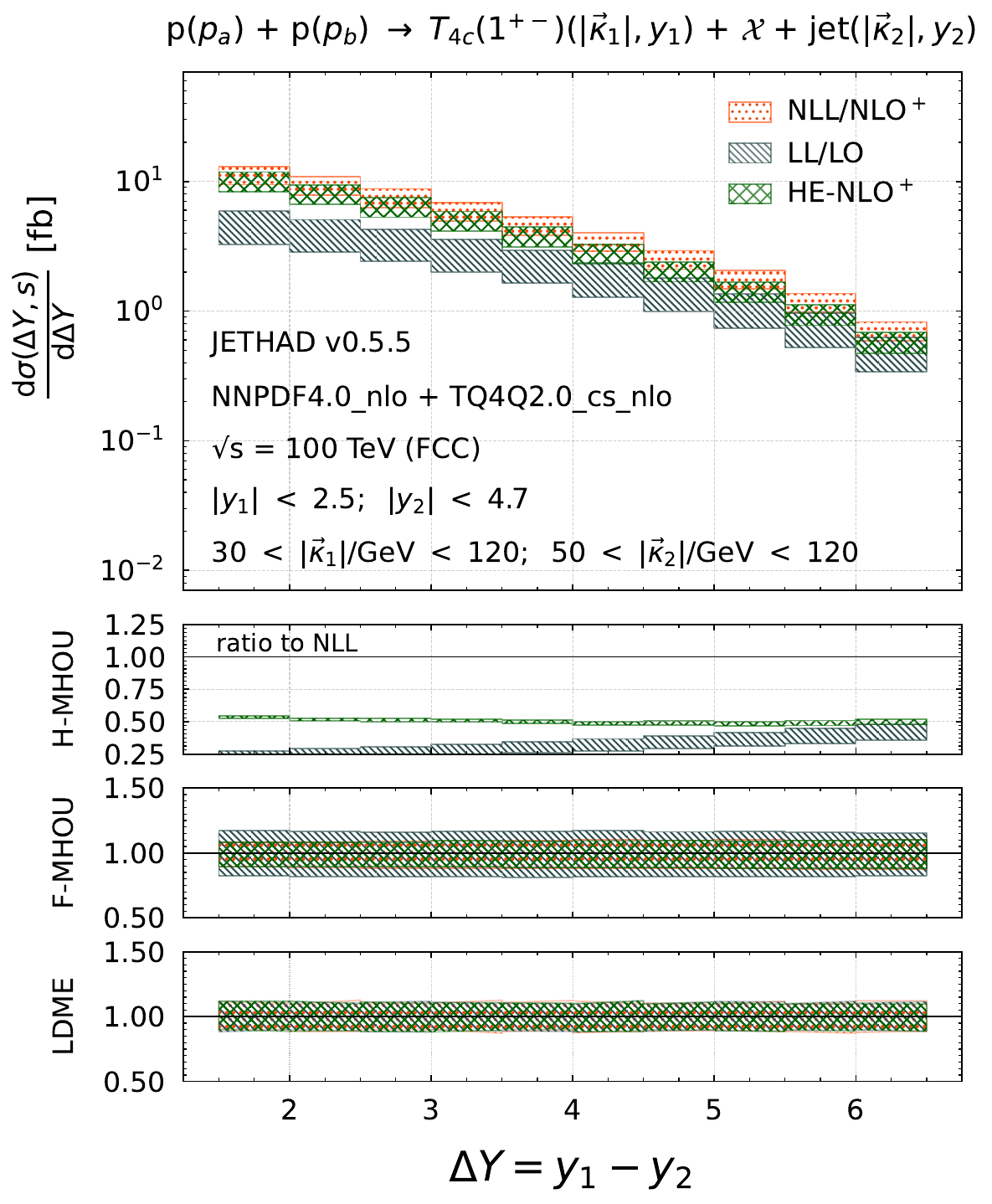}

\caption{
\justifying
\noindent
Distributions in rapidity for scalar $\TQcOpm$ tetraquarks produced in association with a jet at $\sqrt{s} = 13$ TeV (HL-LHC, left) and $100$ TeV (nominal FCC, right). 
Shaded bands in the main panels represent the total uncertainty, obtained by combining H-MHOUs, F-MHOUs, LDME variations, and phase-space integration effects. 
Ancillary panels display: $(i)$ the ratios of $\LL$ and $\HENLOp$ predictions to the $\NLLp$ baseline, including H-MHOUs only; $(ii)$ F-MHOUs shown as the replica envelope normalized to the central prediction; $(iii)$ LDME uncertainties given as ratios to the central value.
}
\label{fig:I_TQ1}
\end{figure*}

\begin{figure*}[!t]
\centering

   \hspace{0.00cm}
   \includegraphics[scale=0.415,clip]{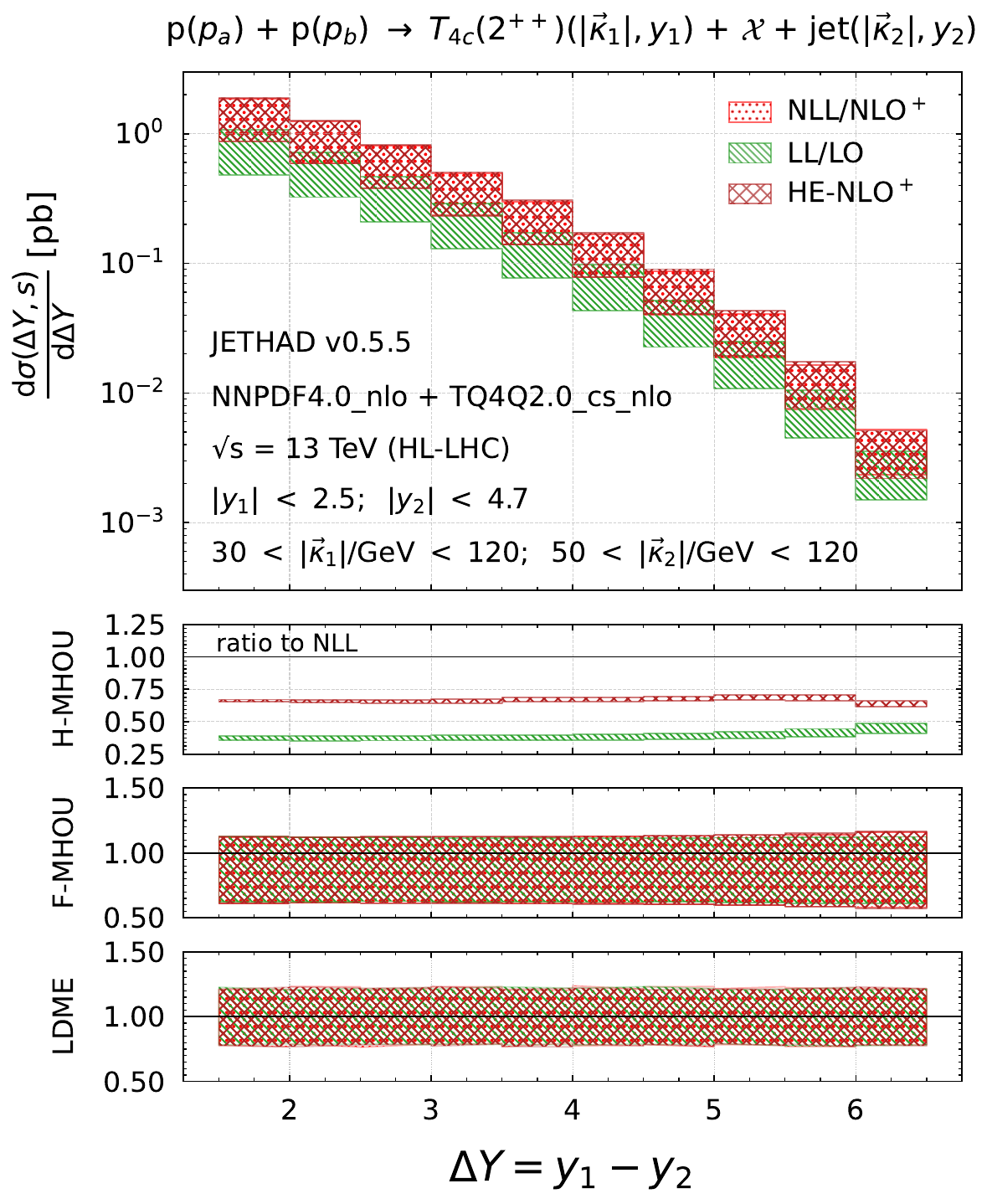}
   \hspace{-0.00cm}
   \includegraphics[scale=0.415,clip]{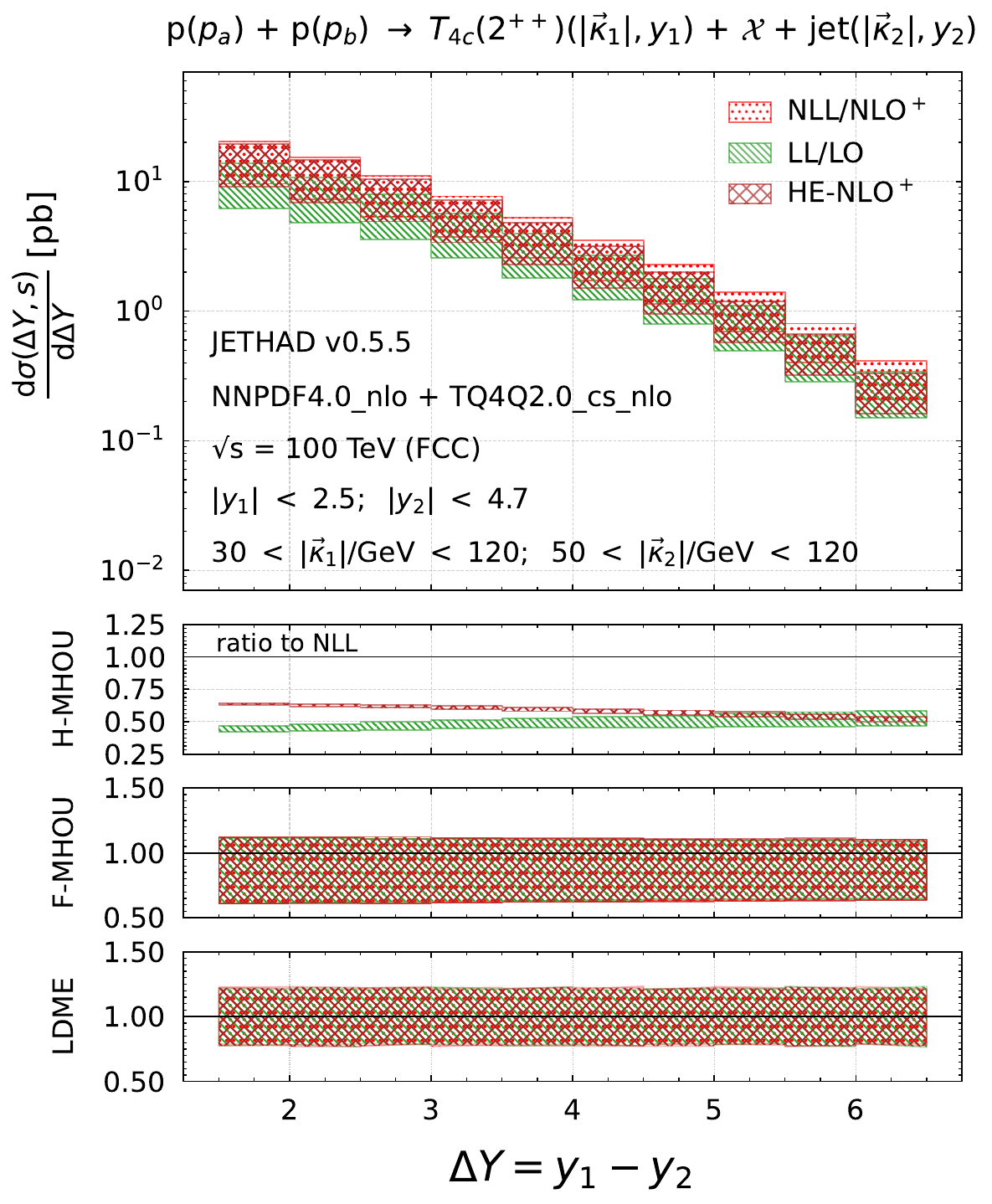}

\caption{
\justifying
\noindent
Distributions in rapidity for scalar $\TQcTpp$ tetraquarks produced in association with a jet at $\sqrt{s} = 13$ TeV (HL-LHC, left) and $100$ TeV (nominal FCC, right). 
Shaded bands in the main panels represent the total uncertainty, obtained by combining H-MHOUs, F-MHOUs, LDME variations, and phase-space integration effects. 
Ancillary panels display: $(i)$ the ratios of $\LL$ and $\HENLOp$ predictions to the $\NLLp$ baseline, including H-MHOUs only; $(ii)$ F-MHOUs shown as the replica envelope normalized to the central prediction; $(iii)$ LDME uncertainties given as ratios to the central value.
}
\label{fig:I_TQ2}
\end{figure*}

Figures~\ref{fig:I_TQ0}, \ref{fig:I_TQ1}, and~\ref{fig:I_TQ2} show predictions for $\drv \sigma / \drv \DY$ in the production of scalar ($0^{++}$), axial-vector ($1^{+-}$), and tensor ($2^{++}$) all-charm tetraquarks in association with a light jet.
The left (right) panels correspond to HL-LHC ($\sqrt{s} = 13$~TeV) [FCC ($\sqrt{s} = 100$~TeV)] configurations.
Main panels display absolute cross sections, while lower panels show ratios of $\LL$ and $\HENLOp$ predictions to the $\NLLp$ baseline.
Uncertainty bands combine H-MHOUs, F-MHOUs, LDME variations, and multidimensional phase-space integration errors in quadrature.
Additional panels disentangle these contributions: $(i)$ H-MHOUs only; $(ii)$ F-MHOU effects, shown as replica envelopes normalized to the central prediction; $(iii)$ LDME variations, given as ratios to the central value.

The $\DY$ bin width is fixed to $0.5$.

A common pattern is observed across all channels: the cross section decreases with increasing $\DY$.
This behavior reflects the competition between two mechanisms.
On the one hand, the resummed partonic coefficient grows with energy, as predicted by BFKL dynamics.
On the other hand, this enhancement is suppressed by the convolution with collinear PDFs and FFs, which rapidly decrease at large momentum fractions.
As a result, the distribution exhibits a mild decrease at small rapidity intervals, $\DY \sim 2.5$--$3$, followed by a steeper falloff at larger $\DY$.

Despite this universal trend, clear channel-dependent differences emerge when comparing the three spin configurations.

\vspace{1em}
\noindent
\textbf{Scalar state ($0^{++}$).}
Predictions for scalar all-charm tetraquarks, $\TQcZpp$, are shown in Fig.~\ref{fig:I_TQ0}.
Among all considered channels, this configuration exhibits the largest cross sections, spanning from approximately $10^{-2}$~pb up to $\mathcal{O}(10)$~pb, depending on $\DY$ and the collider energy.
This behavior is expected for an $S$-wave scalar state, where the absence of spin suppression and the compactness of the bound-state wave function enhance the fragmentation probability.

Theoretical uncertainties associated with H-MHOUs remain moderate across the entire $\DY$ range, with relative variations well below the $50\%$ level.
The comparison between $\LL$ and $\NLLp$ predictions reveals a characteristic pattern: at small rapidity intervals, the $\LL$ approximation tends to overshoot the $\NLLp$ baseline, while the two approaches gradually converge as $\DY$ increases.
This behavior signals a good perturbative control of the resummation procedure and supports the reliability of the $\NLLp$ framework in this kinematic regime.

In addition, F-MHOUs and LDME uncertainties display a markedly different behavior.
The former, implemented through a replica-based strategy, affect primarily the overall normalization of the cross section while leaving the $\DY$ shape essentially unchanged.
LDME variations, being encoded in constant nonperturbative coefficients, act as a global rescaling of the cross section and therefore do not alter the $\DY$ dependence.
This hierarchy among uncertainty sources indicates that the shape of the observable is largely driven by perturbative dynamics, while nonperturbative inputs mainly control its absolute normalization.

A significant enhancement of the cross section is observed when moving from HL-LHC to FCC energies, with an increase of roughly one order of magnitude.
This scaling reflects the extended phase space available at higher energies and points to favorable conditions for the observation of scalar tetraquark states at future colliders.

Further insight is provided by the comparison between fixed-order $\HENLOp$ predictions and the resummed $\NLLp$ result.
At small $\DY$, the two approaches are in close agreement at both collider energies, indicating that high-energy logarithms play a limited role in this region.
However, their relative behavior changes markedly as the rapidity interval increases.
At $\sqrt{s}=13$~TeV, the $\HENLOp$ prediction tends to lie above the $\NLLp$ one at large $\DY$, while at $\sqrt{s}=100$~TeV the trend is reversed, with $\HENLOp$ systematically falling below the resummed result.

This inversion highlights the growing importance of high-energy logarithmic corrections as both the center-of-mass energy and the available rapidity span increase.
At FCC energies, the enhancement of the resummed prediction over the fixed-order one at large $\DY$ indicates that BFKL dynamics becomes a dominant component of the cross section.
In this regime, resummation effects are not only theoretically required, but also manifest themselves in a way that is potentially accessible to experimental observation.

Overall, these results provide clear evidence that high-energy resummation effects, when combined with collinear evolution and fragmentation dynamics, can be probed in realistic collider conditions.
This reinforces the view that the high-energy limit of QCD is not merely a formal construct, but a phenomenologically relevant regime that can be systematically explored at current and next-generation hadron colliders.

\vspace{1em}
\noindent
\textbf{Axial-vector state ($1^{+-}$).}
Predictions for the axial-vector all-charm tetraquark, $\TQcOpm$, are shown in Fig.~\ref{fig:I_TQ1}.
The overall normalization is significantly reduced with respect to the scalar case, with cross sections ranging from $\mathcal{O}(10^{-4})$~pb to $\mathcal{O}(10^{-2})$~pb, depending on $\DY$ and $\sqrt{s}$.
This suppression originates from the reduced overlap between the fragmenting parton and the spin-$1$ tetraquark configuration, as discussed in Sec.~\ref{ssec:FFs_initial_scale}, and from the absence of leading-order gluon fragmentation channels.

Despite the smaller rates, the axial-vector channel exhibits a remarkably clean theoretical structure.
In contrast to scalar and tensor states, all partonic contributions---including gluon as well as nonconstituent light- and heavy-quark channels---are generated radiatively through DGLAP evolution, with no direct initial-scale input.
As a consequence, the $\DY$ dependence is largely driven by perturbative dynamics, leading to a smooth and stable behavior across the entire rapidity range.

This feature is reflected in the uncertainty pattern.
H-MHOUs remain moderate and well controlled, while F-MHOUs, implemented through a replica-based strategy, primarily affect the overall normalization without distorting the $\DY$ shape.
LDME uncertainties, being encoded in constant nonperturbative coefficients, act as a global rescaling and therefore do not modify the distribution profile.
Altogether, these properties result in the narrowest uncertainty bands among all spin configurations, highlighting the robustness of the prediction.

The comparison between fixed-order $\HENLOp$ and resummed $\NLLp$ results further emphasizes the distinctive nature of this channel.
At $\sqrt{s}=13$~TeV, the two predictions remain relatively close over the full $\DY$ range, indicating that the unresummed high-energy approximation already captures a significant fraction of the relevant dynamics.
Nevertheless, the persistent separation between $\LL$ and $\NLLp$ curves signals the importance of higher-order logarithmic contributions.
At $\sqrt{s}=100$~TeV, the gap between fixed-order and resummed results becomes more pronounced, confirming that high-energy resummation effects grow with both energy and rapidity interval.

Overall, while the axial-vector signal is intrinsically suppressed, it provides the cleanest environment to isolate perturbative high-energy dynamics.
Its radiatively generated structure, combined with reduced model dependence and controlled uncertainties, makes it an ideal channel for precision studies and for testing the predictive power of the HyF framework at current and future collider energies.

\vspace{1em}
\noindent
\textbf{Tensor state ($2^{++}$).}
Predictions for the tensor all-charm tetraquark, $\TQcTpp$, are displayed in Fig.~\ref{fig:I_TQ2}.
The overall normalization is comparable to, and in some regions slightly larger than, that of the scalar channel, placing the $2^{++}$ configuration among the most favorable candidates for experimental observation.
This behavior is consistent with the rich Lorentz structure of the tensor state, which enhances the number of contributing short-distance configurations.

The $\DY$ dependence closely follows the pattern observed in the scalar case, featuring a mild decrease at moderate rapidity intervals and a steeper falloff at larger $\DY$.
This similarity reflects the presence of a gluon-initiated fragmentation channel at the initial scale, which drives the overall shape of the distribution.

Theoretical uncertainties remain well controlled across the entire kinematic range.
H-MHOUs exhibit a moderate impact, while F-MHOUs, implemented through a replica-based approach, primarily affect the overall normalization without altering the $\DY$ profile.
As in the other channels, LDME variations act as a global rescaling and do not modify the shape of the observable.
This pattern confirms that the distribution is largely governed by perturbative dynamics, with nonperturbative inputs determining its absolute normalization.

The comparison between fixed-order $\HENLOp$ and resummed $\NLLp$ predictions reveals a more uniform behavior than in the scalar case.
Across both HL-LHC and FCC energies, the $\HENLOp$ curves consistently lie below the $\NLLp$ baseline, with no evidence of the inversion pattern observed for the $0^{++}$ state.
This indicates a smoother interplay between fixed-order and resummed contributions and points to a more stable perturbative structure.
At the same time, the persistent separation between $\LL$ and $\NLLp$ results highlights the continued importance of higher-order logarithmic corrections.

Altogether, the tensor channel combines a sizable production rate with a regular and robust perturbative behavior.
In this sense, it provides an optimal balance between phenomenological relevance and theoretical control, complementing the scalar and axial-vector configurations in the study of all-heavy tetraquark production.

\vspace{1em}
\noindent
\textbf{Overall comparison.}
A combined inspection of the rapidity-interval distributions across all spin configurations reveals a coherent and robust phenomenological picture.
Resummation effects are clearly visible in all channels and remain under control, with the $\NLLp$ predictions providing a stable and well-motivated baseline.
The role of spin manifests both in the overall normalization, through the structure of the underlying FFs, and in the detailed $\DY$ dependence, shaped by the interplay between initial-scale inputs and perturbative evolution.

Among the considered configurations, the axial-vector channel stands out as the cleanest probe of high-energy dynamics.
Its radiatively generated structure, combined with reduced sensitivity to initial-state modeling, enhances its capability to isolate subleading logarithmic effects.
In contrast, the scalar and tensor channels exhibit larger cross sections and therefore represent the most favorable configurations for experimental observation, while still retaining sensitivity to resummation effects.
From an experimental perspective, these results indicate that rapidity-interval distributions constitute a particularly effective observable for the study of all-charm tetraquark production.
The combination of sizable rates in the scalar and tensor channels and enhanced theoretical control in the axial-vector case provides complementary handles for both discovery-oriented and precision-driven analyses.

The behavior of theoretical uncertainties further supports this picture.
H-MHOUs remain moderate across the full kinematic range, while F-MHOUs and LDME uncertainties primarily affect the overall normalization without altering the $\DY$ shape.
In particular, F-MHOUs tend to dominate over LDME variations, although both remain at a comparable level.
A clear reduction of F-MHOU effects is observed when moving from HL-LHC to FCC energies.
This trend can be traced back to the scale dependence of FF evolution: higher center-of-mass energies probe larger factorization scales, where the strong coupling is smaller, and longer DGLAP evolution paths effectively smooth out variations introduced at the initial scale.
In addition, the FCC kinematics accesses smaller momentum fractions, where gluon-driven evolution is more stable, further reducing the relative impact of scale variations.

Altogether, these features demonstrate that rapidity-interval observables provide a theoretically robust and phenomenologically relevant framework to investigate high-energy QCD dynamics in all-charm tetraquark production.

%==========================
\subsection{Expected event yields}
\label{ssec:I-yld}
%==========================

A key outcome of our analysis is the estimate of expected event yields, which directly connect theoretical predictions to experimental observability.
In contrast to differential distributions, yields provide an immediate benchmark to assess the feasibility of detecting all-heavy tetraquark signals in realistic collider conditions.

Event yields are obtained by integrating the rapidity-differential cross sections and rescaling them with the integrated luminosity.
Assuming the full CMS Run~2 dataset at $\sqrt{s}=13$~TeV, $\mathcal{L}^{\rm (CMS)} = 138.6~\text{fb}^{-1}$~\cite{Giraldi:2022mwf,Radl:2024dvn}, we compute
\begin{equation}
\label{eq:event_yields}
 N_{\rm events}(s) \; = \; \mathcal{L}^{\rm (CMS)} \int \drv \DY \, \frac{\drv \sigma^{\rm NLL}}{\drv \DY} \;.
\end{equation}
The $\DY$ distributions are evaluated at $\NLLp$ accuracy and integrated over $|\DY|<6.5$.
Since $\DY$ is symmetric under forward-backward exchange and the detector acceptance is approximately symmetric, both hemispheres are included.

Compared to the setup for rapidity-differential distributions, we adopt symmetric transverse-momentum cuts for both the tetraquark and the jet, $30 < |\vec{\kappa}_{1,2}| < 120$~GeV, while keeping $|y_1|<2.5$ and $|y_2|<4.7$.
This choice maximizes statistics while avoiding the moderate-to-low transverse-momentum region, where both the VFNS description and the hybrid factorization framework become less reliable.

The same luminosity is conservatively used at $\sqrt{s}=100$~TeV to enable a baseline comparison of energy scaling, independently of FCC projections.
Uncertainties are propagated from the $\DY$ distributions and include scale variations, FF evolution, and LDME inputs.

The resulting yields, shown in Tables~\ref{tab:event_yields_LHC} and~\ref{tab:event_yields_FCC} for $\sqrt{s}=13$~TeV (LHC Run~2) and $\sqrt{s}=100$~TeV (nominal FCC), respectively, display clear hierarchies across spin configurations and heavy-flavor sectors.
The last column reports the relative variation $\Delta = (R-1)\times 100$, with $R = \texttt{TQ4Q2.0}/\texttt{TQ4Q1.1}$.

Scalar states ($J^{PC}=0^{++}$) yield among the largest rates.
In the $\TQc$ sector, event counts exceed $2\times10^6$ at 13~TeV and approach $2\times10^7$ at 100~TeV, while $\TQb$ yields grow from $\sim 6\times10^3$ to $\sim 8\times10^4$.
Axial-vector states ($J^{PC}=1^{+-}$) are strongly suppressed, with $\mathcal{O}(10^4)$ events for $\TQc$ and $\mathcal{O}(10)$--$\mathcal{O}(10^2)$ for $\TQb$, but remain valuable due to their clean theoretical structure.
Tensor states ($J^{PC}=2^{++}$) provide the largest yields, confirming scalar and tensor channels as the most promising for experimental searches.

\begin{table}[t]
\centering
\begin{tabular}{c|c|r|r|r}
\toprule
\multicolumn{5}{c}{\textbf{Expected events yields [13 TeV Run 2]}} \\
\midrule
$\TQQ$ & $J^{PC}$ & {\tt TQ4Q2.0} & {\tt TQ4Q1.1} & $\Delta$\,(\%) \\
\midrule
$T_{4c}$ & $0^{++}$ & $2016031 \pm 399168$ & $1768764 \pm 309078$ & +13.98\% \\
$T_{4c}$ & $1^{+-}$ & $8619 \pm 1770$ & $8619 \pm 1612$ & +0\% \\
$T_{4c}$ & $2^{++}$ & $3071882 \pm 602910$ & $2721362 \pm 480942$ & +12.88\% \\
$T_{4b}$ & $0^{++}$ & $6046 \pm 1168$ & $5151 \pm 945$ & +17.38\% \\
$T_{4b}$ & $1^{+-}$ & $17 \pm 3$ & $17 \pm 2$ & +0\% \\
$T_{4b}$ & $2^{++}$ & $8095 \pm 1552$ & $7108 \pm 828$ & +13.88\% \\
\bottomrule
\end{tabular}
\caption{
\justifying
\noindent
Expected event yields for all-heavy tetraquark production at $\sqrt{s} = 13$~TeV, obtained by integrating the NLL-resummed $\DY$ distributions over the rapidity range $|\DY| < 6.5$.
The yields assume an integrated luminosity of $\mathcal{L}^{\rm (CMS)} = 138.6~\text{fb}^{-1}$, corresponding to the total dataset collected by CMS during Run~2~\protect\cite{Giraldi:2022mwf,Radl:2024dvn}.
The last column shows the relative variation $\Delta = (R-1)\times 100$, with $R = \texttt{TQ4Q2.0}/\texttt{TQ4Q1.1}$.
Uncertainties are propagated from the $\DY$ distributions and reflect the combined effect of scale variations, FF evolution, and LDME inputs.
}
\label{tab:event_yields_LHC}
\end{table}

\begin{table}[t]
\centering
\begin{tabular}{c|c|r|r|r}
\toprule
\multicolumn{5}{c}{\textbf{Expected events yields [100 TeV FCC]}} \\
\midrule
$\TQQ$ & $J^{PC}$ & {\tt TQ4Q2.0} & {\tt TQ4Q1.1} & $\Delta$\,(\%) \\
\midrule
$T_{4c}$ & $0^{++}$ & $22140238 \pm 4433814$ & $19465674 \pm 3373524$ & +13.74\% \\
$T_{4c}$ & $1^{+-}$ & $23463 \pm 1094$ & $23463 \pm 1053$ & +0\% \\
$T_{4c}$ & $2^{++}$ & $32784967 \pm 6421338$ & $29144313 \pm 5467493$ & +12.49\% \\
$T_{4b}$ & $0^{++}$ & $23424 \pm 1982$ & $20237 \pm 1552$ & +15.75\% \\
$T_{4b}$ & $1^{+-}$ & $50 \pm 2$ & $50 \pm 3$ & +0\% \\
$T_{4b}$ & $2^{++}$ & $27894 \pm 1972$ & $24493 \pm 1323$ & +13.89\% \\
\bottomrule
\end{tabular}
\caption{
\justifying
\noindent
Expected event yields for all-heavy tetraquark production at $\sqrt{s} = 100$~TeV, obtained by integrating the NLL-resummed $\DY$ distributions over the rapidity range $|\DY| < 6.5$.
The yields assume an integrated luminosity of $\mathcal{L}^{\rm (CMS)} = 138.6~\text{fb}^{-1}$, corresponding to the full Run~2 CMS dataset~\protect\cite{Giraldi:2022mwf,Radl:2024dvn}.
This value is conservatively adopted at $\sqrt{s}=100$~TeV to enable a baseline comparison of energy scaling, independently of specific FCC projections.
The last column shows the relative variation $\Delta = (R-1)\times 100$, with $R = \texttt{TQ4Q2.0}/\texttt{TQ4Q1.1}$.
Uncertainties are propagated from the $\DY$ distributions and reflect the combined effect of scale variations, FF evolution, and LDME inputs.
}
\label{tab:event_yields_FCC}
\end{table}

Comparing {\tt TQ4Q2.0} with {\tt TQ4Q1.1}, we observe a systematic enhancement in scalar and tensor channels driven by nonconstituent quark fragmentation.
In contrast, axial-vector yields remain unchanged, since these channels are absent at the initial scale and generated only through evolution.
The replica-based treatment of F-MHOUs leads to a more realistic, moderately enlarged uncertainty estimate across all channels.

In the $\TQb$ sector, absolute rates are lower but show a comparable relative increase, underscoring the relevance of the central rapidity region. This area is experimentally advantageous due to optimal detector acceptance and the feasibility of reconstructing final states like double-$\Jpsi$ decays.

Finally, we note that high pileup conditions at HL-LHC may affect the reconstruction of double-quarkonium final states, increasing combinatorial backgrounds and complicating vertex association.
Continuum double-quarkonium production represents an additional irreducible background.
Nevertheless, improvements in pileup mitigation, vertexing, and detector performance are expected to preserve sensitivity to all-heavy tetraquark production.

Altogether, these results provide realistic benchmarks for future searches and highlight the role of event-yield observables as a bridge between theory and experiment.
This marks the transition toward a data-driven exploration of exotic multiquark production.

%==========================
\section{Summary and outlook: a high-precision perspective}
\label{sec:conclusions}
%==========================

We have advanced the study of exotic matter production through the derivation and public release of the {\tt TQ4Q2.0} collinear FFs for all-charm tetraquarks~\cite{Celiberto:2026_TQ4Q20}, covering scalar ($0^{++}$), axial-vector ($1^{+-}$), and tensor ($2^{++}$) states.
These functions are constructed within a leading-power NRQCD fragmentation framework, where distinct color-spin configurations enter via color-composite LDMEs.
For the first time in exotic systems, all partonic channels are consistently included at the initial scale, in close analogy with the {\tt NRFF1.0} approach~\cite{Celiberto:2025euy}, and evolved through DGLAP equations in a VFNS using the threshold-matched {\HFNRevo} scheme~\cite{Celiberto:2025euy,Celiberto:2024mex,Celiberto:2024bxu,Celiberto:2024rxa,Celiberto:2025xvy,Celiberto:2026rzi,Celiberto:2026zss}.

The {\tt TQ4Q2.0} sets provide the first unified and systematic treatment of theoretical uncertainties in exotic-hadron fragmentation, combining LDME inputs with perturbative contributions from fragmentation-scale variations (F-MHOUs).
A central element is the replica-based implementation of these uncertainties, which delivers dynamically correlated estimates of missing higher-order effects and establishes a robust baseline for future data-driven extractions.
This strategy naturally interfaces with \emph{multimodal} approaches and modern statistical tools, including machine-learning-assisted analyses, and extends to the exotic sector methodologies developed in hadron-structure studies~\cite{Kassabov:2022orn,Harland-Lang:2018bxd,Ball:2021icz,McGowan:2022nag,NNPDF:2024dpb,Pasquini:2023aaf}.
Altogether, these developments mark a decisive step toward a data-driven characterization of tetraquark fragmentation dynamics.

As a phenomenological application, we have computed tetraquark-plus-jet cross sections at the HL-LHC and FCC using the {\psymJethad} framework~\cite{Celiberto:2020wpk,Celiberto:2022rfj,Celiberto:2023fzz,Celiberto:2024mrq,Celiberto:2024swu,Celiberto:2025csa,Celiberto:2026ooh}, at full $\NLLp$ accuracy within HyF factorization.
Event yields, obtained by integrating the rapidity distributions over $|\Delta Y|<6.5$ and rescaling with the CMS Run~2 luminosity, provide direct benchmarks for experimental searches, with all-bottom states included for completeness.

A consistent pattern emerges.
Nonconstituent quark fragmentation enhances scalar and tensor cross sections by $15$--$20\%$, in agreement with Ref.~\cite{Bai:2024flh}, and propagates to event yields.
This identifies subleading channels as an essential ingredient for precision studies and elevates {\tt TQ4Q2.0} beyond an incremental improvement.

These results also broaden the scope of fragmentation-based studies.
Although used here within HyF factorization, the {\tt TQ4Q2.0} functions constitute a general-purpose tool that can be deployed across complementary approaches, enabling a more comprehensive exploration of all-heavy tetraquark production dynamics across kinematic regimes and factorization schemes.

Looking ahead, the inclusion of color-octet configurations will further refine the framework, while all-bottom tetraquarks, despite their suppressed rates, offer a sensitive probe of fragmentation dynamics at high scales.
Upcoming facilities, including the FCC~\cite{FCC:2025lpp,FCC:2025uan,FCC:2025jtd}, other planned accelerators~\cite{Chapon:2020heu,LHCspin:2025lvj,Anchordoqui:2021ghd,Feng:2022inv,AlexanderAryshev:2022pkx,LinearCollider:2025lya,LinearColliderVision:2025hlt,Arbuzov:2020cqg,Accettura:2023ked,InternationalMuonCollider:2024jyv,MuCoL:2024oxj,MuCoL:2025quu,Black:2022cth,InternationalMuonCollider:2025sys,Accardi:2023chb,Bose:2022obr,Gessner:2025acq,Altmann:2025feg}, together with complementary lepton-hadron environments such as the EIC~\cite{AbdulKhalek:2021gbh,Khalek:2022bzd,Hentschinski:2022xnd,Amoroso:2022eow,Abir:2023fpo,Allaire:2023fgp}, will provide valuable opportunities to investigate gluon-driven dynamics and heavy-flavor hadronization mechanisms relevant to exotic-hadron production.

Axial-vector tetraquarks emerge as particularly sensitive probes of intrinsic charm.
As shown in Ref.~\cite{Celiberto:2025vra}, their radiative production mechanism enhances sensitivity to \emph{intrinsic heavy-quark components}, linking exotic-hadron production to proton structure~\cite{Brodsky:1980pb,Brodsky:2015fna,Jimenez-Delgado:2014zga,Ball:2016neh,Hou:2017khm,Ball:2022qks,Guzzi:2022rca,NNPDF:2023tyk} and suggesting a two-way connection between hadron structure and spectroscopy~\cite{Vogt:2024fky}.
Additional insight may arise from tetraquark-in-jet observables, which provide a novel probe of the \emph{dead-cone effect}~\cite{Dokshitzer:1991fd,ALICE:2021aqk}.
Future studies will also address states such as the $Z_c(3900)$~\cite{Guo:2013ufa}, whose production mechanism remains unresolved.

Finally, the interplay among scalar, axial-vector, and tensor configurations provides complementary diagnostic information on fragmentation and hadronization dynamics, now anchored by recent CMS measurements~\cite{CMS:2023owd,CMS:2025fpt,CMS:2026tiu}.

The {\tt TQ4Q2.0} release marks the transition to precision studies of exotic-hadron formation within collinear fragmentation.
With a complete partonic structure and a controlled treatment of uncertainties, it provides a robust baseline for future data-driven analyses.
This work thus frames the study of all-charm tetraquarks at hadron colliders within a genuinely high-precision fragmentation perspective.

%==========================
\section*{Acknowledgments}
\label{sec:Acknowledgments}
%==========================

We make use of results originally presented in Refs.~\cite{Feng:2020riv,Bai:2024ezn,Bai:2024flh}, which we have independently recomputed within the {\psymJethad} framework~\cite{Celiberto:2020wpk,Celiberto:2022rfj,Celiberto:2023fzz,Celiberto:2024mrq,Celiberto:2024swu,Celiberto:2025csa,Celiberto:2026ooh}. 
These calculations are then adopted as effective inputs to model the fragmentation mechanism at the initial scale.
We are grateful to Marco Bonvini, Angelo Esposito, Alessandro Papa, Fulvio Piccinini, and Alessandro Pilloni for insightful discussions.
We received financial support from the Atracción de Talento Grant No. 2022-T1/TIC-24176 from the Comunidad Autónoma de Madrid, Spain.

%==========================
\section*{Data availability}
\label{sec:data}
%==========================

The datasets underlying the results of this study are publicly accessible~\cite{Celiberto:2026_TQ4Q20}.
The {\tt TQ4Q2.0} fragmentation functions for all-heavy tetraquarks $\TQQ(J^{PC})$~\cite{Celiberto:2026_TQ4Q20} can be retrieved from~\cite{Celiberto:2026_TQ4Q20_url}.
The replica sets include only perturbative uncertainties (F-MHOUs), while LDME variations, entering as overall normalization factors, can be independently applied by the user.
In order to enhance transparency and facilitate independent validation, we also release a stand-alone \textsc{Mathematica} notebook derived from our internal {\symJethad} framework~\cite{Celiberto:2020wpk,Celiberto:2022rfj,Celiberto:2023fzz,Celiberto:2024mrq,Celiberto:2024swu,Celiberto:2025csa,Celiberto:2026ooh}.
This notebook includes the symbolic representation of all short-distance coefficients employed in the present analysis.
%
%The file is fully self-contained, does not rely on external packages, and can be readily used for both numerical evaluations and symbolic manipulations.
It is distributed within the same GitHub repository hosting the {\tt TQ4Q2.0} grids.

\clearpage

\appendix
\onecolumngrid

\counterwithin*{equation}{section}
\renewcommand\theequation{\thesection\arabic{equation}}

\counterwithin*{figure}{section}
\renewcommand\thefigure{\thesection\arabic{figure}}

\counterwithin*{table}{section}
\renewcommand\thetable{\thesection\arabic{table}}

%==========================
\hypertarget{app:A}{
\section{Analytic SDC expressions}
}

In this appendix, we present the explicit expressions for all nonvanishing dimensionless SDCs relevant to the $\TQQ$ fragmentation channels.

\vspace{1em}
\noindent
\textbf{Scalar channel ($0^{++}$).}

The $[g \to \TQQ(0^{++})]$ SDCs are~\cite{Feng:2020riv}
\begin{equation}
\begin{split}
 \label{Dg_FF_SDC_0pp_33}
\hspace{-0.00cm}
 \tilde{\cal D}^{(0^{++})}_g&(z,[3,3]) \,=\, 
 \frac{\pi^{2} \alpha_{s}^{4}(4m_Q)}{497664 \, d^{\TQQ}_g(z)}\left[186624-430272 z+511072 z^2-425814 z^3\right. \\
 & +\, 217337 z^4-61915 z^5+7466 z^6+42(1-z)(2-z)(3-z)(-144+634 z\\
 & \left.-\, 385 z^2+70 z^3\right) \ln (1-z)+36(2-z)(3-z)\left(144-634 z+749 z^2-364 z^3\right. \\
 & \left.+\, 74 z^4\right) \ln \left(1-\frac{z}{2}\right)+12(2-z)(3-z)\left(72-362 z+361 z^2-136 z^3+23 z^4\right) \\
 & \left.\times\, \ln \left(1-\frac{z}{3}\right)\right]
 \;,
\end{split}
\end{equation}
\\[-0.35cm]
\begin{equation}
\begin{split}
 \label{Dg_FF_SDC_0pp_66}
\hspace{-0.00cm}
 \tilde{\cal D}^{(0^{++})}_g&(z,[6,6]) \,=\,  
 \frac{\pi^{2} \alpha_{s}^{4}(4m_Q)}{331776 \, d^{\TQQ}_g(z)}\left[186624-430272 z+617824 z^2-634902 z^3\right. \\
 & +\, 374489 z^4-115387 z^5+14378 z^6-6(1-z)(2-z)(3-z)(-144-2166 z\\
 & \left.+\, 1015 z^2+70 z^3\right) \ln (1-z)-156(2-z)(3-z)\left(144-1242 z+1693 z^2-876 z^3\right. \\
 & \left.+\, 170 z^4\right) \ln \left(1-\frac{z}{2}\right)+300(2-z)(3-z)\left(72-714 z+953 z^2-472 z^3+87 
 z^4\right) \\
 & \left.\times\, \ln \left(1-\frac{z}{3}\right)\right]
 \;,
\end{split}
\end{equation}
\\[-0.35cm]
\begin{equation}
\begin{split}
 \label{Dg_FF_SDC_0pp_36}
\hspace{-0.00cm}
 \tilde{\cal D}^{(0^{++})}_g&(z,[3,6]) \,=\,  
 \frac{\pi^{2} \alpha_{s}^{4}(4m_Q)}{165888 \, d^{\TQQ}_g(z)}\left[186624-430272 z+490720 z^2-394422 z^3\right. \\
 & +\, 199529 z^4-57547 z^5+7082 z^6+6(1-z)(2-z)(3-z)(-432+3302 z \\
 & \left.-\, 1855 z^2+210 z^3\right) \ln (1-z)-12(2-z)(3-z)\left(720-2258 z+2329 z^2-1052 z^3\right. \\
 & \left.+\, 226 z^4\right) \ln \left(1-\frac{z}{2}\right)+12(2-z)(3-z)\left(936-4882 z+4989 z^2-1936 z^3+331 z^4\right) \\
 & \left.\times\, \ln \left(1-\frac{z}{3}\right)\right]
 \;,
\end{split}
\end{equation}
where $d^{\TQQ}_g(z) = z(2-z)^{2}(3-z)$.

Then, the $[Q \to \TQQ(0^{++})]$ SDCs are~\cite{Bai:2024ezn}
\begin{equation}
\begin{split}
 \label{DQ_FF_SDC_0pp_33}
 \hspace{-0.00cm}
 \tilde{\cal D}&^{(0^{++})}_Q(z,[3,3]) \,=\, 
 \frac{\pi^2 \as^4(5m_Q)}{559872 \, d^{\TQQ}_Q(z)} \left[ -264 (z-4) (11 z-12) (z^2-16 z+16) \right. \\
 & \times\, (13 z^4-57 z^3-656 z^2+1424z-512) (3 z-4)^5 \log (z^2-16 z+16) + 6 (11 z-12)(z^2-16 z+16) \\
 & \times\, (1273 z^5-16764 z^4+11840 z^3 + 247808z^2-472320 z+171008) (3 z-4)^5 \log (4-3 z) \\
 & -\, 3 (11 z-12)(z^2-16 z+16) (129 z^5 - 7172 z^4+49504 z^3-108416z^2 + 73984 z-9216) (3 z-4)^5  \\
 & \times\, \log\left[\left(4-\frac{11z}{3}\right)(4-z)\right] + 16 (z-1) (657763 z^{12}-10028192z^{11} + 188677968 z^{10}-2600899712 z^9 \\
 & +\, 18018056448 z^8-71685000192z^7 + 179414380544 z^6-294834651136 z^5 \\
 & +\, 321642168320z^4-229388845056 z^3 + 102018056192 z^2-25480396800z \left. + 2717908992)\right]
 \;,
\end{split}
\end{equation}
\\[-0.35cm]
\begin{equation}
\begin{split}
 \label{DQ_FF_SDC_0pp_66}
 \hspace{-0.00cm}
 \tilde{\cal D}&^{(0^{++})}_Q(z,[6,6]) \,=\, 
 \frac{\pi^2 \as^4(5m_Q)}{373248 \, d^{\TQQ}_Q(z)} \left[ -120 (z-4) (11 z-12) (z^2-16 z+16) \right. \\
 & \times\, (35 z^4-535 z^3+3472 z^2-4240z+512) (3 z-4)^5 \log (z^2-16 z+16)  \\
 & -\, 30 (11 z-12)(z^2-16 z+16) (3395 z^5-48020 z^4+126144 z^3 \\
 & -\, 75776^2-38656 z+62464) (3 z-4)^5 \log (4-3 z) + 75 (11 z-12)(z^2-16 z+16) \\
 & \times\, (735 z^5 - 10684 z^4+34208 z^3-44160z^2 + 20224 z+9216) (3 z-4)^5 \log\left[\left(4-\frac{11z}{3}\right)(4-z)\right]\\
 & +\, 16 (z-1) (7916587 z^{12}-263987840z^{11} + 3125201872 z^{10}-16993694336 z^9 \\
 & +\, 51814689024 z^8-99638283264^7 + 133459423232 z^6-140136398848 z^5 \\
 & +\, 127161204736z^4-96695746560 z^3 + 53372518400 z^2-17930649600z \left. + 2717908992)\right]
 \;,
\end{split}
\end{equation}
\\[-0.35cm]
\begin{equation}
\begin{split}
 \label{DQ_FF_SDC_0pp_36}
 \hspace{-0.00cm}
 \tilde{\cal D}&^{(0^{++})}_Q(z,[3,6]) \,=\, 
 \frac{\pi^2 \as^4(5m_Q)}{186624 \sqrt{6} \, d^{\TQQ}_Q(z)} \left[ 24 (z-4) (11 z-12) (z^2-16 z+16) \right. \\
 & \times\, (225 z^4-3085 z^3+17456 z^2 - 19760z+1536) (3 z-4)^5 \log (z^2-16 z+16)  \\
 & -\, 6 (11 z-12)(z^2-16 z+16) (555 z^5+52428 z^4-363328 z^3 + 616448z^2-270080 z+70656) \\
 & \times\, (3 z-4)^5 \log (4-3 z) + 75 (11 z-12)(z^2-16 z+16) (1245 z^5 -84308 z^4 \\
 & +\, 601696z^3-1333120z^2 + 914688 z-119808) (3 z-4)^5 \log\left[\left(4-\frac{11z}{3}\right)(4-z)\right]\\
 & +\, 16 (z-1) (1829959z^{12}-44960912 z^{11} + 285792656 z^{10}-1090093952z^9 \\
 & +\, 5123084544 z^8-24390724608 z^7 + 77450817536 z^6-153897779200z^5 \\
 & +\, 194102034432 z^4-155643543552 z^3 + 77091307520 z^2-21705523200z \left. + 2717908992)\right]
 \;,
\end{split}
\end{equation}
where $d^{\TQQ}_Q(z) = (4-3 z)^6(z-4)^2 z(11 z-12)(z^2-16 z+16)$.

Finally, the nonconstituent (light or heavy) quark-channel $[\tilde{q}, \tilde{Q} \to \TQQ(0^{++})]$ SDCs read~\cite{Bai:2024flh}
\begin{equation}
\begin{split}
 \label{DqtQt_FF_SDC_0pp_33}
 \hspace{-0.00cm}
 \tilde{\cal D}^{(0^{++})}_{\tilde{q},\tilde{Q}}(z,[3,3]) \,&=\, 
 \frac{\pi^2 \as^4(4m_Q+m_{\tilde{q},\tilde{Q}})}{1728 z \, m_Q^4} 
 \left\{
 m_{\tilde{q},\tilde{Q}}^{2}\left[ m_{\tilde{q},\tilde{Q}}^2z-2m_Q^2(4-z)\right]
 \log\left[\frac{8m_Q^{2}(z-2)(z-1)+m_{\tilde{q},\tilde{Q}}^{2} z^2}{m_{\tilde{q},\tilde{Q}}^{2} z^2-16m_Q^{2} (z-1)}\right]
 \right.
 \\
 & +\,
 \frac{8m_Q^{2}(z-1)}{(16m_Q^{2}(z-1)-m_{\tilde{q},\tilde{Q}}^{2}z^2)(8m_Q^{2}(z-2)(z-1)+m_{\tilde{q},\tilde{Q}}^{2}z^2)}\left[32m_Q^{6}(z^2-5z+4)^2\right. 
 \\
 & -\,
 2m_Q^{4} m_{\tilde{q},\tilde{Q}}^{2}(z-1)z(z^3+32z-64)\left.+2m_Q^{2} m_{\tilde{q},\tilde{Q}}^{4}(z-4)z^2(3z-2)+m_{\tilde{q},\tilde{Q}}^{6}z^4\right]
 \Bigg\}
 \;.
\end{split}
\end{equation}
\\[-0.35cm]
\begin{equation}
\begin{split}
 \label{DqtQt_FF_SDC_0pp_66}
 \hspace{-0.00cm}
 \tilde{\cal D}&^{(0^{++})}_{\tilde{q},\tilde{Q}}(z,[6,6]) \,=\, 
 \frac{3}{2} \,
 \tilde{\cal D}^{(0^{++})}_{\tilde{q},\tilde{Q}}(z,[3,3])
 \;,
\end{split}
\end{equation}
\\[-0.35cm]
\begin{equation}
\begin{split}
 \label{DqtQt_FF_SDC_0pp_36}
 \hspace{-0.00cm}
 \tilde{\cal D}&^{(0^{++})}_{\tilde{q},\tilde{Q}}(z,[3,6]) \,=\, 
 \frac{3}{\sqrt{6}} \,
 \tilde{\cal D}^{(0^{++})}_{\tilde{q},\tilde{Q}}(z,[3,3])
 \;.
\end{split}
\end{equation}

\vspace{1em}
\noindent
\textbf{Axial-vector channel ($1^{+-}$).}

As outlined in Sec.~\ref{ssec:FFs_initial_scale}, the combined effect of Fermi-Dirac statistics and the symmetry restrictions of the $S$-wave configuration permits only the $[3,3]$ color-spin channel to contribute to the axial-vector state.
In addition, the $[g \to \TQQ(1^{+-})]$ fragmentation channel is absent at LO as a consequence of the Landau-Yang selection rule, while the nonconstituent (light or heavy) quark channels $[\tilde{q}, \tilde{Q} \to \TQQ(1^{+-})]$ are likewise forbidden by charge-conjugation ($C$-parity) conservation.
One has

\begin{equation}
\begin{split}
\label{sDQ_FF_SDC_1pm_33}
 \tilde{\cal D}^{(1^{+-})}_Q&(z,[3,3])
 \,=\, \frac{\pi^{2}\left[\alpha_{s}^4(5m_Q)\right]}{279936 \, d^{\TQQ}_Q(z)}
 \left[ 480 (z-4)
 (11 z-12)\left(z^2-16 z+16\right) \right.
 \\
 \,&\times\, 
 \left(4 z^4+115 z^3-316 z^2+112z+64\right) (3 z-4)^5 \log \left(z^2-16 z+16\right)
 \\
 \,&+\, 6 (11 z-12)\left(z^2-16 z+16\right) (4825 z^5-56232 z^4+378480z^3
 \\
 \,&-\, 942528 z^2+672768 z-60416) (3 z-4)^5 \log (4-3 z)-3 (11z-12) \left(z^2-16 z+16\right) (5465 z^5
 \\
 \,&-\, 40392 z^4+254320z^3-722368 z^2+611328 z-101376) (3 z-4)^5 \log\left[\left(\frac{11z}{3}-4\right)(z-4)\right]
 \\
 \,&+\, 16 (z-1) z(476423 z^{11}+32559240 z^{10}-934590720 z^9+8015251776z^8
 \\
 \,&-\, 35393754624 z^7+94265413632 z^6-160779010048 z^5+177897046016z^4
 \\
 \,&-\,  \left. 124600254464 z^3 + 51223461888 z^2-10217324544z+490733568) \right] \;,
\end{split}
\end{equation}

\vspace{1em}
\noindent
\textbf{Tensor channel ($2^{++}$).}

As highlighted in Sec.~\ref{ssec:FFs_initial_scale}, the interplay between Fermi-Dirac statistics and the $S$-wave configuration constrains the nonvanishing SDCs for tensor states to the $[3,3]$ color-spin structures only.
For the $[g \to \TQQ(2^{++})]$ channel, one obtains~\cite{Feng:2020riv}
\begin{equation}
\begin{split}
 \label{Dg_FF_SDC_2pp_33}
\hspace{-0.00cm}
 \tilde{\cal D}^{(2^{++})}_g&(z,[3,3]) \,=\, 
 \frac{\pi^{2} \alpha_{s}^{4}(4m_c)}{622080 \, z \, d^{\TQQ}_g(z)}\left[\left(46656-490536 z+1162552 z^2-1156308 z^3\right.\right. \\
 & \left.+\, 595421 z^4-170578 z^5+21212 z^6\right) 2z+3(1-z)(2-z)(3-z)(-20304-31788 z) \\
 & \left.\left.\times\, (1296+1044 z + 73036 z^2-36574 z^3+7975 z^4\right)\right. \\
 & \left.\times\, \ln (1-z)+33(2-z)(3-z)(1296+25)\right] \\
 & \left.\left.\, -9224 z^2+9598 z^3-3943 z^4+725 z^5\right) \ln \left(1-\frac{z}{3}\right)\right]
  \;.
\end{split}
\end{equation}

Then, the $[Q \to \TQQ(2^{++})]$ SDC are given by~\cite{Bai:2024ezn}
\begin{equation}
\begin{split}
 \label{DQ_FF_SDC_2pp_33}
 \hspace{-0.00cm}
 \tilde{\cal D}&^{(2^{++})}_Q(z,[3,3]) \,=\, 
 \frac{\pi^2 \as^4(5m_Q)}{2799360 \, z \, d^{\TQQ}_Q(z)} \left[ 672 (z-4) (11 z-12) (z^2-16 z+16) \right. \\
 & \times\, (47z^5+12186 z^4-44608 z^3 + 40000 z^2 -7936 z+4608) \\
 & \times\, (3 z-4)^5 \log (z^2-16 z+16) + 6 (11 z-12)(z^2-16 z+16) \\
 & \times\, (107645 z^6-1088988 z^5+7805536 z^4 - 20734976 z^3 +8933504z^2 - 6013952 z+1695744) \\
 & \times\, (3 z-4)^5 \log (4-3 z) - 33 (11 z-12)(z^2-16 z+16) (3581 z^5 - 53216 z^4-326176 z^3+419456z^2 \\
 & -\, 6912 z+55296) (3 z-4)^6 \log\left[\left(4-\frac{11z}{3}\right)(4-z)\right]\\
 & +\, 16 (z-1) (96449507 z^{12} - 158520388z^{11} - 26228206896 z^{10} + 281743037888 z^9 \\
 & -\, 1355257362432 z^8 + 1355257362432 z^7 - 6637452959744 z^6 + 7595797282816 z^5 \\
 & -\, 5643951472640 z^4 + 2662988513280 z^3 - 788934950912 z^2 + 161828831232 z \left. - 24461180928)\right]
 \;.
\end{split}
\end{equation}

Finally, the nonconstituent (light or heavy) quark-channel $[\tilde{q}, \tilde{Q} \to \TQQ(2^{++})]$ SDCs read~\cite{Bai:2024flh}
\begin{equation}
\begin{split}
 \label{DqtQt_FF_SDC_2pp_33}
 \hspace{-0.00cm}
 \tilde{\cal D}&^{(2^{++})}_{\tilde{q},\tilde{Q}}(z,[3,3]) \,=\, 
 \frac{\pi^{2}\alpha_{s}^{4}}{3240 m_Q^{4} z^2}\bigg\{-\left[12m_Q^{4}((z-2)^2z-4)-m_Q^{2} m_{\tilde{q},\tilde{Q}}^{2}(z-1)z(3z-16)+2m_{\tilde{q},\tilde{Q}}^{4}z^2\right] \\
 & \times \, 
 \log\left[\frac{m_{\tilde{q},\tilde{Q}}^{2} z^2-16 m_Q^{2}(z-1)}{8m_Q^{2}(z-2)(z-1)+m_{\tilde{q},\tilde{Q}}^{2}z^2}\right]
 +\frac{8m_Q^{2}(z-1)z}{[16m_Q^{2}(z-1)-m_{\tilde{q},\tilde{Q}}^{2}z^2][8m_Q^{2}(z-2)(z-1)+m_{\tilde{q},\tilde{Q}}^{2}z^2]}\\
 & \times \, 
 \left[16m_Q^{6}(z-1)[z((z-33)z+72)-16]+m_Q^{4} m_{\tilde{q},\tilde{Q}}^{2}z[z(z((97-13z)z-440)\right.\\
 & + \,
 \left.576)-256]-m_Q^{2}m_{\tilde{q},\tilde{Q}}^{4}(z-1)z^2[3(z-8)z+32]+2m_{\tilde{q},\tilde{Q}}^{6}z^4\right]\bigg\}
 \;.
\end{split}
\end{equation}

\clearpage

%==========================
\hypertarget{app:B}{
\section{{\tt TQ4Q2.0} FFs for all-bottom tetraquarks}
}
%==========================
\label{app:B}
%\addcontentsline{toc}{section}{\nameref{app:V}}

\begin{figure*}[!t]
\centering

   \hspace{-0.00cm}
   \includegraphics[scale=0.410,clip]{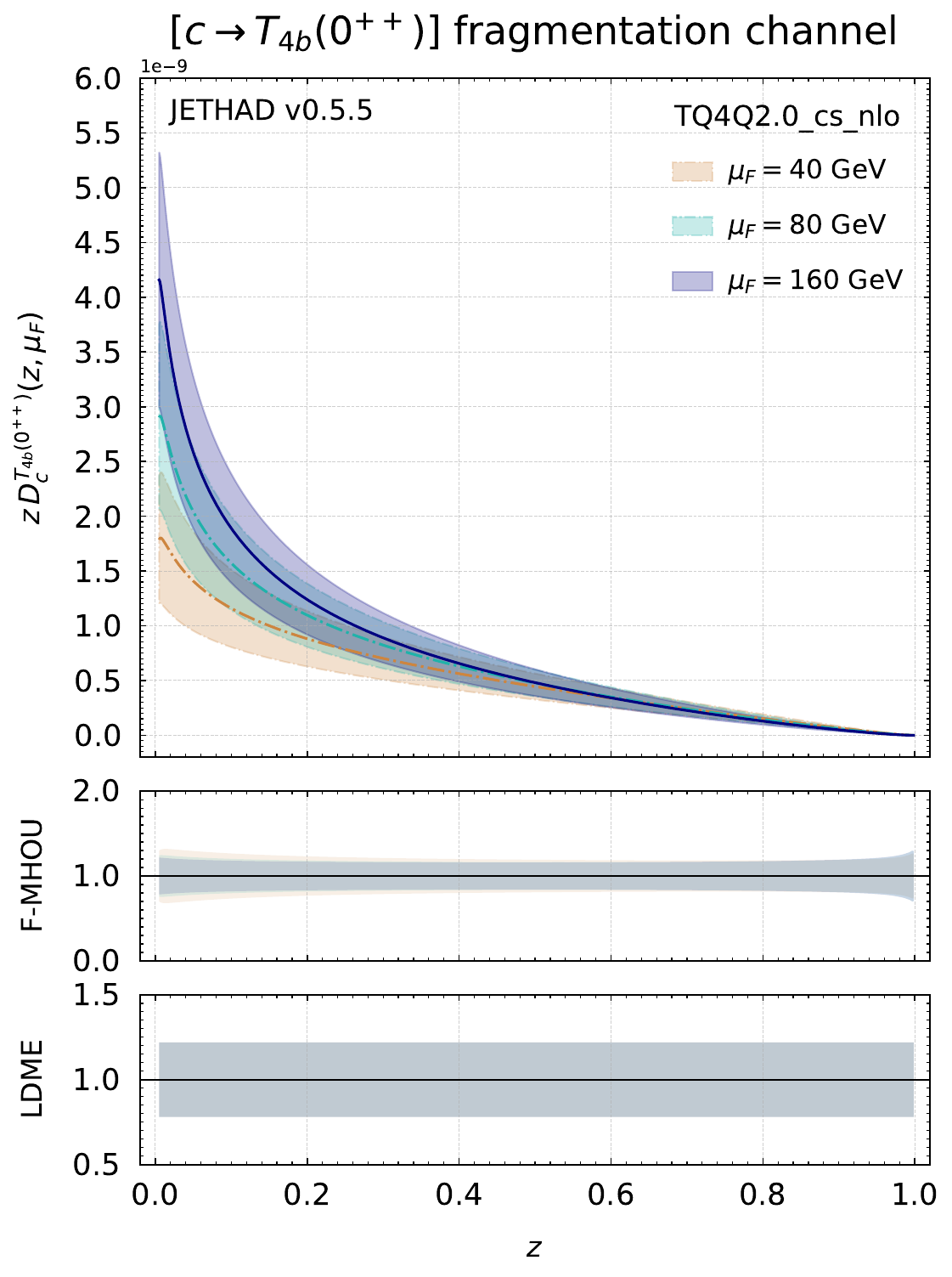}
   \hspace{0.90cm}
   \includegraphics[scale=0.410,clip]{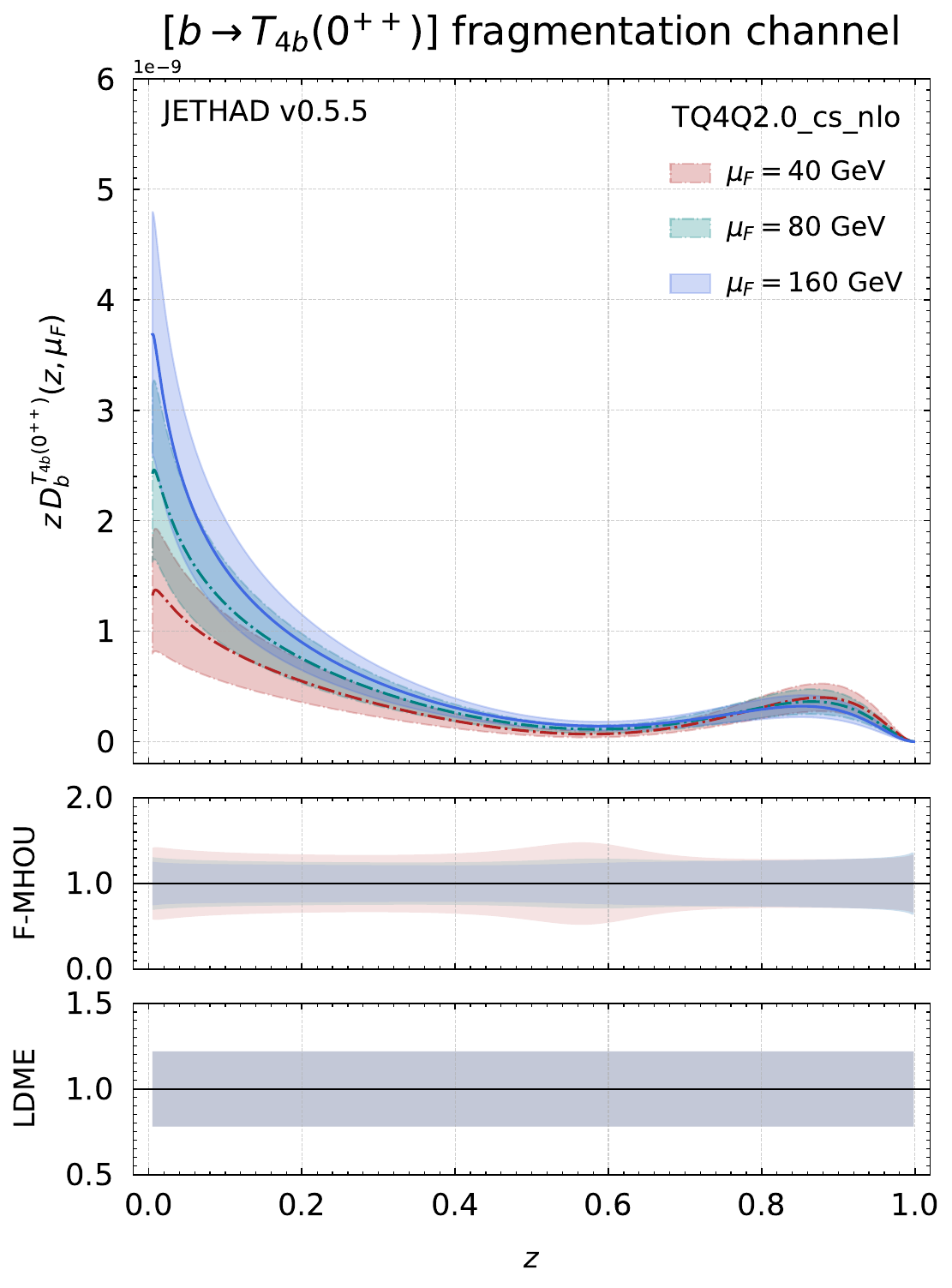}
%   \hspace{0.05cm}

   \vspace{0.25cm}

   \hspace{-0.00cm}
   \includegraphics[scale=0.410,clip]{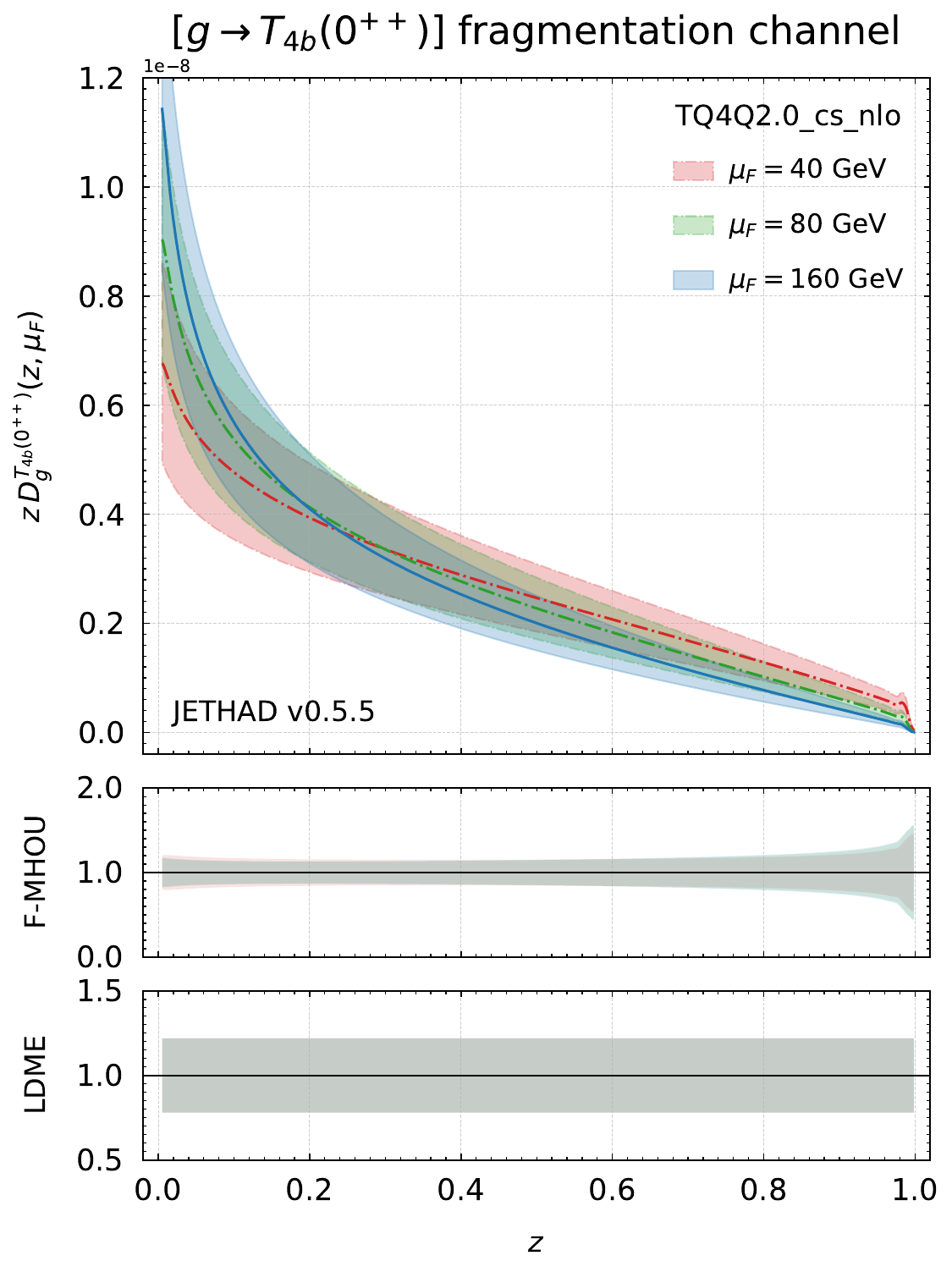}
   \hspace{0.90cm}
   \includegraphics[scale=0.410,clip]{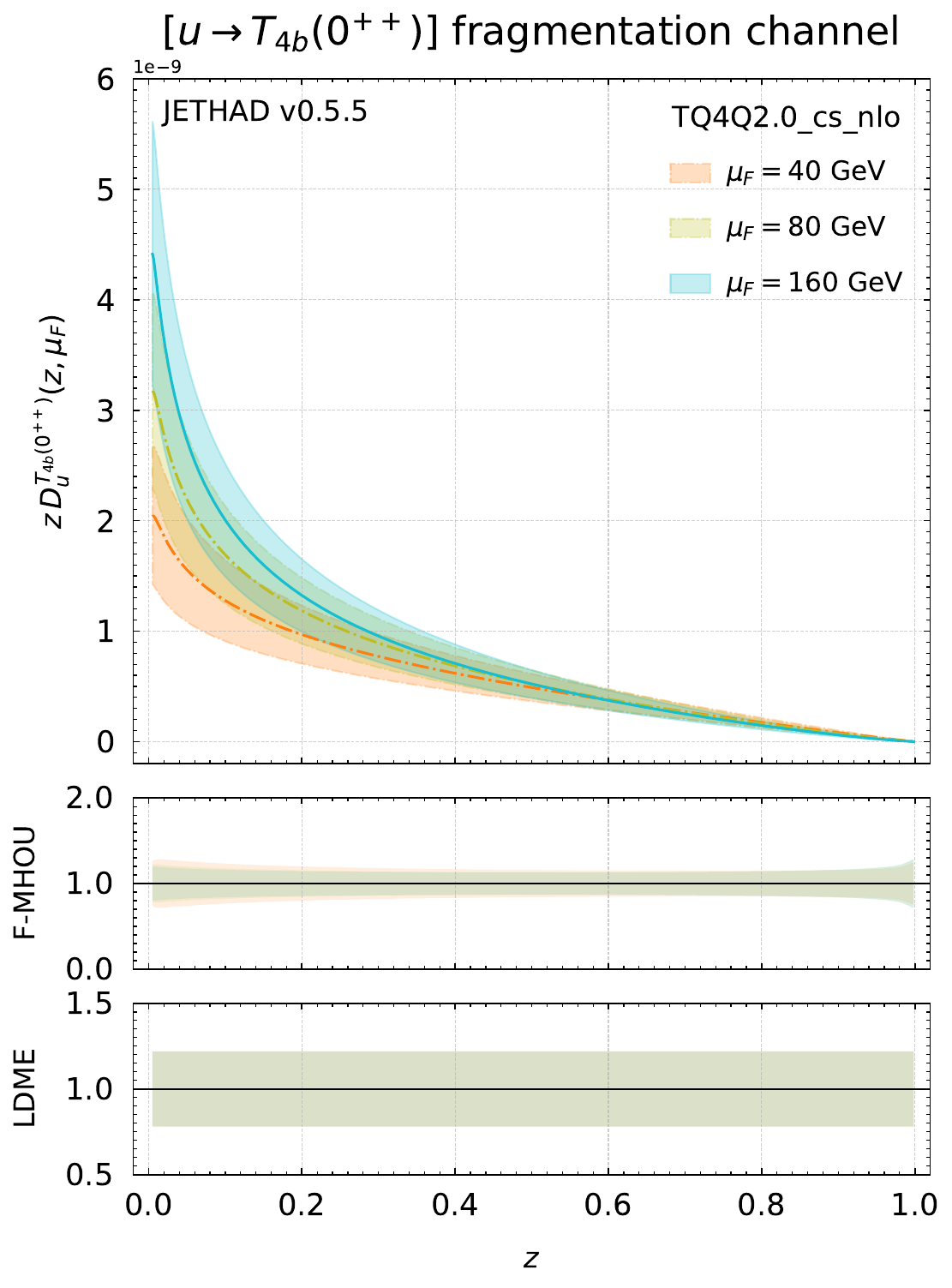}
%   \hspace{0.05cm}

\caption{
\justifying
\noindent
Momentum dependence of the {\tt TQ4Q2.0} FFs for all-bottom scalar tetraquarks, $\TQbZpp$, at different energy scales.
Shaded bands in the main panels denote the total uncertainty, obtained by combining F-MHOUs and LDME variations.
The first auxiliary panel highlights the effect of F-MHOUs through the replica envelope normalized to the central prediction, while the second isolates LDME uncertainties as ratios to the central curve.}
\label{fig:FFs-z_Tb0}
\end{figure*}

\begin{figure*}[!t]
\centering

   \hspace{-0.00cm}
   \includegraphics[scale=0.410,clip]{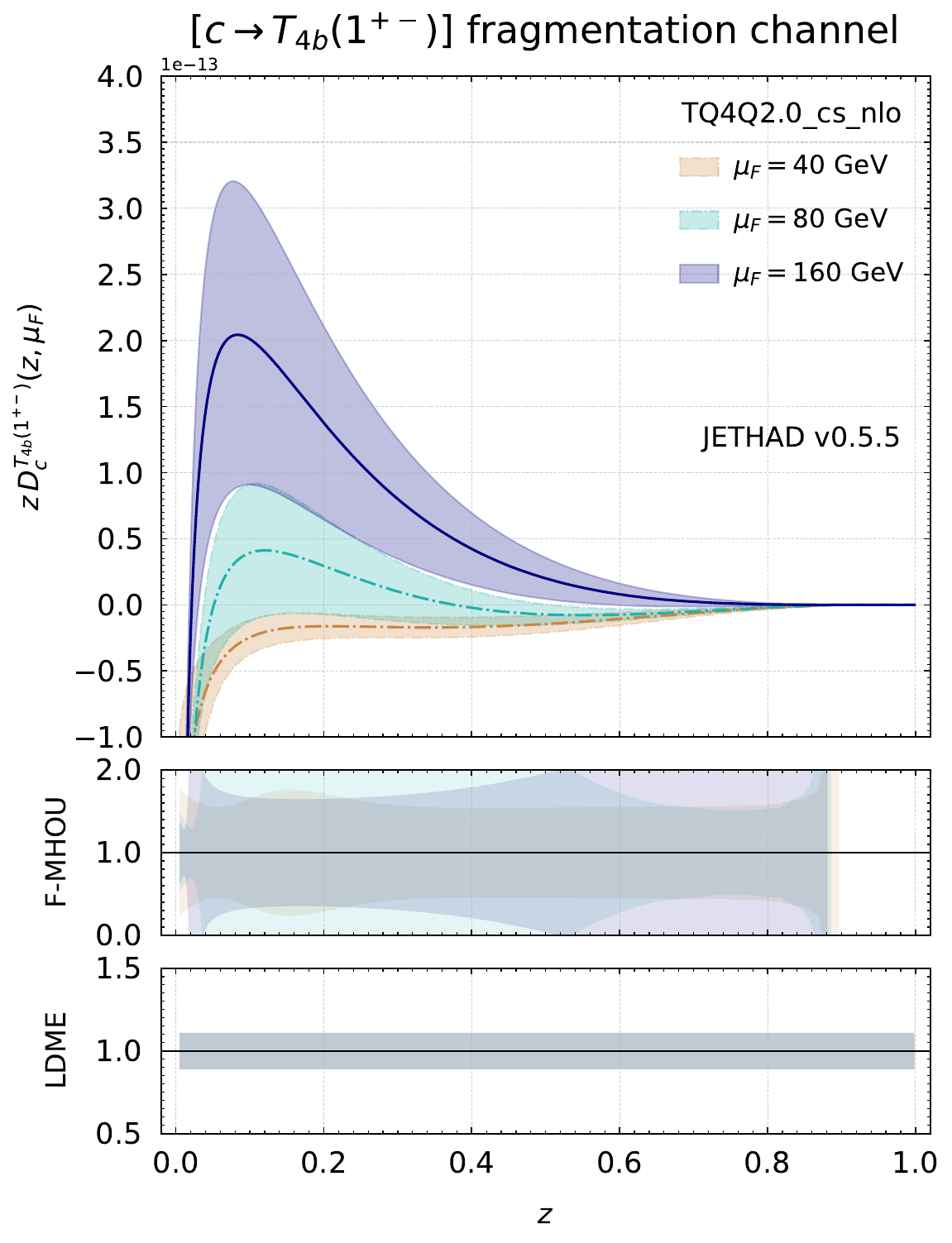}
   \hspace{0.90cm}
   \includegraphics[scale=0.410,clip]{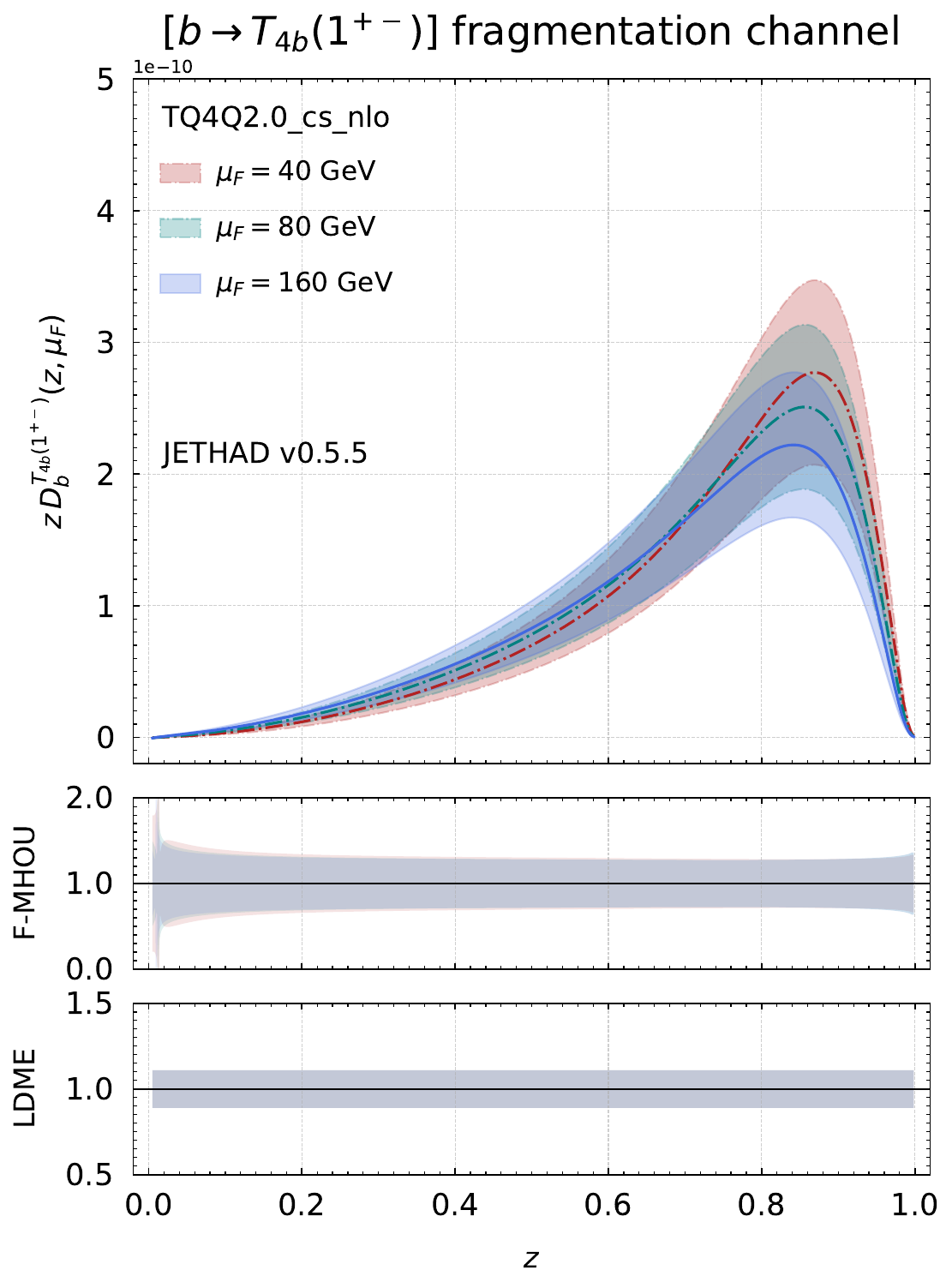}
%   \hspace{0.05cm}

   \vspace{0.25cm}

   \hspace{-0.00cm}
   \includegraphics[scale=0.410,clip]{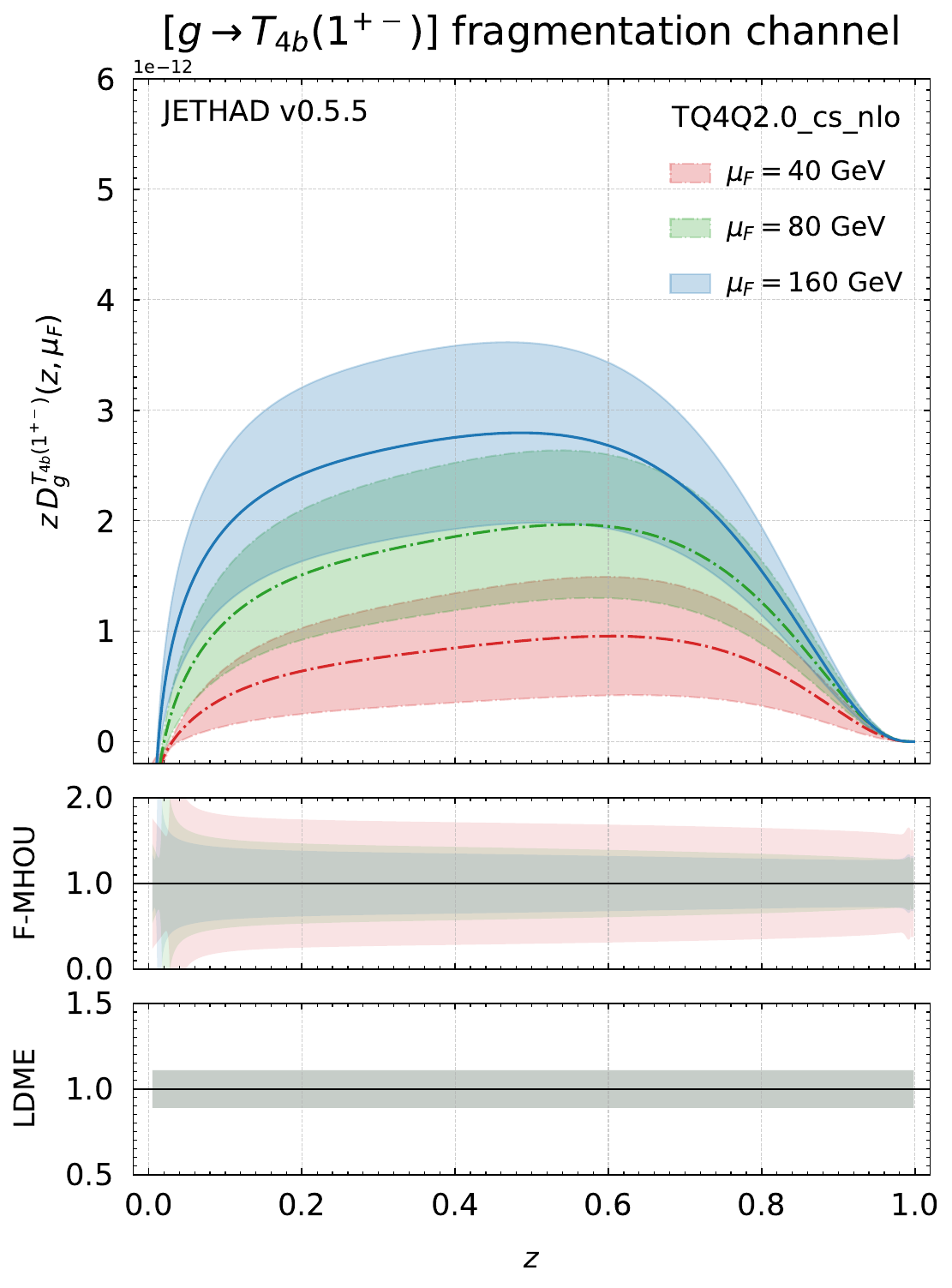}
   \hspace{0.90cm}
   \includegraphics[scale=0.410,clip]{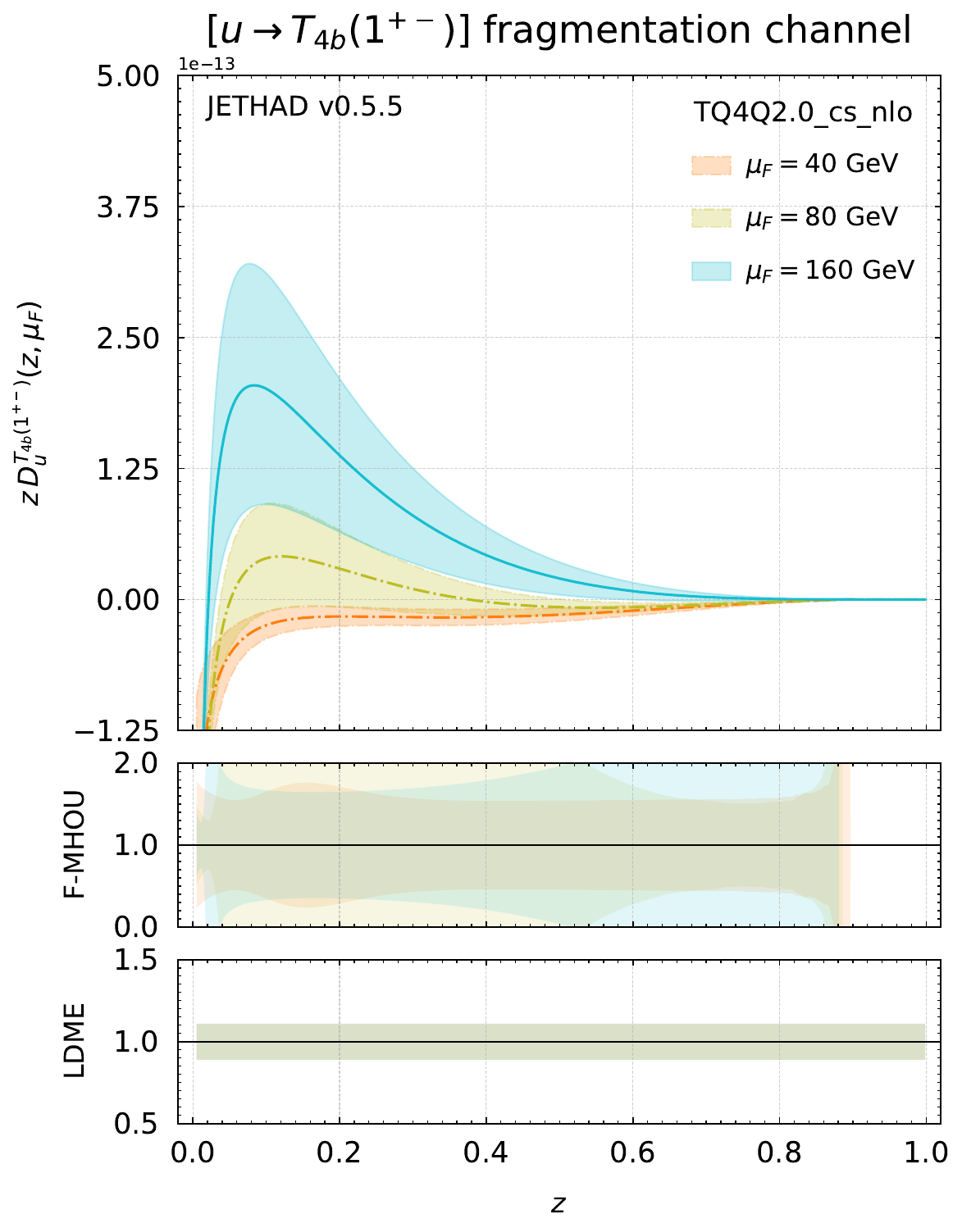}
%   \hspace{0.05cm}

\caption{
\justifying
\noindent
Momentum dependence of the {\tt TQ4Q2.0} FFs for all-bottom axial-vector tetraquarks, $\TQbOpm$, at different energy scales.
Shaded bands in the main panels denote the total uncertainty, obtained by combining F-MHOUs and LDME variations.
The first auxiliary panel highlights the effect of F-MHOUs through the replica envelope normalized to the central prediction, while the second isolates LDME uncertainties as ratios to the central curve.}
\label{fig:FFs-z_Tb1}
\end{figure*}

\begin{figure*}[!t]
\centering

   \hspace{-0.00cm}
   \includegraphics[scale=0.410,clip]{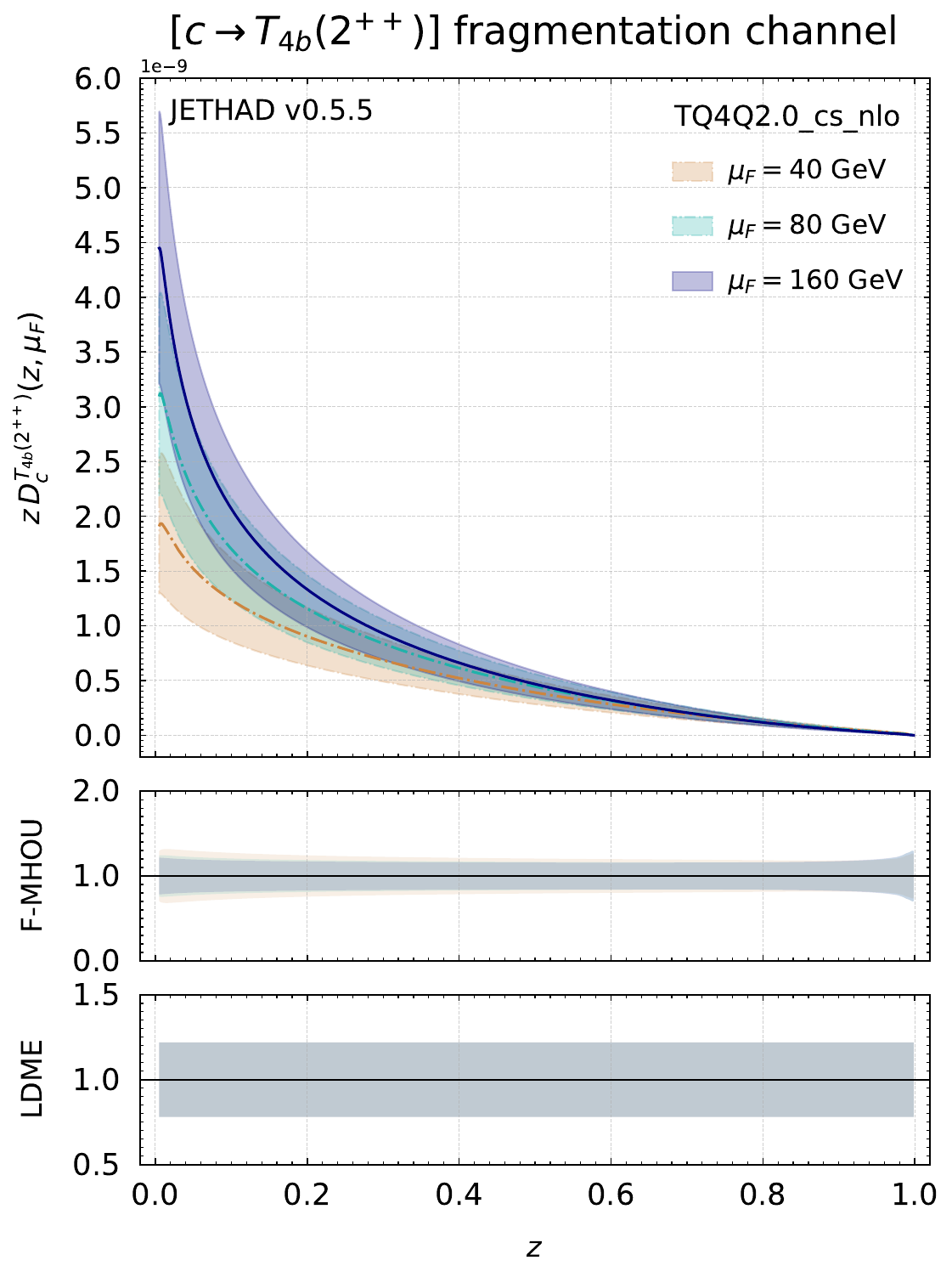}
   \hspace{0.90cm}
   \includegraphics[scale=0.410,clip]{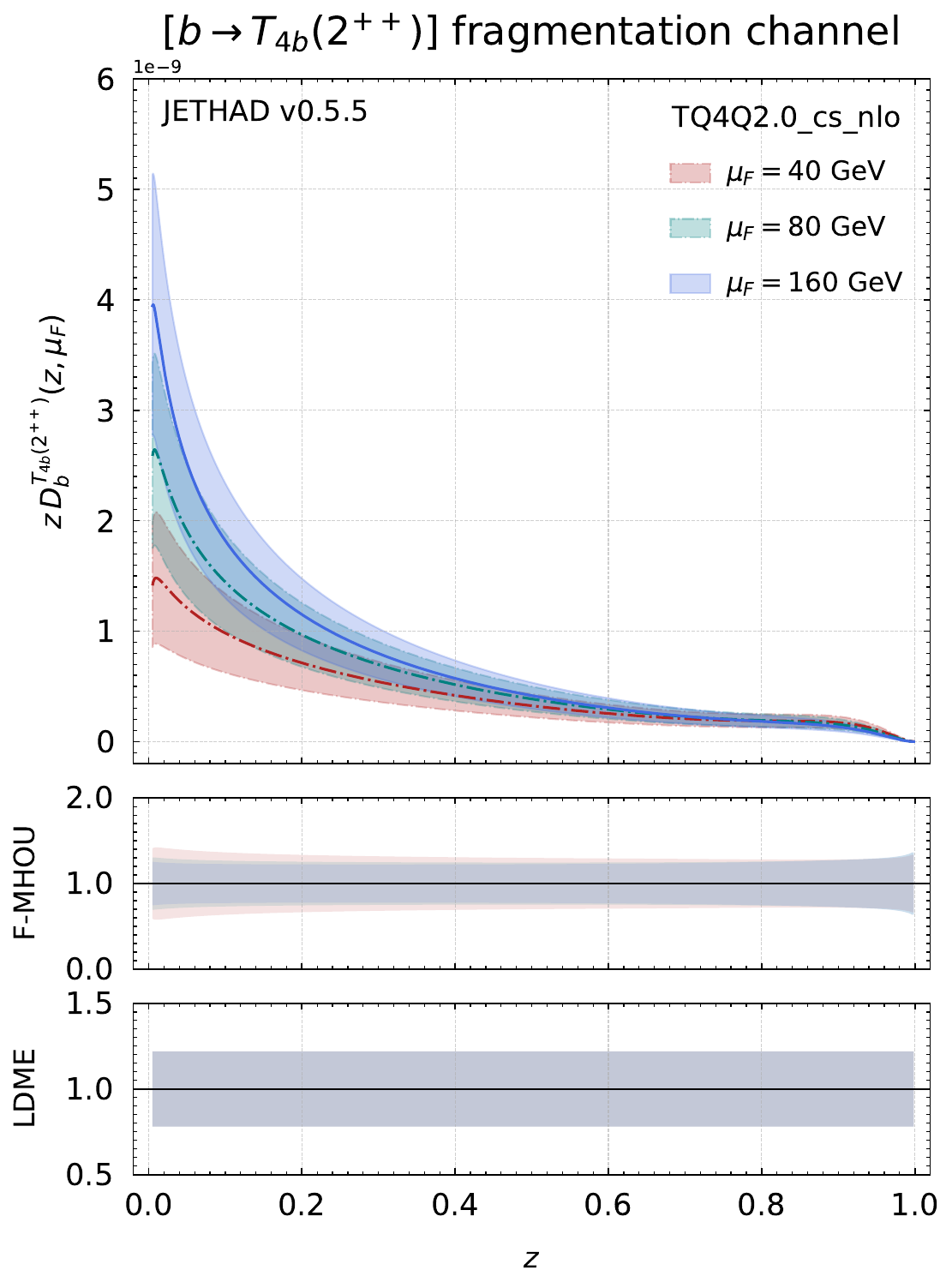}
%   \hspace{0.05cm}

   \vspace{0.25cm}

   \hspace{-0.00cm}
   \includegraphics[scale=0.410,clip]{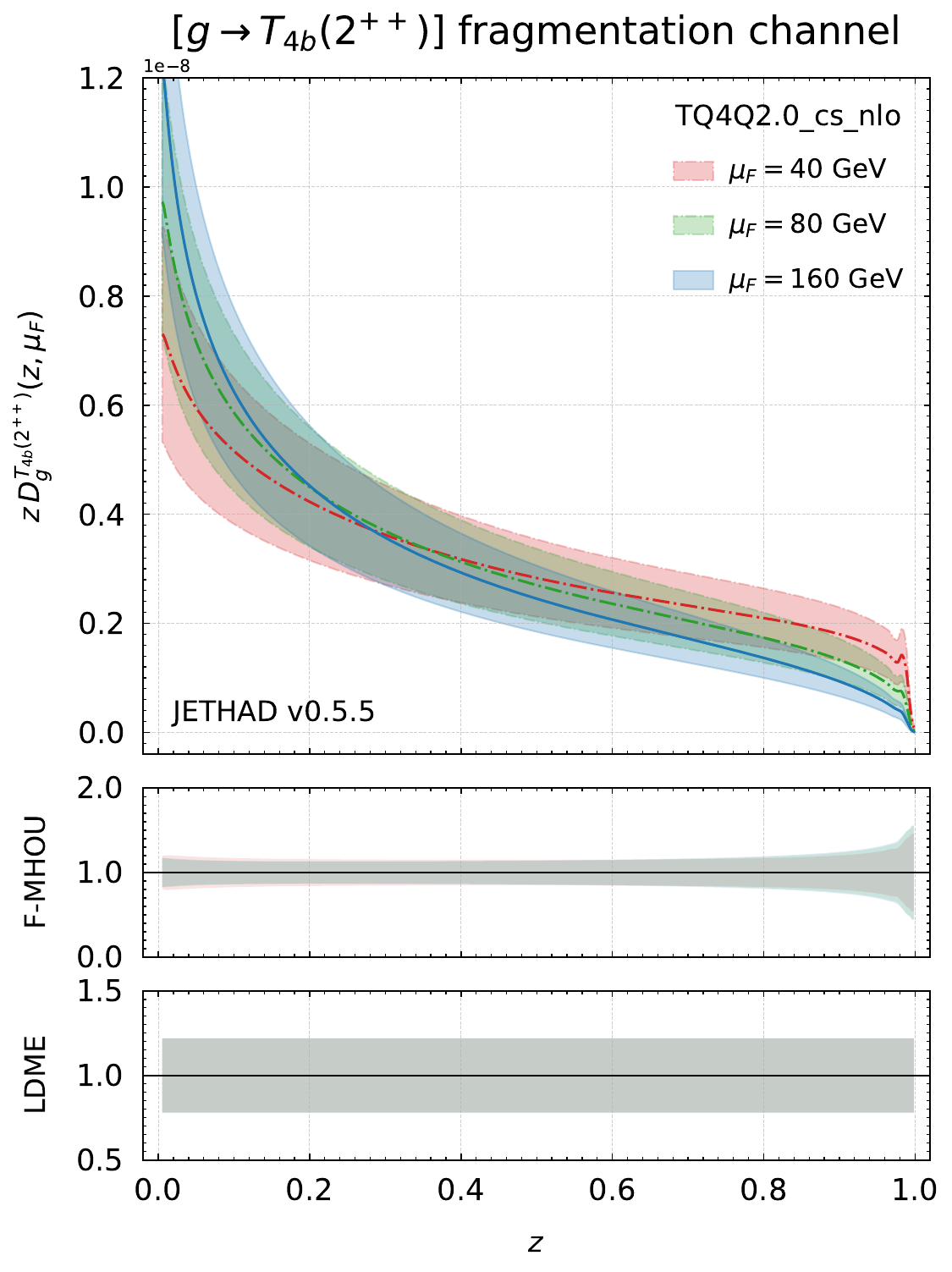}
   \hspace{0.90cm}
   \includegraphics[scale=0.410,clip]{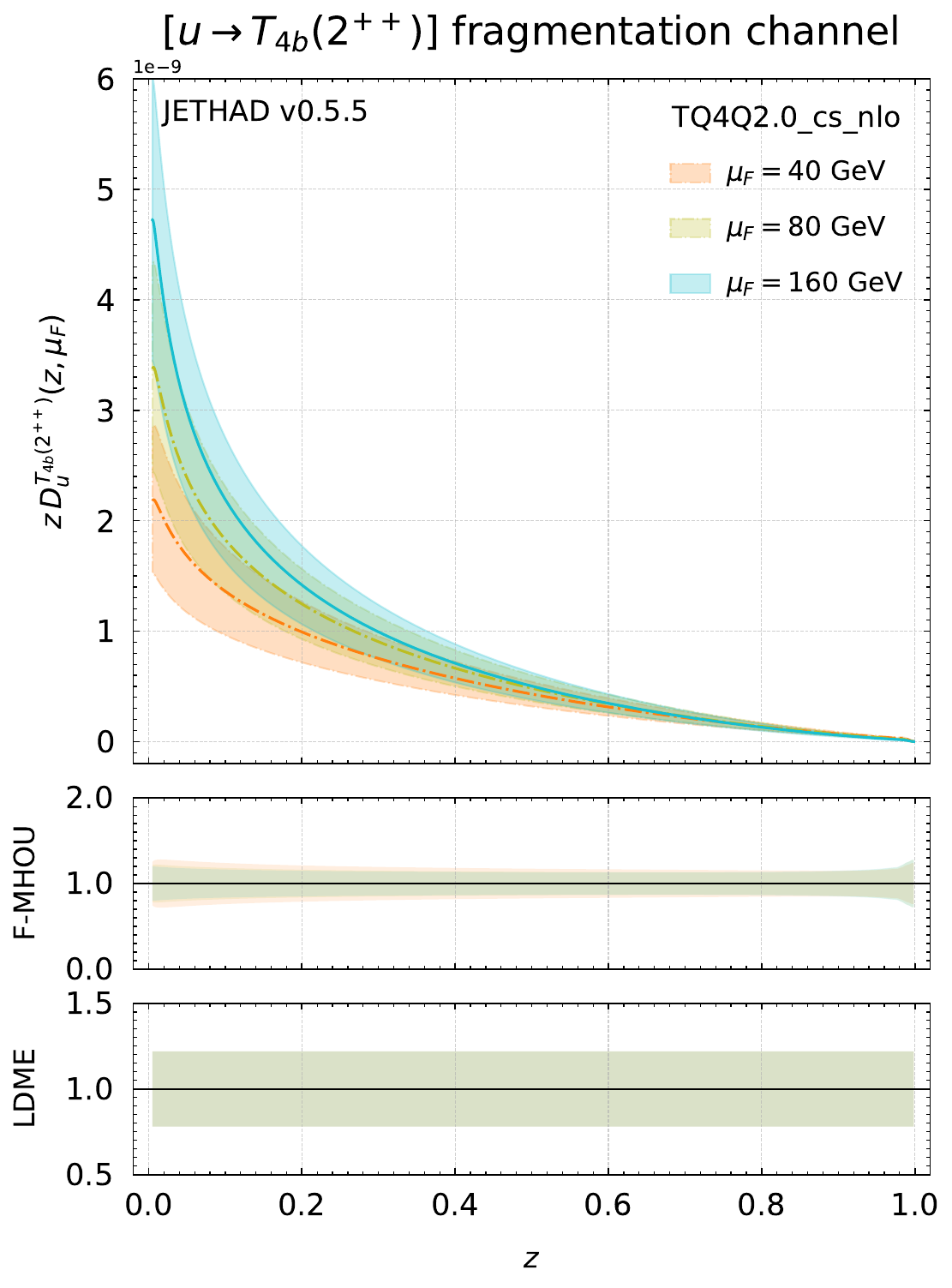}
%   \hspace{0.05cm}

\caption{
\justifying
\noindent
Momentum dependence of the {\tt TQ4Q2.0} FFs for all-bottom scalar tetraquarks, $\TQbTpp$, at different energy scales.
Shaded bands in the main panels denote the total uncertainty, obtained by combining F-MHOUs and LDME variations.
The first auxiliary panel highlights the effect of F-MHOUs through the replica envelope normalized to the central prediction, while the second isolates LDME uncertainties as ratios to the central curve.}
\label{fig:FFs-z_Tb2}
\end{figure*}

The $z$-dependence of the {\tt TQ4Q2.0} FFs, multiplied by $z$, is displayed in Figs.~\ref{fig:FFs-z_Tb0},~\ref{fig:FFs-z_Tb1}, and~\ref{fig:FFs-z_Tb2} for the $0^{++}$, $1^{+-}$, and $2^{++}$ all-bottom tetraquarks.
These distributions represent the direct analog of the all-charm case discussed in Sec.~\ref{ssec:FFs_TQ4Q20}.
For simplicity, only the up-quark channel is shown among light flavors, as mass differences induce negligible effects in both SDCs and evolution.

A general feature emerging from Figs.~\ref{fig:FFs-z_Tb0} to~\ref{fig:FFs-z_Tb2} is the strong suppression of the FFs for all-bottom tetraquarks with respect to their all-charm counterparts.
Across the entire $z$ range, all partonic channels are reduced by approximately three orders of magnitude.

This suppression originates from the larger heavy-quark mass, which decreases the overall fragmentation probability and shifts the dynamics toward harder kinematic configurations.
It overcomes the enhancement expected from the corresponding LDMEs, which are larger for bottomed states by roughly a factor of $400$ (see Sec.~\ref{ssec:FFs_initial_scale}), thus confirming that the overall normalization is primarily controlled by the perturbative SDCs rather than by nonperturbative inputs.

A similar pattern is observed not only for gluon- and heavy-quark-initiated channels, but also for nonconstituent quark contributions, whose relative behavior closely follows that of the all-charm case.
This indicates that mass-driven suppression acts uniformly across all partonic channels, preserving the qualitative hierarchy while significantly reducing the overall normalization.

%\clearpage

%==========================
\hypertarget{app:C}{
\section{NLL/NLO$^+$ HyF description of tetraquark-jet systems}
}
%==========================
\label{app:C}
%\addcontentsline{toc}{section}{\nameref{app:C}}

We study the process
\begin{eqnarray}
\label{process}
 {\rm p}(p_a) + {\rm p}(p_b) \to \TQQ(\kappa_1, y_1) + {\cal X} + {\rm jet}(\kappa_2, y_2) \;,
\end{eqnarray}
where a fully heavy tetraquark, either $\TQc$ or $\TQb$, is tagged in a semi-inclusive configuration together with a light jet, while ${\cal X}$ denotes the accompanying undetected gluon radiation.
The observed final-state objects carry large transverse momenta, $|\vec \kappa_{1,2}| \gg \Lambda_{\rm QCD}$, with $\Lambda_{\rm QCD}$ the characteristic hadronization scale of QCD.
They are also widely separated in rapidity, with $\DY = y_1 - y_2$.

We parametrize the momenta $\kappa_{1,2}$ via a Sudakov decomposition in terms of the incoming proton momenta $p_{a,b}$,
\begin{eqnarray}
\label{sudakov}
\kappa_{1,2} = x_{1,2} p_{a,b} + \frac{\vec \kappa_{1,2}^{\,2}}{x_{1,2} s}p_{b,a} + \kappa_{1,2\perp} \, \qquad
\kappa_{1,2\perp}^2 = - \vec \kappa_{1,2}^{\,2} \;,
\end{eqnarray}
which directly relates longitudinal momentum fractions and transverse components.
In the center-of-mass frame, the rapidities $y_{1,2}$ are connected to the longitudinal fractions through
\begin{eqnarray}
\label{y-vs-x}
y_{1,2} = \pm \ln  \frac{x_{1,2} \sqrt{s}}{|\vec \kappa_{1,2}|} \;.
\end{eqnarray}

Within a purely collinear-factorization setup, the LO differential cross section would take the form of a one-dimensional convolution involving partonic hard factors, proton PDFs, and tetraquark FFs,
\begin{eqnarray}
\label{sigma_collinear}
%\begin{split}
\frac{\drv\sigma^{\rm LO}_{\rm [collinear]}}{\drv x_1\drv x_2\drv ^2\vec \kappa_1\drv ^2\vec \kappa_2}
= \hspace{-0.25cm} \sum_{i,j=q,{\bar q},g}\int_0^1 \hspace{-0.20cm} \drv x_a \!\! \int_0^1 \hspace{-0.20cm} \drv x_b\ f_i\left(x_a\right) f_j\left(x_b\right) 
\int_{x_1}^1 \hspace{-0.15cm} \frac{\drv \zeta}{\zeta} \, D^{\TQQ}_i\left(\frac{x_1}{\zeta}\right) 
\frac{\drv {\hat\sigma}_{i,j}\left(\hat s\right)}
{\drv x_1\drv x_2\drv ^2\vec \kappa_1\drv ^2\vec \kappa_2} \;.
\end{eqnarray}
Here, $i,j$ run over all parton species except the top quark, which does not hadronize.
For brevity, the dependence on the factorization scale $\mu_F$ is not explicitly shown in Eq.~\eqref{sigma_collinear}.
The functions $f_{i,j}(x_{a,b}, \mu_F)$ denote the proton PDFs, while $D^{\TQQ}i(x_1/\zeta, \mu_F)$ encode the fragmentation of parton $i$ into the tetraquark.
The variables $x{a,b}$ represent the incoming parton momentum fractions, whereas $\zeta$ is the fraction carried by the fragmenting parton.
The quantity $\drv\hat\sigma_{i,j}(\hat s)$ is the partonic cross section, with $\hat s = x_a x_b s$.

The differential cross section can be conveniently decomposed in terms of azimuthal harmonics $C_{n \ge 0}$,
\begin{eqnarray}
 \label{dsigmaFourier}
 \frac{\drv \sigma}{\drv \DY \, \drv \varphi \, \drv |\vec \kappa_1| \, \drv |\vec \kappa_2|} 
 =
 \frac{1}{\pi} \left[ \frac{1}{2} C_0 + \sum_{n=1}^\infty \cos (n \varphi)\,
 C_n \right]\;,
\end{eqnarray}
where $\varphi = \phi_1 - \phi_2 - \pi$ and $\phi_{1,2}$ standing for the azimuthal angles of the two outgoing particles.

Adopting the HyF framework and working in the $\MSb$ scheme~\cite{PhysRevD.18.3998}, we derive a master representation for the $C_n$ coefficients that achieves NLO accuracy while resumming NLL high-energy logarithms.
The resulting expression reads
\begin{equation}
\begin{split}
\label{CnNLL}
 C_n^{\NLLp} \;&=\; 
 \int_{\kappa_1^{\rm min}}^{\kappa_1^{\rm max}} \drv |\vec \kappa_1|
 \int_{\kappa_2^{\rm min}}^{\kappa_2^{\rm max}} \drv |\vec \kappa_2|
 \int_{y_1^{\rm min}}^{y_1^{\rm max}}
 \drv y_1
 \int_{y_2^{\rm min}}^{y_2^{\rm max}} 
 \drv y_2
 \; \delta(y_1 - y_2 -\DY)
 \int_{-\infty}^{+\infty} \drv \nu \, e^{\bar \alpha_s \DY \chi^{\rm NLL}(n,\nu)}
\\[0.20cm]
 &\times \,
 \frac{e^{\DY}}{s}
 \alpha_s^2(\mu_R)
 \left\{ 
 \F_1^{\rm NLO}(n,\nu,|\vec \kappa_1|, x_1)[\F_2^{\rm NLO}(n,\nu,|\vec \kappa_2|,x_2)]^*
 + \bar \alpha_s^2 \frac{\beta_0 \DY}{4 N_c}\chi(n,\nu)\upsilon(\nu)
 \right\} \;,
\end{split}
\end{equation}
where $\bar \alpha_s(\mu_R) = \alpha_s(\mu_R) N_c/\pi$ and $\beta_0 = 11N_c/3 - 2 n_f/3$.
We adopt a two-loop running coupling with $\alpha_s\left(M_Z\right)=0.118$ and a scale-dependent number of active flavors $n_f$.

The function $\chi(n,\nu)$ entering the exponent is the BFKL kernel~\cite{Fadin:1975cb,Balitsky:1978ic}, which resums energy logarithms at NLL accuracy,
\begin{eqnarray}
 \label{chi}
 \chi^{\rm NLL}(n,\nu) = \chi(n,\nu) + \bar\alpha_s \hat \chi(n,\nu) \;,
\end{eqnarray}
with LO eigenvalues
\begin{eqnarray}
\chi\left(n,\nu\right) = -2\left\{\gamma_{\rm E}+{\rm Re} \left[\psi\left( (n + 1)/2 + i \nu \right)\right] \right\}
\label{chiLO}
\end{eqnarray}
and NLO corrections
\begin{align}
\label{chiNLO}
\hat \chi\left(n,\nu\right) \;&=\; \bar\chi(n,\nu)+\frac{\beta_0}{8 N_c}\chi(n,\nu)
%\\ \nonumber &\times \,
\left\{-\chi(n,\nu)+2\ln\left(\mu_R^2/\hat{\mu}^2\right)+\frac{10}{3}\right\} \;,
\end{align}
Here $\psi(z)=\Gamma^\prime(z)/\Gamma(z)$ and $\hat{\mu}=\sqrt{|\vec \kappa_1| |\vec \kappa_2|}$.
The full expression for $\bar\chi(n,\nu)$ can be found in Sec.2.1.1 of Ref.~\cite{Celiberto:2020wpk}.

The quantities
\begin{eqnarray}
\label{EFs}
\F_{1,2}^{\rm NLO}(n,\nu,|\vec \kappa|,x) =
\F_{1,2}(n,\nu,|\vec \kappa|,x) +
\alpha_s(\mu_R) \, \hat \F_{1,2}(n,\nu,|\vec \kappa|,x)
\end{eqnarray}
represent the NLO singly off-shell emission functions, namely forward-production impact factors in the BFKL formalism.
Tetraquark production is described through the forward-hadron impact factor computed in~\cite{Ivanov:2012iv}.
Although derived for light hadrons, its use remains justified in our VFNS framework~\cite{Mele:1990cw,Cacciari:1993mq}, provided that transverse masses are well above heavy-quark thresholds.

At LO, the tetraquark emission function reads
\begin{equation}
\begin{split}
\label{LOHEF}
\F_{\TQQ}(n,\nu,|\vec \kappa|,x) \;&=\; 2 \sqrt{\frac{C_F}{C_A}} \; |\vec \kappa|^{2i\nu-1}\int_{x}^1 \frac{\drv \zeta}{\zeta} \left( \frac{x}{\zeta} \right)^{1-2i\nu} 
%\nonumber 
\\
 &\times \, 
 \left[ \frac{C_A}{C_F} f_g(\zeta, \mu_F)D_g^{\TQQ}\left( \frac{x}{\zeta}, \mu_F \right)
 +\sum_{i=q,\bar q}f_i(\zeta, \mu_F)D_i^{\TQQ}\left( \frac{x}{\zeta}, \mu_F \right) \right] \;,
\end{split}
\end{equation}
where $C_F=(N_c^2-1)/(2N_c)$ and $C_A=N_c$.
The full NLO correction is given in~\cite{Ivanov:2012iv}.

The LO jet emission function is instead
\begin{eqnarray}
 \label{LOJEF}
 \hspace{-0.09cm}
 \F_J(n,\nu,|\vec \kappa|,x) = 2 \sqrt{\frac{C_F}{C_A}} \;
 |\vec \kappa|^{2i\nu-1}\,\hspace{-0.05cm} \left[ \frac{C_A}{C_F} f_g(x, \mu_F)
 +\hspace{-0.15cm}\sum_{j=q,\bar q}\hspace{-0.10cm}f_j(x, \mu_F) \right] \;.
\end{eqnarray}
with NLO corrections taken from~\cite{Colferai:2015zfa}.
We adopt a small-cone approximation with $r_J=0.5$, consistent with CMS analyses~\cite{Khachatryan:2016udy,Khachatryan:2020mpd,CMS:2021maw}.

The remaining term in Eq.~\eqref{CnNLL} is
\begin{eqnarray}
 \upsilon(\nu) = \frac{1}{2} \left[ 4 \ln \hat{\mu} + i \frac{\drv}{\drv \nu} \ln\frac{\F_1(n,\nu,|\vec \kappa_1|, x_1)}{\F_2[(n,\nu,|\vec \kappa_1|, x_1)]^*} \right] \;.
\label{fnu}
\end{eqnarray}

Equations~\eqref{CnNLL}--\eqref{LOJEF} fully specify our hybrid-factorization setup.
In the high-energy limit, the cross section factorizes into the convolution of the BFKL Green's function with two off-shell emission functions, which embed collinear dynamics through PDF--FF convolutions.
The collinear ingredients are treated at NLO accuracy, including DGLAP evolution for both PDFs and FFs.
The high-energy sector is described within NLL BFKL resummation, namely through the inclusion of the NLO BFKL kernel, which resums next-to-leading logarithms in energy, together with NLO forward-production impact factors entering the off-shell emission functions.
Accordingly, the $\NLLp$ notation denotes a NLL high-energy resummation consistently matched to NLO collinear dynamics, while the `$+$' indicates additional subleading contributions generated by products of NLO impact factors.

For comparison, we also consider the LL approximation in the $\MSb$ scheme, obtained by neglecting NLO corrections in both the kernel and emission functions,
\begin{eqnarray}
\label{CnLL}%\nonumber
 C_n^{\LL} \propto 
 \frac{e^{\DY}}{s} 
 \int_{-\infty}^{+\infty} \drv \nu \, 
 e^{\bar \alpha_s \DY \chi(n,\nu)} %\,
 \alpha_s^2(\mu_R) \, \F_{\TQQ}(n,\nu,|\vec \kappa_1|, x_1)[\F_J(n,\nu,|\vec \kappa_2|,x_2)]^* \;.
\end{eqnarray}
Here, integrations over phase space are understood implicitly.

A quantitative assessment of high-energy resummation requires comparison with fixed-order predictions.
However, no available numerical tool currently provides full NLO predictions for observables involving two identified final-state particles.
We therefore construct a fixed-order baseline by expanding Eq.~(\ref{CnNLL}) up to ${\cal O}(\alpha_s^3)$, defining an effective $\HENLOp$ approximation.
This retains the dominant high-energy logarithms while discarding subleading power-suppressed contributions.

The corresponding $\MSb$-scheme expression reads
\begin{align}
\label{CnHENLO}%\nonumber
 C_n^{\HENLOp} &\propto %\frac{x_1 x_2}{|\vec \kappa_1| |\vec \kappa_2|} 
 \frac{e^{\DY}}{s} 
 \int_{-\infty}^{+\infty} \drv \nu \, 
 %e^{{\DY} \bar \alpha_s(\mu_R) \chi^{\rm NLO}(m,\nu)}
 \alpha_s^2(\mu_R) \,
 \left[ 1 + \bar \alpha_s(\mu_R) \DY \chi(n,\nu) \right] \,
% \end{equation}
%\[
 %\\ \nonumber
 %&\times
 \F_{\TQQ}^{\rm NLO}(n,\nu,|\vec \kappa_1|, x_1)[\F_J^{\rm NLO}(n,\nu,|\vec \kappa_2|,x_2)]^* \;.
% \]
\end{align}
The exponentiated kernel is expanded to first order in $\alpha_s$, while integrations over kinematics are again left implicit.

\twocolumngrid

%%%%%%%%%%%%%%%%%%

\bibliographystyle{apsrev4-1} %apsrev4-2 gives an error when "journal" entry is missing in a reference
\bibliography{bibliography}

\end{document}